%% file: Paper.tex
\documentclass[aps,pre,reprint,twocolumn,nofootinbib]{revtex4-1}

\usepackage{subfigure}
\usepackage{amsmath}    % need for subequations
\usepackage{amssymb}
\usepackage{graphicx}   % need for figures
\usepackage{verbatim}   % useful for program listings
\usepackage{color}      % use if color is used in text
\usepackage{subfigure}  % use for side-by-side figures
\usepackage{hyperref}   % use for hypertext links, including those to external documents and URLs
\usepackage{algpseudocode,algorithm,algorithmicx}

\newcommand*\aivd{\text{AIVD}}
\newcommand*\sivd{\text{SIVD}}

\newcommand*\fit{\text{fit}}
\newcommand*\err{\text{err}}
\newcommand*\avg{\text{avg}}
\newcommand*\stat{\text{stat}}
\newcommand*\syst{\text{syst}}
\newcommand*\emp{\text{emp}}

\algrenewcommand\algorithmicrequire{\textbf{Precondition:}}
\algrenewcommand\algorithmicensure{\textbf{Postcondition:}}

\newcommand\eqd{\mathrel{\stackrel{\makebox[0pt]{\mbox{\normalfont\tiny d}}}{=}}}

\begin{document}

\title{Evidence for mixed rationalities in preference formation}

\author{Alexandru-Ionu\c{t} B\u{a}beanu}
\author{Diego Garlaschelli}
%\affil[1]{Lorentz Institute for Theoretical Physics, Leiden University, The Netherlands}
\affiliation{Lorentz Institute for Theoretical Physics, Leiden University, The Netherlands}
\date{\today}

\begin{abstract}

Understanding the mechanisms underlying the formation of cultural traits, such as preferences, opinions and beliefs is an open challenge. 
Trait formation is intimately connected to cultural dynamics, which has been the focus of a variety of quantitative models.
Recently, some studies have emphasized the importance of connecting those models to snapshots of cultural dynamics that are empirically accessible. 
By analyzing data obtained from different sources, it has been suggested that culture has properties that are universally present, 
and that empirical cultural states differ systematically from randomized counterparts.
Hence, a question about the mechanism responsible for the observed patterns naturally arises.
This study proposes a stochastic structural model for generating cultural states that retain those robust, empirical properties.
One ingredient of the model, already used in previous work, assumes that every individual's set of traits is partly dictated by one of several, universal ``rationalities'', informally postulated by several social science theories. 
The second, new ingredient taken from the same theories assumes that, apart from a dominant rationality, each individual also has a certain exposure to the other rationalities. 
It is shown that both ingredients are required for reproducing the empirical regularities. 
This key result suggests that the effects of cultural dynamics in the real world can be described as an interplay of multiple, mixing rationalities, 
and thus provides indirect evidence for the class of social science theories postulating such mixing. 
The model should be seen as a static, effective description of culture, while a dynamical, more fundamental description is left for future research. 

\end{abstract}

\maketitle

\section{Introduction}
\label{Intr}

A solid theoretical understanding of how preferences form is currently lacking.
There is little doubt that preferences, opinions, values and beliefs, which are generically referred to as ``cultural traits'', are dynamical entities, 
and that interpersonal social influence plays an important role in driving their dynamics, among other factors. 
Moreover, a complete theoretical understanding should account for the fact that the dynamics of traits takes place in parallel along multiple dimensions, 
namely that opinions and preferences can develop in relation to multiple topics or aspects of life. 
Along these lines, various dynamical models been developed and studied~\cite{Castellano}, such as the Axelrod model~\cite{Axelrod}, 
which is very representative for studies of multidimensional dynamics, commonly referred to as ``cultural dynamics'',
in contrast to studies of unidimensional dynamics, commonly referred to as ``opinion dynamics''. 
Various studies of cultural dynamics extending the Axelrod model can be found in the literature~\cite{Klemm_1, Klemm_2, Kuperman, Flache, Gonzalez-Avella, Centola, Pfau, Battiston, Stivala_2}.
Recent studies~\cite{Valori, Stivala, Babeanu} have shown that models of cultural dynamics are sensitive to the initial conditions, namely to how the initial vectors of agents' traits are chosen:
initial cultural states constructed from empirical data show systematic deviations from their shuffled and random counterparts. 
In fact, Ref.~\cite{Babeanu} argues that such deviations point towards universal structural properties inherent in empirical cultural states.
More insights about the formation of cultural traits should be achievable by studying these states, 
since they can be regarded as partial snapshots of cultural dynamics in the real world.

The universal properties mentioned above are expressed in terms of the effects the empirical cultural state has on social influence models using it for their initial conditions --
here, a ``cultural state'' is a set of cultural vectors (SCV), where each cultural vector encodes the sequence of cultural traits associated to one agent in the model.
On one hand, an Axelrod-type model~\cite{Axelrod} of (multi-dimensional) cultural dynamics is used to evaluate the propensity of the cultural state to long-term cultural diversity (LTCD).
On the other hand, a Count-Bouchaud-type model~\cite{Cont} of (one-dimensional) opinion dynamics is used to evaluate the propensity of the cultural state to short-term collective behavior (STCB).
Both measures are functions of a common parameter $\omega$, controlling for the range of social influence in cultural space, which allows for an LTCD-STCB correspondence to be drawn for a given cultural state. 
It turns out that an empirical cultural state generally induces an LTCD-STCB curve that is close to the second diagonal ($\text{LTCD}(\omega) \approx 1-\text{STCB}(\omega), \forall \omega$), 
while exhibiting, for a given STCB value, higher LTCD values than a trait-shuffled cultural state, which in turn exhibits higher LTCD values than a randomly generated counterpart~\cite{Valori, Babeanu}.
These results seem universal~\cite{Babeanu}, namely independent of the data set used for constructing the cultural vectors composing the empirical cultural state, 
suggesting that real-world cultural dynamics is governed by universal laws. 
Moreover, as argued in Ref.~\cite{Babeanu}, this type of analysis suggests that inter-agent social influence, the essential ingredient of cultural dynamics models, 
is insufficient for explaining the observed structure.
Although it is meaningful to incorporate additional ingredients into social influence models, while attempting to give rise to empirical-like structure in a dynamical setting,
this study does not aim for that.
Instead, it aims at providing an effective, phenomenological, static description of the observed structure, which should provide additional insights before developing a more fundamental, dynamical description. 

The purpose of this study is to develop a structural stochastic model that would generate realistic cultural states, while incorporating plausible ingredients from social science.
Specifically, these states should retain the universal properties inherent to empirical cultural states that are observed in Ref.~\cite{Babeanu}.
In fact, Ref.~\cite{Stivala} has already investigated various ways of generating sets of cultural vectors in random, but non-uniform ways. 
A method that appeared particularly promising relied on the notion of ``cultural prototypes'':
a few underlying, abstract sequences of logically compatible, self-enforcing cultural traits, which govern the way the generated vectors are distributed in cultural space.
According to the method, each cultural vector is partly a copy of one of the prototypes and partly random.
The implicit claim is that each cultural prototype is induced by one of a few (3 to 5) fundamental and universal ``principles of social life'', or ``rationalities'', that would strongly affect any process of trait formation in any social system. 
Such entities are postulated, under different names and in slightly different numbers, by several theoretical frameworks in social science~\cite{Thompson, Fiske, Triandis, Shweder, Graham}.
The exact number of such entities depends on the exact theory that is considered, as different theories are built on somewhat different arguments and pieces of evidence.
It is important that the number is larger than 1 but not too large, while independent of system size. 
From a natural science perspective, such ideas are attractive, since they exhibit a certain reductionist tendency of trying to understand the observed socio-cultural variability in terms of combinations of a few, elementary and universal building blocks. 
Various parallels and similarities between these theories are discussed in the literature~\cite{Verweij, Bruce, Verweij_2}. 
For the purpose of the current study, all these theories are equivalent.
Still, for creating an instructive and compact context, the discussion is restricted to one of them, namely to Plural Rationality Theory, chosen for reasons discussed in Sec.~\ref{Disc}. 

Plural Rationality Theory (PRT), also referred to as ``(Grid-Group) Cultural Theory''~\cite{Thompson}, is a qualitative description of socio-cultural structure and dynamics as an interplay between a small number of irreducible ``ways of life'', or ``rationalities''.
These ways of life are understood as abstract, ``elementary building blocks'' of societies and are supposedly recognizable regardless of the geographical context, of the historical context or of the scale of the system that is studied.
It is believed that the ways of life go along with different perceptions of risk~\cite{Rayner, Tansey} and, interestingly,
that they always coexist, although either of them is often dominant for a given period of time, for a given (part of) the system that one studies\footnote
{It may be useful to think of the ways of life as being the elements of a complete, orthogonal basis of some abstract vector space. 
One may then associate a vector in this space to a certain part of a certain socio-cultural system, at a given moment in time.
It is not clear to what extent such vectors would be related to the cultural vectors used in this study.
This is only a semi-formal analogy that is not exploited further here, nor in any other study so far, to the extent that the authors are aware of.}.
Such ideas appear compatible with recent empirical findings concerning the existence of a small number of behavioural phenotypes in dyadic games~\cite{Poncela-Casasnovas}. 
In PRT, each way of life is understood as a self-enforcing combination of a ``pattern of (social) relations'' and a ``cultural bias''.
On one hand, a pattern of relations is often understood as a tendency of organizing the social ties between people in a certain way, thus a connectivity pattern in the social graph. 
On the other hand, a cultural bias is a combination of preferences, opinions, values and beliefs that are compatible with each other and with the associated pattern of relations. 
By comparison to the definitions in Ref.~\cite{Babeanu}, one can easily realize that a cultural bias can be thought of as a point or a region in ``cultural space'' that is representative for the respective ``way of life''.
A cultural bias is formally represented here by the notion of ``cultural prototype'', previously used in Ref.~\cite{Stivala}.

This notion is at the core of two stochastic, structural models of culture that are defined and studied here.
The first model, called ``Prototype Generation'' (PG), postulates that each cultural vector is partly a copy of one of the $k$ prototypes and partly random. 
This generation method is similar to the ``Prototype Evolution'' method of Ref.~\cite{Stivala},
although with small technical differences.
The second model, called ``Mixed Prototype Generation'' (MPG), postulates that each cultural vector is an asymmetric mixture (or combination) of all the prototypes.
From the perspective of PRT, this ``mixing'' is a formal realization of the idea that every person combines the ways of life in a unique way,
such that preferences and opinions related to different aspects of life -- cultural traits of different cultural features (or variables) -- are due to the ``influence'' of different cultural biases,
though at any given moment in time one cultural bias is usually dominating. 
In the literature concerned with PRT and the other, similar, theories, this mixing aspect often goes under the name of ``the multiple self'', 
and was not implemented in Ref.~\cite{Stivala}.
The importance of mixing for correctly interpreting (and testing) PRT has been already stressed out~\cite{Tansey},
while the general importance of multiple selves for social science has also been extensively discussed~\cite{Elster}.
Moreover, research on preferences in economic contexts also suggests that the multiple self is important~\cite{Fehr_1, Fehr_2, Fehr_3}.
On the other hand, research in cross-cultural psychology appears to be divided: 
some studies seem to ignore the multiple self~\cite{Schwartz}, while others seem to acknowledge it~\cite{Nisbett, Kuhnen}.
This study provides further insights on this matter, by directly comparing the PG and MPG models with each other and with empirical data,

Sec.~\ref{ModDesc} explains the models in detail,
while Sec.~\ref{ModFit} describes how the free parameters are tuned, as to reproduce some lower-order properties of one empirical cultural state.
Cultural states generated with the two models are then evaluated, in Sec.~\ref{ModOut}, by means of the LTCD-STCB analysis of Refs.~\cite{Valori,Babeanu}.
It is shown that cultural states generated by PG are structurally dissimilar to the empirical ones, as they do not exhibit the universal LTCD-STCB behavior, after tuning the free parameters to empirical data in terms of simpler, but meaningful quantities.
On the other hand, cultural states generated with MPG are structurally similar to the empirical ones, as they reproduce the universal LTCD-STCB behavior, after applying an analogous tuning procedure.
This suggests that the mixing, multiple self ingredient is crucial for describing the effects of preference formation in terms of cultural prototypes,
and that MPG should be regarded as the successful model. 
Sec.~\ref{Disc} further discusses the results, their limitations, as well as extensions of this work and questions that are worth investigating in the future.
The manuscript is concluded in Sec.~\ref{Conc}. 

\section{Model description}                  
\label{ModDesc}

This section describes the two stochastic models of culture: the Prototype Generation (PG) model and the Mixed Prototype Generation (MPG) model,
which are used below for generating sets of cultural vectors (SCVs) that can be quantitatively studied with the LTCD-STCB tool, 
previously applied to empirical SCVs in Refs.~\cite{Valori, Babeanu}.
Both models rely on the concept of cultural prototype introduced above.

\begin{figure*} 
\centering
    \def\svgwidth{16.0cm}
    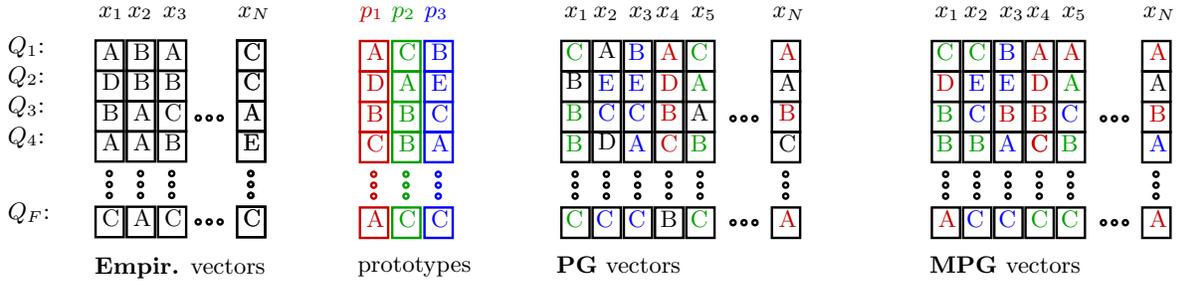 
    \caption{Schematic illustration of the two stochastic models, showing (from left to right): 
    an empirical SCV with $N$ vectors ($x_1$ to $x_N$) and $F$ nominal variables ($Q_1$ to $Q_F$); 
    a set of $k=3$ cultural prototypes for the same $F$ variables;
    a SCV with $N$ vectors generated, from the prototypes, using the PG model;
    a SCV with $N$ vectors generated, from the same prototypes,  using the MPG model.
    For the PG and MPG sketches, red, green and blue denote the copies of cultural traits from one of the first, second and third prototype respectively, while black denotes the explicitly random generation of traits. 
}
    \label{FigSketch} 
\end{figure*}

An SCV can be visualized as a table of cultural traits, where the columns correspond to cultural vectors (or sequences) and the rows correspond to cultural features (or variables). 
If the SCV is constructed from empirical data, the columns correspond to real people that are sampled by a social survey, while the rows correspond to questions that are asked in the social survey.
This is illustrated by Fig.~\ref{FigSketch}, which is explained in detail below. 
Consistently with Ref.~\cite{Babeanu}, a ``cultural space'' is the set of all possible cultural vectors (or combinations of traits) allowed by the given set of cultural features:
one combination of traits is one point in this discrete space. 
For the purpose of this work, the general set-up is restricted to cultural spaces defined in terms of features that are exclusively nominal.
In this setting, distances between points in the cultural space are given by Eq. \eqref{CultDist} of Sec.~\ref{ModFit}. 
Disregarding ordinal features makes the modeling paradigm compatible with the (arguably strong) assumption that one prototype corresponds to one point in cultural space,
meaning that a prototype picks up one and only one trait of any given feature. 
Other limitations of this assumptions are extensively discussed in Sec.~\ref{Disc}, together with possible ways of relaxing it, for the purpose of generalizing the current modeling paradigm in future work.

The two models are schematically illustrated in Fig.~\ref{FigSketch}.
The figure first shows a sketch of an empirical SCV, where the rows correspond to cultural features, the columns correspond to cultural vectors and the letters correspond to cultural traits -- 
the $n$'th row shows the traits of the $N$ agents that are expressed (or formulated) with respect to the $n$'th feature.
Then, it shows a set of 3 cultural prototypes (their number could have been different), in 3 different colors, all of them spanning over all features (or questions) relevant for the empirical set of vectors.
Finally, it illustrates a typical set of vectors generated using the PG method, followed by one generated using the MPG method.
The colors distinguish between the prototypes, while indicating how the traits are copied from the prototypes to the cultural vectors,
while black denotes traits that generated in an explicitly random way (uniform distribution, independently of the prototypes).

There are several things worth noting in relation to Fig.~\ref{FigSketch}.
First, the possibility that two or more prototypes pick the same trait for a certain feature is allowed by the current modeling paradigm 
(note that any of the traits that can be copied from one of the prototypes can also be generated via explicit randomness).
This is essential for controlling the average prototype-prototype distance, as will become apparent below.
Second, a PG vector is partly copied from one prototype and partly generated in an explicitly random way, 
while a MPG vector is a mixture of copies from all the prototypes, with one dominating prototype and with few traits generated in an explicitly random way.
Third, both models make use of another type of randomness, in addition to the explicitly random trait generation and to the randomness involved in generating the prototypes.
This randomness has to do with assigning every trait of every vector to a ``prototype of origin'',
once the random generation fraction and the influence fractions of the prototypes are specified. 
In the case of MPG, it is mainly this trait-assignment randomness that allows for the generation of a multitude of distinct cultural vectors from a small set of fixed prototypes, 
in the presence of little explicitly random trait generation. 

The procedure for generating the cultural prototypes is the same for both the PG and the MPG models. 
One needs to specify the number of prototypes $k$, as well as the value of another parameter $\alpha \in (0,1)$, which controls for the expected cultural distance between the prototypes.
This parameter governs the expected number of overlaps (or coincidences) between prototypes in terms of how they are distributed over the traits of a specific feature. 
In the extreme case of $\alpha \rightarrow 1$, all prototypes pick the same trait for every feature, 
yielding the smallest possible separation between the prototypes in cultural space (which coincides with the minimum of 0 allowed by the cultural distance definition in Eq. \eqref{CultDist}).
In the other extreme case of $\alpha \rightarrow 0$, the prototypes are distributed as uniformly as possible over the traits of every feature, 
yielding the largest possible separation between the prototypes in cultural space (which only coincides with the maximum of 1 allowed by Eq. \eqref{CultDist} if the number of traits $q$ is larger or equal to the number of prototypes $k$ for every feature). 
This is achieved by a formulation in terms of the set of integer partitions $I_k^q$ describing the possible ways of distributing the $k$ prototypes over the $q$ traits of a certain feature.
The $\alpha$ parameter actually controls the probability distribution over the set $I_k^q$, via the ``compactness'' of the integer partitions in this set. 
Sec.~\ref{AppIntPartGen} precisely describes how these probabilities are assigned and how the set $I_k^q$ is computationally generated in the first place, for any combination of $k$ and $q$.
Once the prototypes are chosen, everything else is conditional on them, for both models. 

According to the {\bf Prototype Generation (PG)} model, each cultural vector is a partial realization of one of the prototypes.
Each of the $N$ cultural vectors is generated by copying a random sequence of traits from one of the $k$ prototypes, while generating the other traits in a uniformly random way 
-- choosing the prototype is done randomly for every vector.
Then, a subset of the $F$ features of length $\text{round}(\beta \cdot F)$ is randomly and independently selected for each vector and the traits of these features are copied from the prototype to the vector.
Here, ``round'' returns the integer that is closest to its argument, while $\beta \in [0,1]$ is a third model parameter, 
in addition to $k$ and $\alpha$ (which are already needed for the purpose of specifying the prototypes, in the manner described above).
The $\beta$ parameter specifies the fraction of traits that are directly copied from the prototype, thus controlling for the expected distance between a vector and its prototype.
The traits for the remaining features are generated randomly and independently, according to uniform feature-level probability distributions -- the explicit random generation mentioned above.
Thus, $\beta$ also controls for the amount of explicitly random generation of traits.
The PG method effectively specifies that there are $k$ ``classes'' of cultural vectors and those of a certain class are located at a certain, $\beta$-controlled average distance from the associated cultural prototype.
This is similar to the ``Prototype Evolution'' method of Ref.~\cite{Stivala},
although there are small differences in how exactly the vectors are generated in the two cases.  
Moreover, the method of Ref.~\cite{Stivala} did not allow for controlling the expected cultural distance between the prototypes.

According to the {\bf Mixed Prototype Generation (MPG)} model, each cultural vector is a combination of all prototypes, although an unbalanced combination, 
meaning that the numbers of traits copied from the different prototypes are deliberately unequal.
The extent of this discrepancy is explicitly controlled via the third model parameter, which, like for PG, is called $\beta$.
Although the exact definition and usage of the $\beta \in (0,1)$ parameter is different in MPG than in PG, its role is quite similar.
Specifically, also in the context of MPG, $\beta$ (indirectly) controls for the fraction of traits copied from the dominating prototype to the vector:
more traits are copied from the dominating prototype if the discrepancy between the prototypes is higher.
In addition to traits copied from the prototypes, there are traits that are generated in an explicitly random way, but in a small number.
For each generated vector, this number is by construction not higher than the number of traits copied from the lowest-contributing prototype.
Consequently, if there are $k$ prototypes, the number of traits generated via explicit randomness does not exceed $F/(k+1)$. 
Thus, $1/(k+1)$ is an upper bound for the fraction of explicit randomness in an entire set of cultural vectors generated with MPG.
It is also important to note that, like for PG, this fraction is controlled by $\beta$ and that the upper bound is reached when $\beta$ is in the limit of minimal imbalance.
The limited usage of explicitly random trait generation by MPG means that cultural vectors are more strongly constrained by the prototypes, compared to PG. 
Still, MPG allows for generating a large variety of possible cultural vectors, since the $k$ prototypes can mix in many different ways.

The MPG model needs a procedure of specifying, for each generated vector, the $k$ values of the numbers of traits that are to be copied from the $k$ prototypes, along with the number associated to explicitly random generation. 
These $k+1$ positive, integer numbers should add up to $F$ and have their discrepancy controlled by the $\beta$ parameter.
Moreover, there is no reason to believe that the sequence of numbers associated to one $\beta$ value should be the same across all generated vectors,  
so randomness should be involved in choosing these numbers.
Therefore, the model needs a probabilistic way of drawing $k+1$ random, positive integers $\{t_1(\beta),..,t_{k+1}(\beta)\}$ satisfying $\sum_{l=1}^{k+1} t_l(\beta) = F$, such that their expected discrepancy is controlled via a single parameter $\beta$. 
The procedure chosen for this purpose is described below. 

This procedure heavily relies on isometrically mapping the discrete $\{0,1,..,F\}$ set of integers to the $[0,1]$ interval of the real axis.
For each generated vector, the latter interval is split into $k+1$ parts, by performing ``cuts'' in $k$ randomly chosen points.
In this manner, a sequence of $k+1$ preliminary weights $\{W_1,...,W_{k+1}\}$, subject to $\sum_{l=1}^{k+1} W_l = 1$ is numerically obtained.
These weights are obviously independent of $\beta$ and have a fixed expected discrepancy. 
A $\beta$-dependent transformation (explained below) is applied on the preliminary weights $\{W_1,...,W_{k+1}\}$, 
thus providing a sequence of $\beta$-dependent weights $\{w_1(\beta),...,w_{k+1}(\beta)\}$ satisfying $\sum_{l=1}^{k+1} w_l(\beta) = 1$,
with expected discrepancy controlled by $\beta$.
Finally, the sequence of $\beta$-dependent weights is converted back to the desired sequence $\{t_1(\beta),..,t_{k+1}(\beta)\}$.
This final operation is non-trivial, requiring a self-consistent, joint rounding procedure, which is generally difficult to choose, since one cannot generally ensure that $w_l = \text{round}(t_l/F), \forall l$ -- a non-trivial problem of weight discretization.
Here, a simple, pragmatic choice is made: converting the lowest $k$ weights to the closest, lower integer, while converting the highest weight to the integer needed for satisfying the summation constraint --
this ensures that the highest weight, which should correspond to the dominating prototype, is converted to the highest integer.

The only aspect of MPG remaining to be explained is how the $\beta$-dependent weights $\{w_1(\beta),...,w_{k+1}(\beta)\}$ are obtained from the preliminary weights $\{W_1,...,W_{k+1}\}$.
This is done by raising the latter to a common power $p(\beta)$ and then normalizing: 
\begin{equation}
  w_l(\beta) = \frac{(W_l)^{p(\beta)}}{\sum_{l'=1}^{k+1} (W_{l'})^{p(\beta)}},
\end{equation}
where the common power $p(\beta) \in (0,+\infty)$ controls for the average discrepancy between these weights and maps to $\beta \in (0,1)$ via:
\begin{equation}
	p(\beta) = \tan\left(\beta \frac{\pi}{2}\right),
\end{equation}
where the tangent is a convenient choice of a smooth, continuous function, with the appropriate domain and range.
Thus, a value $\beta > 0.5$ implies a value $p > 1$ and a higher discrepancy of $\{w_1^p,...,w_{k+1}^p\}$ than that of $\{W_1,...,W_{k+1}\}$,
while a value $\beta < 0.5$ implies a value $p < 1$ and a lower discrepancy of $\{w_1^p,...,w_{k+1}^p\}$ than that of $\{W_1,...,W_{k+1}\}$.

Before describing the fitting and the outcomes of the PG and MPG models, it is worth summarizing a few important aspects.
Both models rely on the notion of cultural prototypes, which is currently formalized in a simplistic manner, which is only sensible for cultural spaces defined exclusively in terms of nominal features.  
The procedure for generating the prototypes is the same for both models and relies on two parameters, $k$ and $\alpha$, which specify, respectively, the number of prototypes and the expected distance between them.
The differences between PG and MPG consist in how the cultural vectors are generated conditionally on the prototypes:
for PG, every vector is in part a copy from one of the prototypes and in part explicitly random; 
for MPG, every vector is an imbalanced mixture of all prototypes and explicitly random to a much lower extent, which is how the ``multiple-self'' ingredient is implemented. 
Nonetheless, in both cases, there is a third model parameter, $\beta$, which governs, in different ways, 
the lengths of the randomly selected subsets of features whose traits that are copied from the prototypes.
In both cases, $\beta$ effectively controls for the expected distance between a vector and its (dominating) prototype, as well as for the fraction of explicit randomness. 

\section{Model fitting}
\label{ModFit} 

Before applying the LTCD-STCB analysis on SCVs generated with either the PG or MPG models, it is useful to somehow constrain some of the free model parameters. 
This is done in terms of statistical quantities simpler than the LTCD and the STCB measures, that can be evaluated on both empirical SCVs and on the model SCVs. 
On the empirical side, the quantities are averaged over several, empirical SCVs constructed by randomly selecting $N=500$ cultural vectors from the ~13000 available ones in Eurobarometer data set~\cite{EBM}, 
while restricting to the nominal features -- let ``($\text{EBM}_{n}$)'' stand for the nominal part of the Eurobarometer data set.
The empirical data is formatted according to the procedure explained in Ref.~\cite{Babeanu}.
On the model side, these quantities are averaged over many SCVs, of the same size $N$, that are realizable in the cultural space of ($\text{EBM}_{\text{n}}$), 
for the given combination of parameters -- the prototypes are independently generated upon creating every model SCV.

The two simple quantities in terms of which the models are tuned to empirical data
are the average and the standard deviation of the inter-vector distances in the SCV, which are here denoted by ``AIVD'' and ``SIVD'' respectively:
\begin{align}
    &	\text{AIVD} = \frac{2}{N(N-1)}\sum_{i<j} d_{ij} \label{AIVD}, \\
    &	\text{SIVD} = \sqrt{\frac{2}{N(N-1)}\sum_{i<j} (d_{ij} - \text{AIVD})^2}, \label{SIVD}
\end{align}
where $N$ is the number of cultural vectors and $d_{ij}$ is the cultural distance, as defined and used in Refs.~\cite{Babeanu, Stivala, Valori}. 
The notation $i<j$ denotes that the respective summation is carried out over all distinct pairs $(i,j)$.
In the case of a fully-nominal cultural space, with which this study is dealing, $d_{ij}$ reduces to the Hamming distance between the two sequences of symbols encoding cultural vectors $i$ and $j$: 
\begin{equation}
  \label{CultDist}
  d_{ij} = 1 - \frac{1}{F}\sum_{l=1}^{F} \delta(x_i^l, x_j^l) = \frac{1}{F}\sum_{l=1}^{F} d^l_{ij},
\end{equation}
with, $d_{ij}$ taking values within the $[0,1]$ interval.
Here, $l$ iterates over the $F$ nominal features, $x_i^l, x_j^l$ are the traits of vectors $i$ and $j$ with respect to feature $l$ and $\delta$ stands for the Kroneker-Delta function. 
The second equality shows that the cultural distance can be expressed as an average over feature-level contributions, which becomes useful below.  
Previous work has shown that an empirical SCV is characterized by a lower AIVD than its random counterpart and a higher SIVD than both its random and shuffled counterparts~\cite{Valori, Stivala}.  
The AIVD and SIVD quantities, which incorporate pairwise distance information, are conceptually different than what is often used in the context of cultural dynamics and of the Axelrod model, 
namely the size of the largest connected component, which can be regarded as an overall measure of similarity.
Instead, the latter is somewhat similar to the STCB quantity explained and used in Sec.~\ref{ModOut}. 

It is instructive to see that the expressions of AIVD and SIVD can be rewritten in the following way:
\begin{equation}
  \label{AIVD_reform}
    \text{AIVD} = \frac{1}{F}\sum_{l=1}^{F} \frac{2}{N(N-1)}\sum_{i<j} d^l_{ij},
\end{equation}
\begin{widetext}
\begin{equation}
  \label{SIVD_reform}
    \text{SIVD} = 
      \sqrt{\frac{1}{F^2} \sum_{l=1}^F \sum_{l'=1}^F \left[ \frac{2}{N(N-1)}\sum_{i<j}d_{ij}^{l} d_{ij}^{l'} 
	- \frac{4}{N^2(N-1)^2}\sum_{i<j}d_{ij}^{l} \sum_{i'<j'}d_{i'j'}^{l'} \right]},
\end{equation}
\end{widetext}
by using a feature-level cultural distance $d_{ij}^l$ introduced via Eq. \eqref{CultDist} -- the transition from \eqref{SIVD} to \eqref{SIVD_reform} was suggested by the SI of Ref~\cite{Valori}.

Note that the AIVD can be understood as an average over feature-level AIVD contributions, which are represented by the expression within the $l$-summation of Eq. \eqref{AIVD_reform}.
It can be checked that the (nominal) feature-level AIVD contribution is a measure of how uniformly the $N$ vectors are distributed over the possible traits of that feature.
This is more obvious when expressing the expected value of the AIVD contribution in terms of probabilities associated to the traits, which is shown in Eq. \eqref{IdeaAIVD} below.
Thus, for an empirical SCV containing only nominal features, the AIVD is a measure of average uniformity of the empirical frequency distributions associated to the features. 
Consequently, the AIVD is also a measure of how subjective the questions/topics associated to the features are on average -- 
when the frequencies of possible answers are more similar to each other, there is less justification to talk about a ``better'', a ``more correct'' or a ``more agreed upon'' answer, 
so the question is inherently more subjective. 

Also note that, in Eq. \eqref{SIVD_reform}, the quantity inside the average over pairs of features $(k,l)$ is the covariance between features $k$ and $l$, defined in terms of the feature-level cultural distances.
Given that this quantity is averaged over all possible pairs of features and that the square-root is a monotonous function,
the SIVD encodes information about the pairwise correlations between features, although in a somewhat indirect way. 

\begin{figure*}
\centering
	\subfigure{\includegraphics[width=8cm]{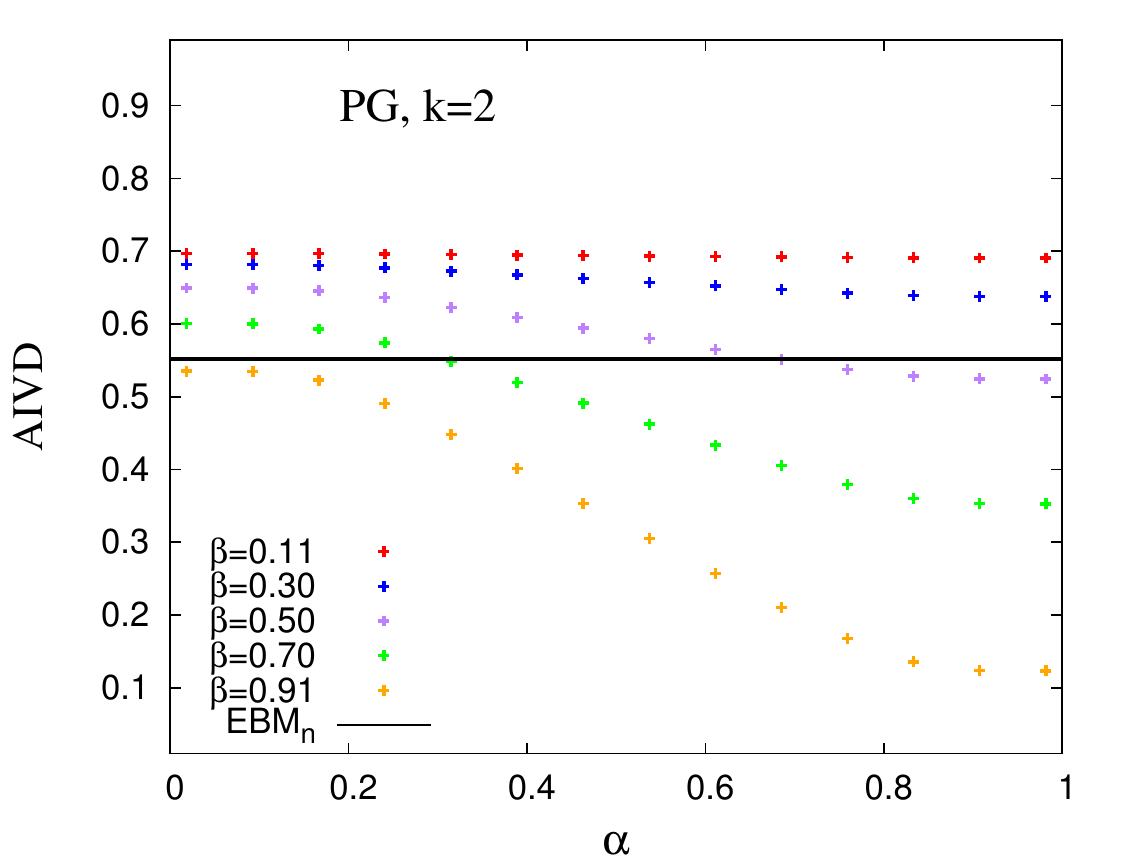}}
	\subfigure{\includegraphics[width=8cm]{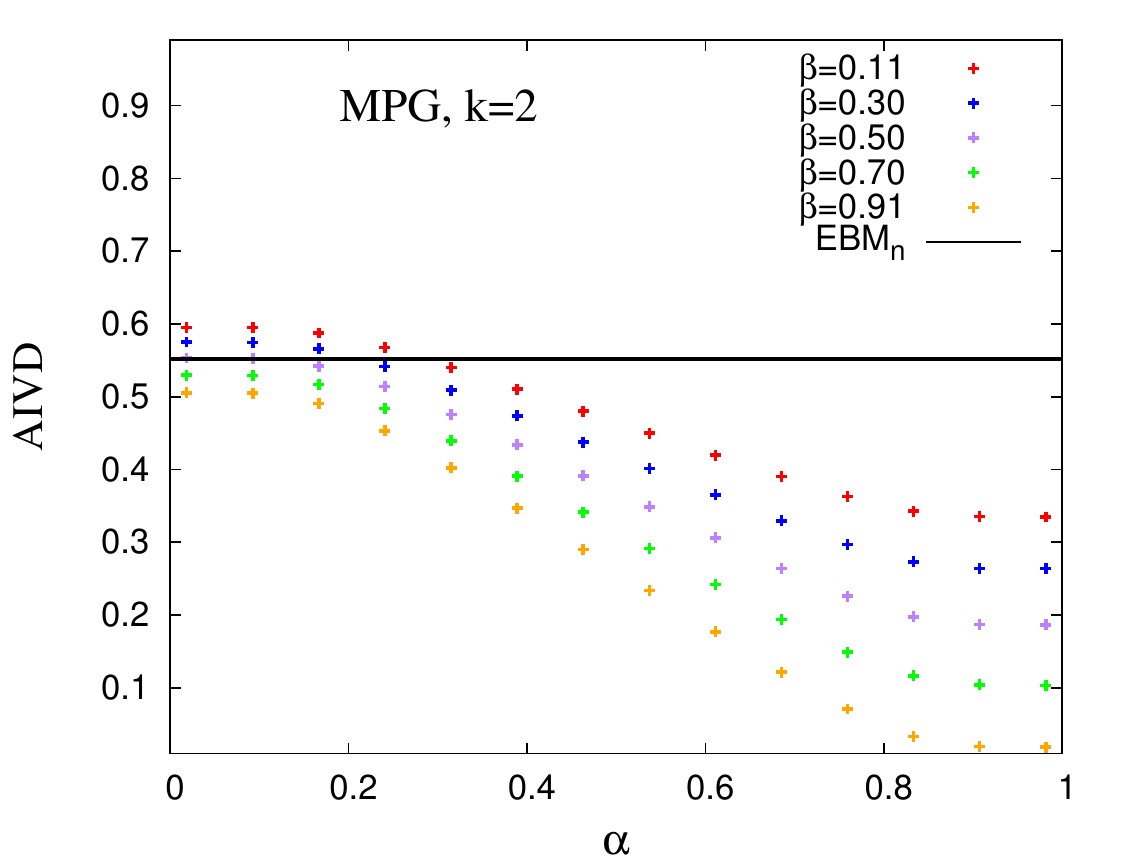}} \\
	\subfigure{\includegraphics[width=8cm]{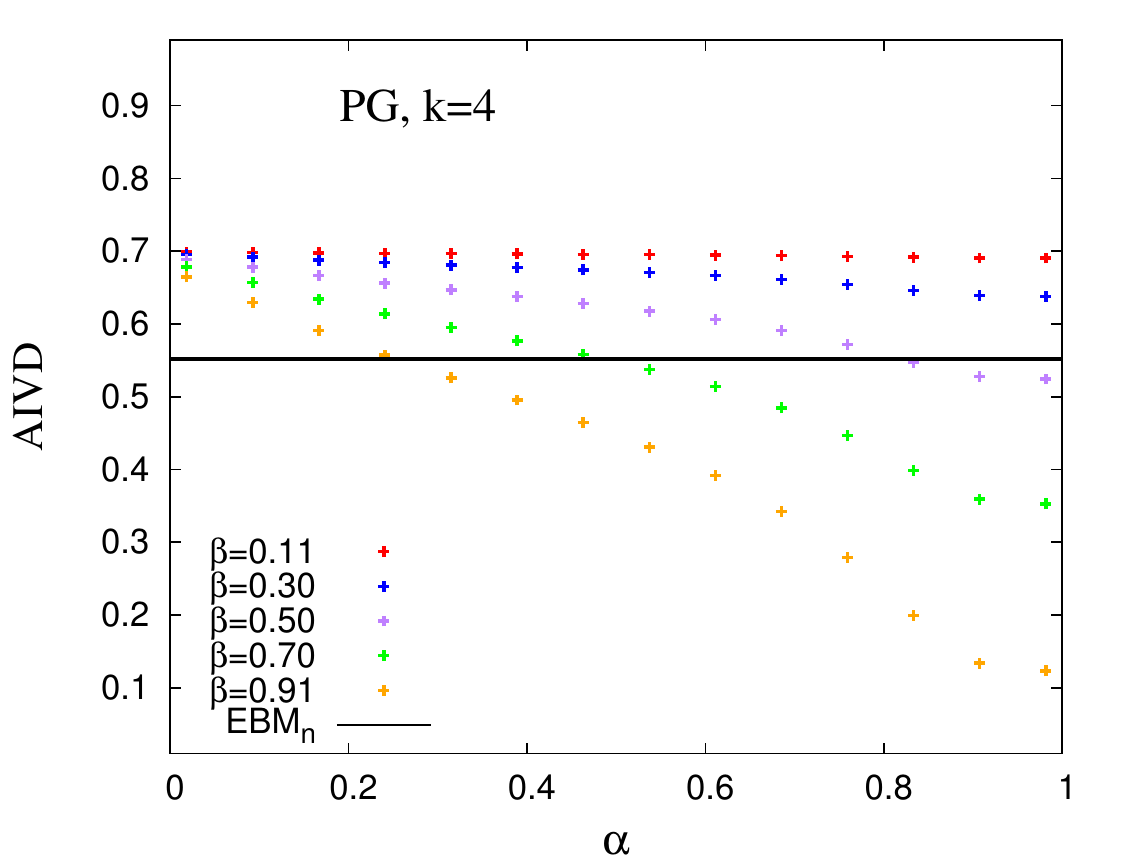}}
	\subfigure{\includegraphics[width=8cm]{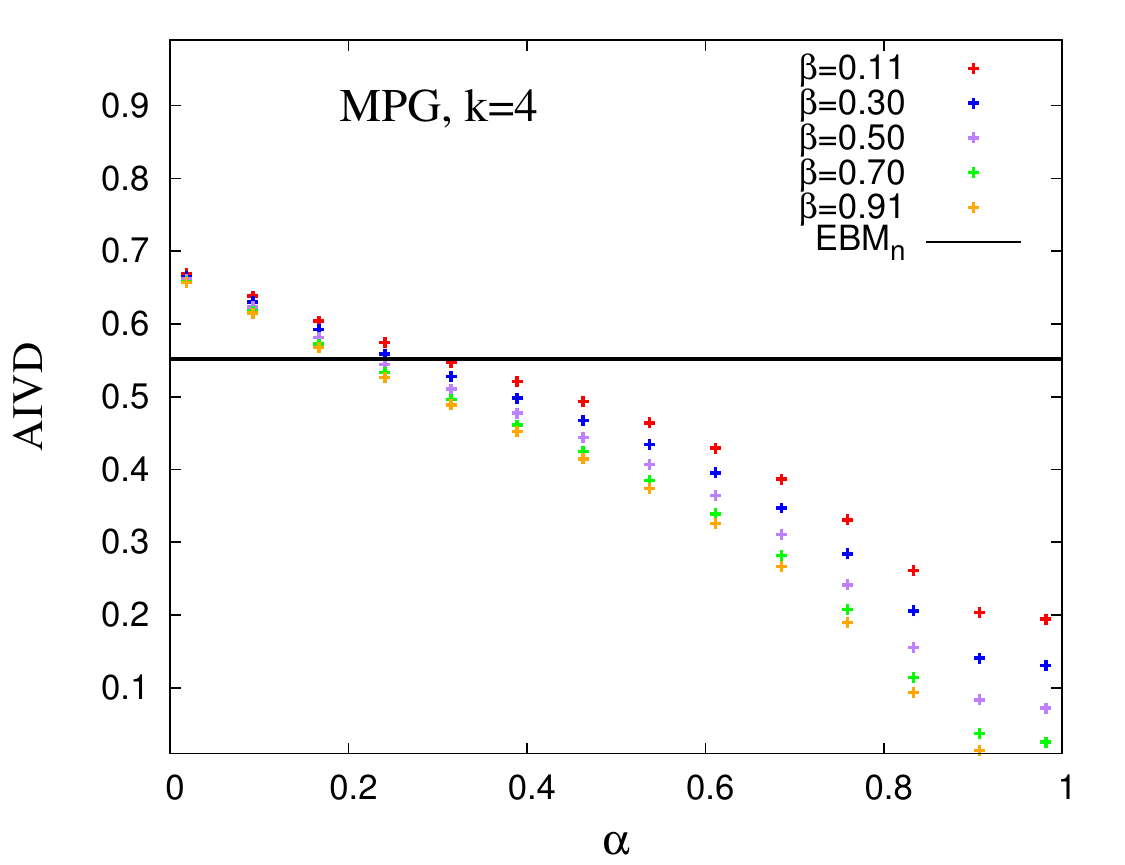}} \\
	\subfigure{\includegraphics[width=8cm]{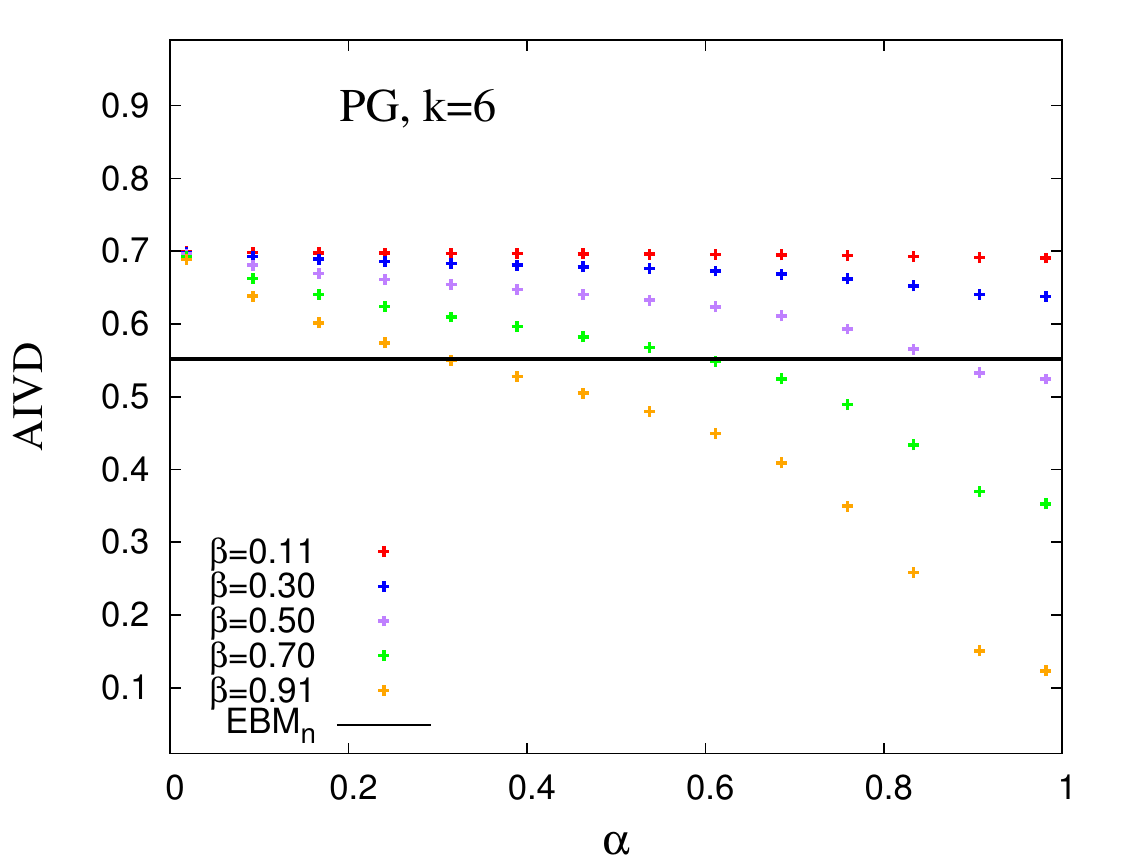}}
	\subfigure{\includegraphics[width=8cm]{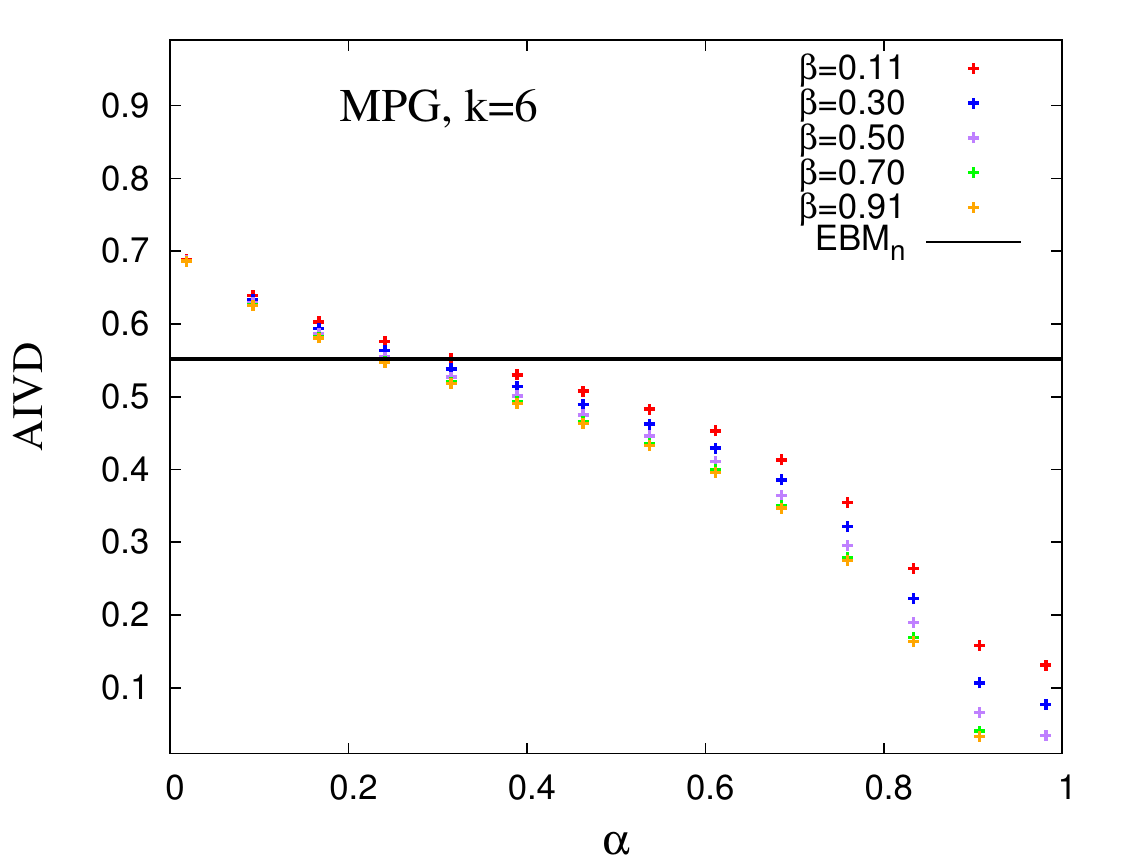}}
	\caption{Dependence on model AIVD on the $\alpha$ parameter, for several values of the $\beta$ parameter (legend),
					for $k=2$ (top), $k=4$ (center) and $k=6$ (bottom) prototypes, for the PG (left) and MPG (right) models.
					The horizontal lines show the empirical AIVD uncertainty range (one standard error on each side of the mean).}
\label{FigAIVD}
\end{figure*} 

\begin{figure*}
\centering
	\subfigure{\includegraphics[width=8cm]{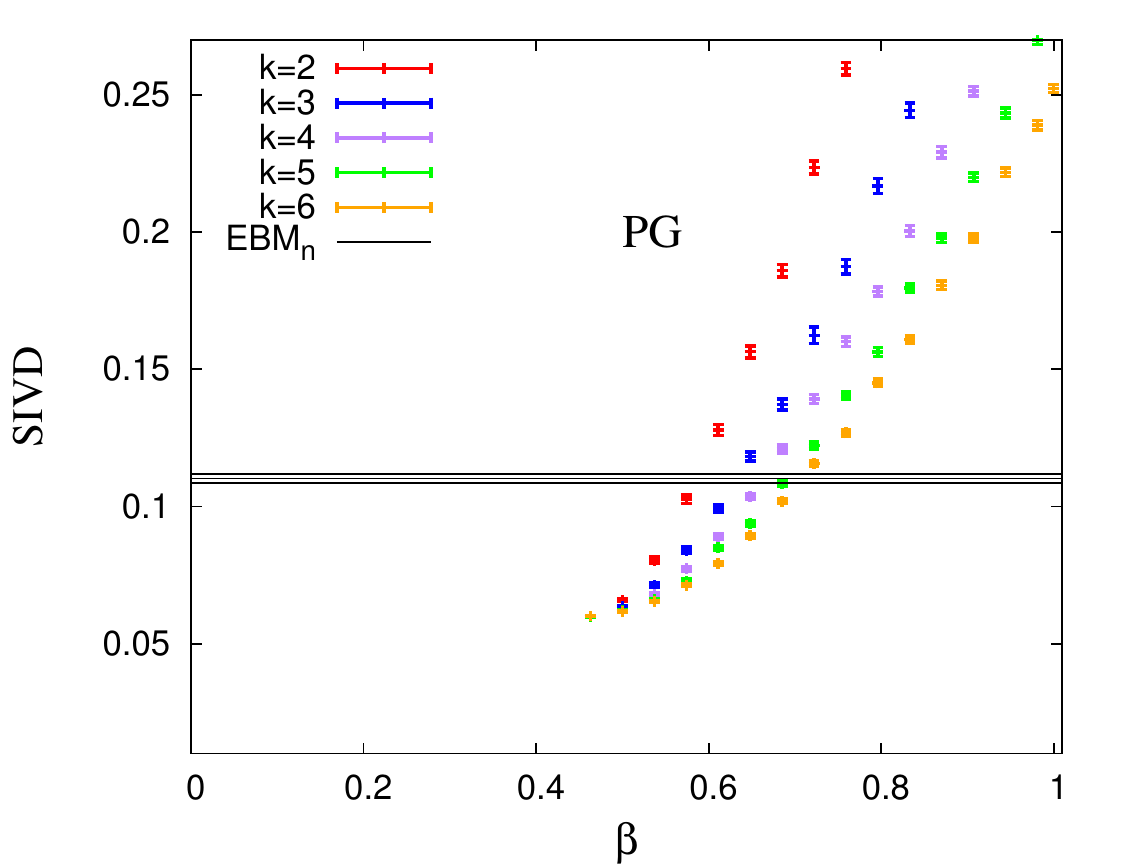}}
	\subfigure{\includegraphics[width=8cm]{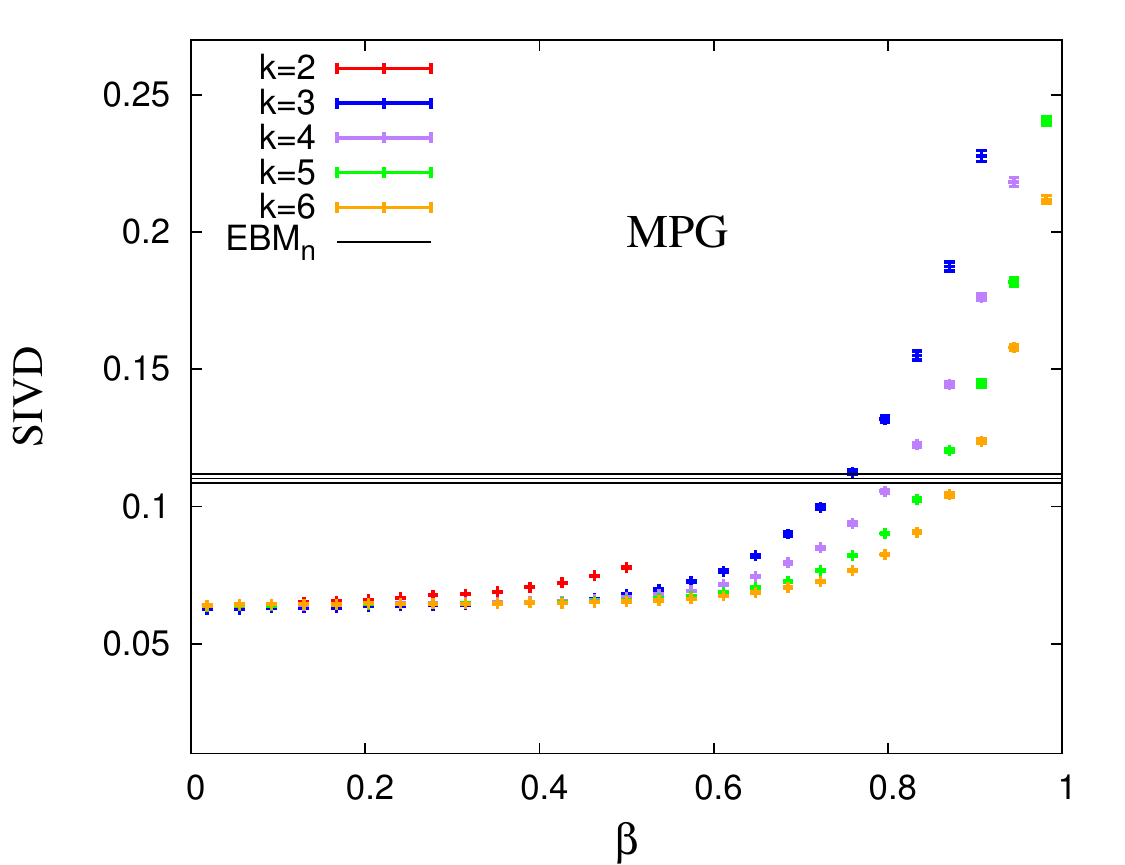}}
	\caption{Dependence of model SIVD on the $\beta$ parameter, for several values of the number of prototypes $k$ (legend),
					for the PG (left) and MPG (right) models, where the $\alpha$ parameter is tuned such that the empirical AIVD is reproduced.
					The error bars of the points show the numerical uncertainty ranges, 
					while the horizontal lines show the empirical SIVD uncertainty range (one standard error on each side of the mean).
}
\label{FigSIVD}
\end{figure*}

For both models, the choice made here is that of:
\begin{itemize}
	\item tuning the $\alpha$ parameter in terms of the AIVD quantity (Eqs. \eqref{AIVD}, \eqref{AIVD_reform}),
				for any combination of values of the $\beta$ and $k$ parameters;
	\item tuning the $\beta$ parameter in terms of the SIVD quantity (Eqs. \eqref{SIVD}, \eqref{SIVD_reform}), 
				for any value of the $k$ parameter, based on the previous fitting of $\alpha$ in terms of AIVD;
	\item simply repeating the tuning (and the LTCD-STCB analysis in Sec.~\ref{ModOut}) for several values of $k$.
\end{itemize}
This implies that, for every value of $k$, the tuning (or fitting) is done at two levels: 
the $\alpha$-AIVD level and the $\beta$-SIVD level, the former being nested into the latter. 
In practice, the fitting is carried out automatically, using a nested, 2-levels algorithm that relies on a modified bisection-type method for each level. 
The algorithm is precisely described in Sec.~\ref{AppFitAlg}.
In order to work, this approach relies on the assumption that there is one, unique solution for the fitting problem, for every value of $k$.
This uniqueness is demonstrated via Figs.~\ref{FigAIVD} and~\ref{FigSIVD},
which are also used for providing a general intuition of how the fitting works and 
of how the AIVD and SIVD quantities depend on $\alpha$, $\beta$ and $k$, for the two models.

Before entering this description, it is worth mentioning that the computer time for the fitting algorithm is greatly reduced by being able to evaluate the average (model) AIVD quantity analytically, 
in a manner that properly accounts for all SCVs that can be generated for any combination of $k$, $\alpha$ and $\beta$.
While the calculation is described in detail in Sec.~\ref{AppAnAIVD}, 
a schematic understanding can already be provided here.
The essential ingredient of the calculation is a simple, exact formula for the expected AIVD contribution of one feature of range $q$:
\begin{equation}
  \label{IdeaAIVD}
    \langle \aivd\left(\{p_1,...,p_q\}\right) \rangle = 1 - \sum_{i=1}^q p_i^2,
\end{equation}
which assumes that the probabilities of its traits $\{p_1,...,p_q\}$ are all known -- see Sec.~\ref{AppAnAIVD} for the proof.
For a discrete probability distribution, Eq. \eqref{IdeaAIVD} is a measure of uniformity very similar to the Shannon entropy.
Conditional on a specific choice of the prototypes, this set of probabilities (thus the feature-level probability distribution)
is fully determined by the integer partition describing how the prototypes are distributed over the traits 
and by the fraction of traits that are randomly generated, the latter being controlled by $\beta$.
In this context, Eq. \eqref{IdeaAIVD} already assumes that an averaging is performed over SCVs generated from the same set of prototypes. 
One still needs to perform an average of this expression over integer partitions (Eq. \eqref{AvAIVD_q} of Appendix Sec.~\ref{AppIntPartProb}), 
according to the probability distribution controlled by $\alpha$ (Eqs. \eqref{IntPartProbNN} and \eqref{IntPartProb} of Appendix Sec.~\ref{AppAnAIVD}),
followed by another average over all features (Eq. \eqref{AvAIVD} of Appendix Sec.~\ref{AppAnAIVD}), since different features will in general have different ranges $q$.
At a superficial inspection, using a similar approach for analytically computing the SIVD quantity appears very complicated, if at all possible.
Numerical calculations are instead employed for computing the (model) SIVD.

Fig.~\ref{FigAIVD} deals with the first-level fitting. 
It shows the dependence of the analytically computed AIVD quantity (see above) on the $\alpha$ parameter, for several $\beta$ values, for several $k$ values
and for both the PG and MPG models. 
Moreover, it shows the empirical AIVD uncertainty range\footnote{An uncertainty range, as defined in Sec.~\ref{AppFitAlg}, is the interval spanned by one standard mean error on each side of the mean.} 
via the horizontal bands in the six panels.
Thus, a solution of the first-level fitting is indicated by an intersection between a model curve of a given combination of $k$ and $\beta$ and the horizontal band.
Note that, for either of the two models and for any combination of $k$ and $\beta$, if a solution exists, this solution is actually unique.
In order to understand the behavior implicit in Fig.~\ref{FigAIVD}, which is explained below, one should keep in mind that AIVD measures the average uniformity of the feature-level probability distributions.

First, it is worth focusing on the AIVD dependence on the $\alpha$ and $\beta$ parameters.
Note, on one hand, that for a given combination of $k$ and $\beta$, the AIVD generally decreases with $\alpha$, or at least remains constant.
This is due to the fact that the AIVD decreases with decreasing distance between prototypes, thus with increasing $\alpha$. 
For PG, this decrease is stronger for higher $\beta$ values, since for low $\beta$ value the uniformity is anyway high, because of the large fraction of randomly generated traits.
For MPG, this $\beta$-dependence of the decrease is not that strong, since the fraction of randomly generated traits cannot exceed $1/(k+1)$.
On the other hand, for a given combination of $k$ and $\alpha$, the AIVD generally decreases with increasing $\beta$.
This is due to the fact that the AIVD decreases with decreasing fraction of randomly generated traits, thus with increasing $\beta$.

Second, it is worth focusing on the AIVD dependence on the number of prototypes $k$.
For PG, for a given $\alpha$, a larger number of prototypes $k$ implies a higher AIVD, since traits copied from prototypes are more uniformly distributed, 
but this has a significant effect only for large $\beta$ values, again due to the uniformity being anyway in place for small $\beta$ values.
For MPG, the corresponding behavior is more subtle. 
While for large, $\beta \rightarrow 1$ values, the AIVD still increases with increasing $k$ at a given $\alpha$ (for the same reason as for PG), 
the AIVD($\alpha$) curves corresponding to small $\beta$ approach the AIVD($\alpha$) curve corresponding to large $\beta \rightarrow 1$ with increasing $k$, 
rather than remaining in place (which is the case for PG).
This is related to the fact that the upper bound on the fraction of randomly generated traits $1/(k+1)$ decreases with increasing $k$, 
thus decreasing the role of $\beta$ in controlling the AIVD via the uniform component of the feature-level probability distributions. 
            
Fig.~\ref{FigSIVD} deals with the second-level fitting.
Everything shown in this figure relies on $\alpha$ already being tuned (at the first level) such that the empirical AIVD is matched  --
as apparent from Fig.~\ref{FigAIVD}, the tuned $\alpha$ value depends on $\beta$ and on $k$.
Fig.~\ref{FigSIVD} shows the dependence of the numerically computed SIVD quantity (with uncertainty ranges) on the $\beta$ parameter, for several $k$ values
and for both the PG and MPG models. 
Moreover, it shows the empirical SIVD uncertainty range via the horizontal bands in the two panels.
Thus, a solution of the second-level fitting is indicated by an intersection between a model curve of a given $k$ and the horizontal band.
Note, again, that for either of the models and either of the $k$ values, 
if a solution exists, this solution is actually unique.
The exact technical procedure employed for producing any of the model points in Fig.~\ref{FigSIVD} is described at the end of Sec.~\ref{AppFitAlg}, 
followed by the explanation of the final choice of values for the $\alpha$ and $\beta$ parameters, for use in the analysis of Sec.~\ref{ModOut}.

Note that the SIVD increases with $\beta$ for both models and for all $k$ values, 
suggesting that the extent of feature-feature correlation increases with decreasing distance between vectors dominated by the same prototype.
For PG, all SIVD($\beta$) curves meet for some $\beta \approx 0.45$, at which point they also end.
No points are plotted for lower $\beta$ because $\alpha$ cannot be tuned in terms of AIVD, 
which can be understood from Fig.~\ref{FigAIVD} when noticing the AIVD($\alpha$) curves of low $\beta$ that do not cross the empirical line.
For MPG, the SIVD($\beta$) curve of $k=2$ ends at a value of $\beta \approx 0.5$, before crossing the empirical line,
meaning that the MPG model cannot be entirely tuned when only 2 prototypes are used. 
No points are plotted for higher $\beta$ because $\alpha$ cannot be tuned in terms of AIVD, 
which can be understood from Fig.~\ref{FigAIVD}, by noticing the AIVD($\alpha$) curves of $k=2$ and high $\beta$ that do not cross the empirical line.
This is due to certain limitations of the current modeling paradigm, which are further discussed in Sec.~\ref{Disc}.

\section{Model Outcomes}
\label{ModOut}

\begin{figure*}
\centering
	\subfigure{\includegraphics[width=8cm]{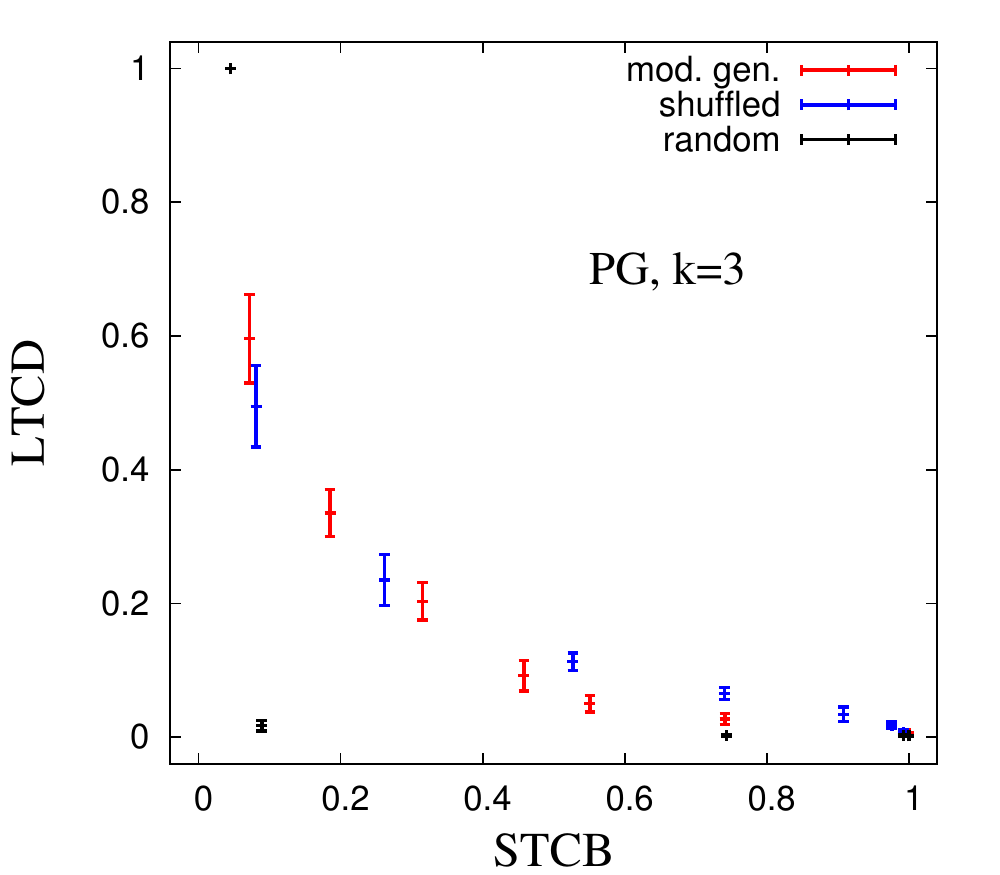}}
	\subfigure{\includegraphics[width=8cm]{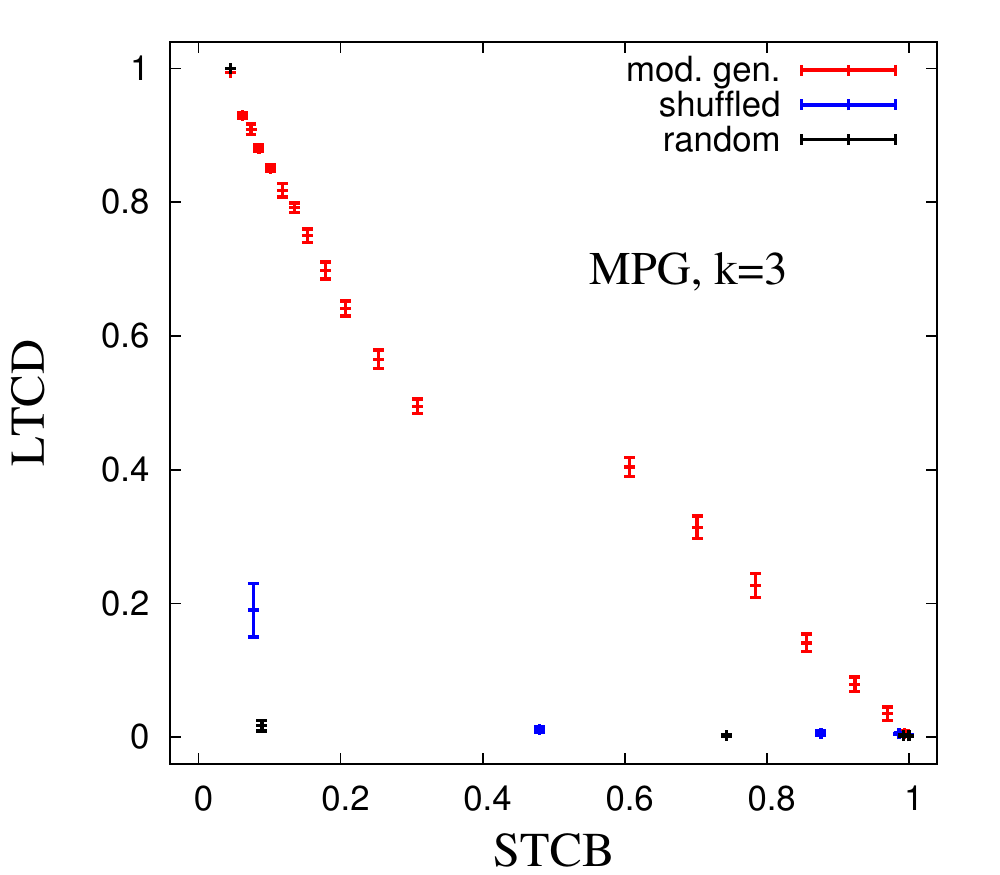}} \\ \vspace{-0.3cm}
	\subfigure{\includegraphics[width=8cm]{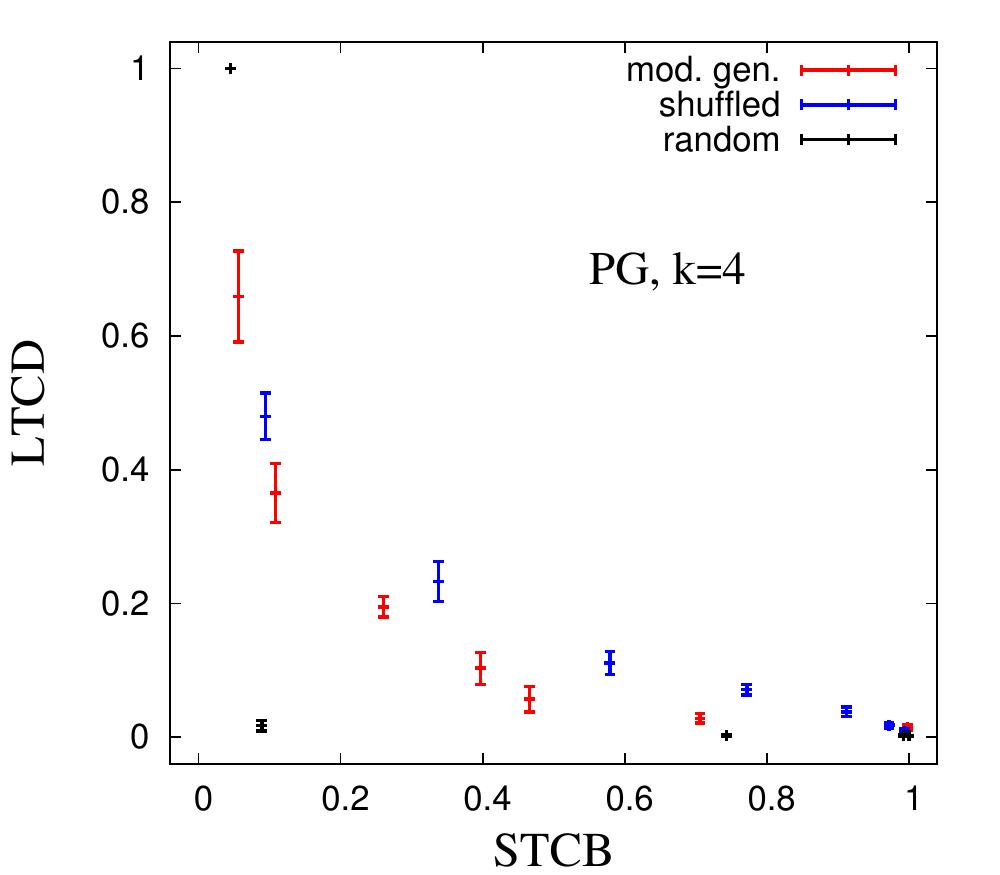}}
	\subfigure{\includegraphics[width=8cm]{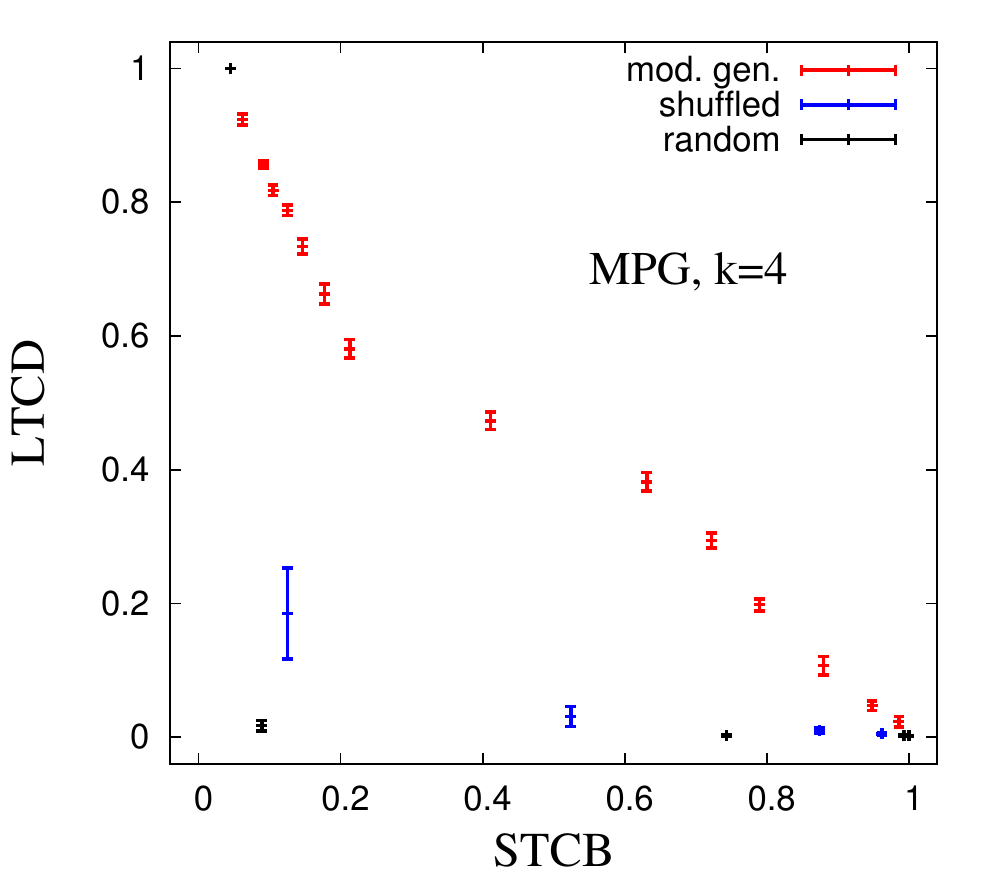}} \\ \vspace{-0.3cm}
	\subfigure{\includegraphics[width=8cm]{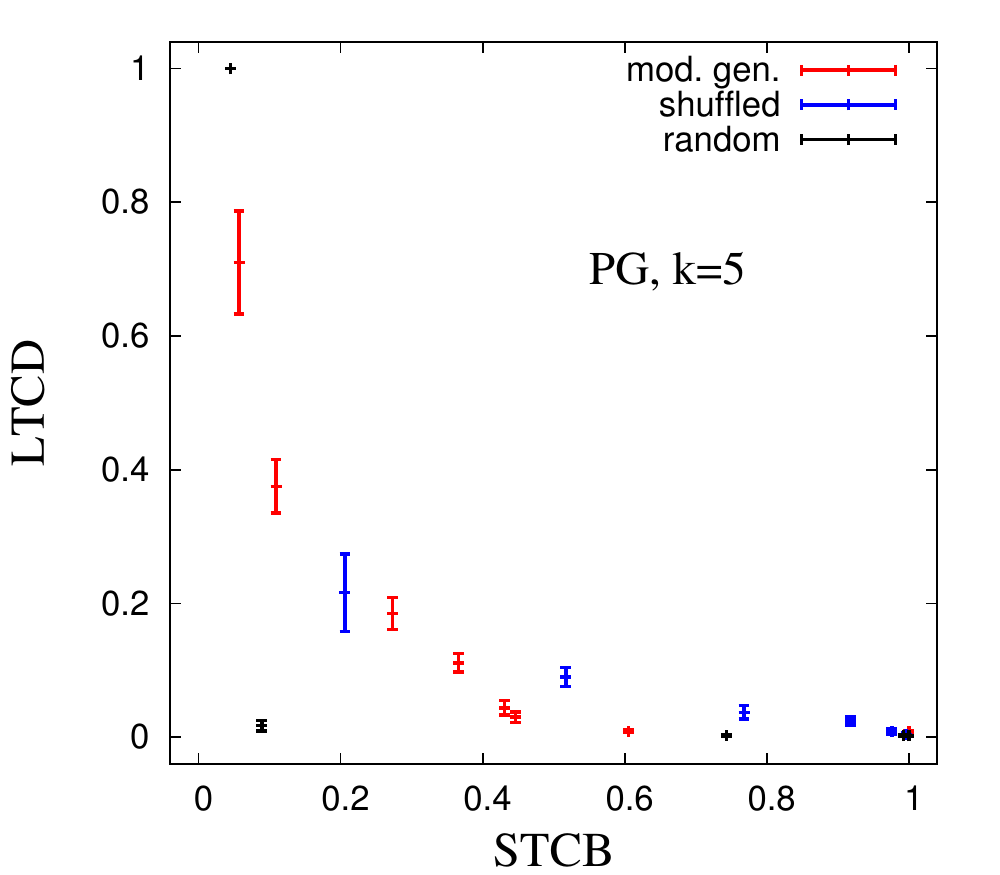}}
	\subfigure{\includegraphics[width=8cm]{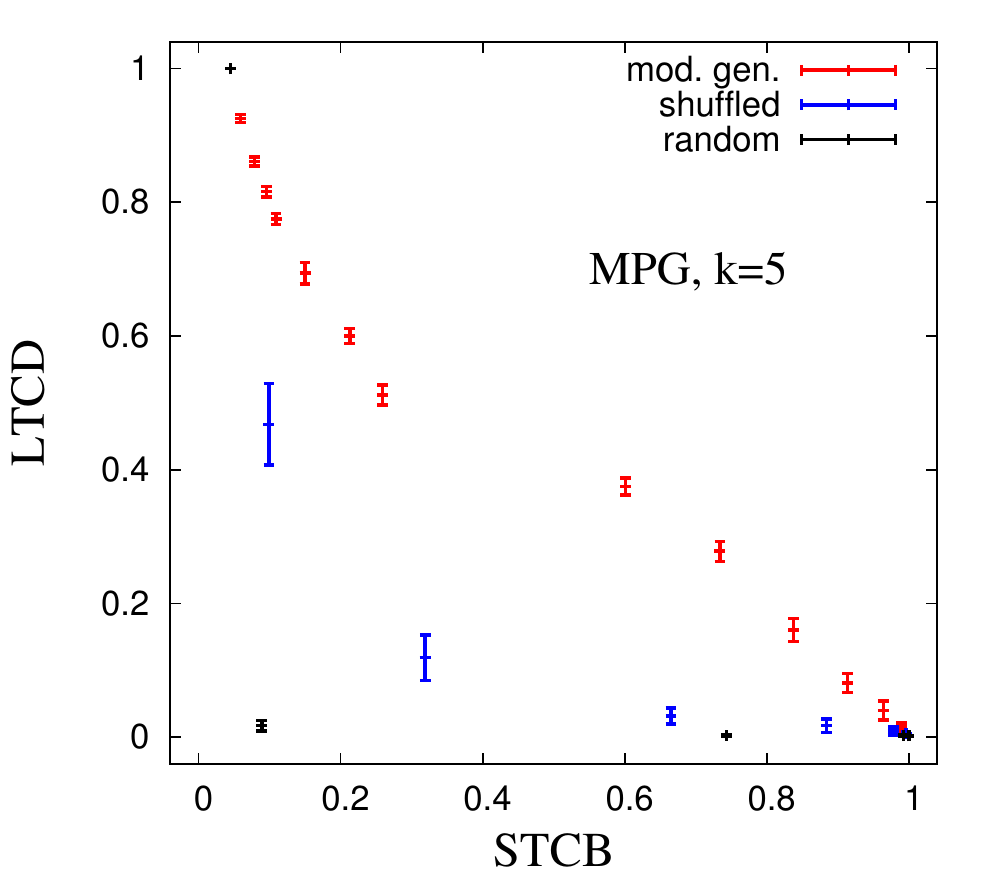}}
	\caption{The correspondence between Long-term cultural diversity (LTCD) and short-term collective behavior (STCB) for a model-generated (red), a shuffled (blue) and a random (black) SCV 
  obtained via the PG model (left) and MPG model (right), for $k=3$ (top), $k=4$ (centre) and $k=5$ (bottom) prototypes. 
  Error bars denote standard deviations over multiple trait dynamics runs.
  There are $N=500$ elements in each set of cultural vectors. 
}
\label{FigArtifDvC}
\end{figure*} 

\begin{figure}
\centering
	\includegraphics[width=8cm]{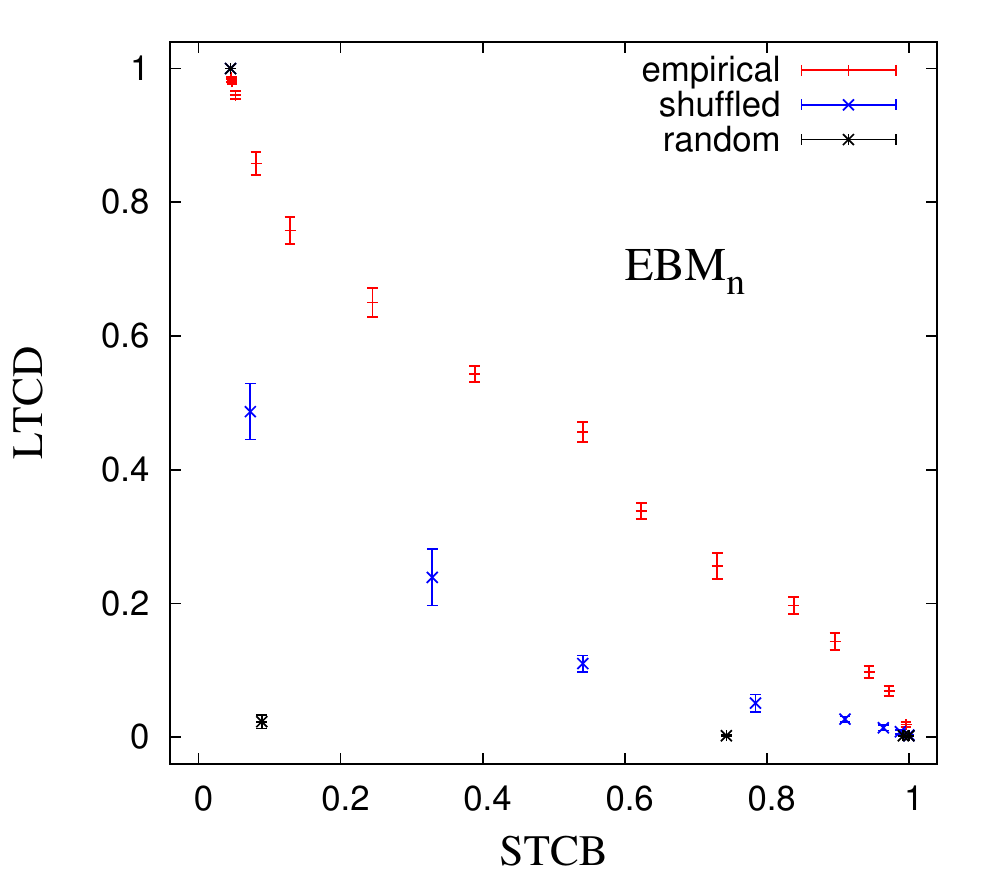}
	\caption{The correspondence between long-term cultural diversity (LTCD) and short-term collective behavior (STCB) for the empirical (red), shuffled (blue) and random (black) sets of cultural vectors,
  for the nominal part of the Eurobarometer data set ($\text{EBM}_{\text{n}}$). 
  Error bars denote standard deviations over multiple cultural dynamics runs.
  There are $N=500$ elements in each set of cultural vectors 
}
\label{FigEBM-Nom}
\end{figure}

Here, the most important results of this work are presented. 
The focus is on the LTCD-STCB analysis, applied to sets of cultural vectors generated with the PG and MPG models.
The aim is to assess how well the two models reproduce the universal empirical patterns described in Ref.~\cite{Babeanu}.
Fig.~\ref{FigArtifDvC} illustrates the results obtained with the two models, whereas Fig.~\ref{FigEBM-Nom} summarizes, for comparison purposes, the empirical results,
focusing on the nominal part of the Eurobarometer dataset ($\text{EBM}_{n}$) -- formatted according to the procedure explained in Ref.~\cite{Babeanu}.

Before describing the results, it is worth recalling the main ingredients of the LTCD-STCB analysis.
This is essentially a two-dimensional plot showing the correspondence between the LTCD quantity vs the STCB quantity, both of them being evaluated on empirical, on shuffled and on random SCVs. 
Drawing the LTCD-STCB correspondence is made possible by the fact that, for each of the three scenarios, both quantities depend on the bounded-confidence threshold $\omega$, 
which controls the maximal cultural distance over which social influence can act.
On one hand, the LTCD quantity is a measure of cultural diversity after a long-term process of cultural dynamics driven by $\omega$-bounded social influence, 
starting from an initial cultural state specified by the respective SCV.
Essentially, it counts the number of distinct points in cultural space (commonly referred to as ``cultural domains'') towards which the agents converge in the final state of a minimalisitic, bounded-confidence Axelrod model. 
The STCB quantity is a measure of collective behavior (or social coordination) after a short-term process of opinion dynamics driven by $\omega$-bounded social influence. 
Essentially, it is the standard deviation of the aggregate opinion distribution of the agent population, 
resulting from a minimalistic Cont-Bouchaud-type model applied to the (cultural) graph obtained by drawing a link for each pair of agents separated by a cultural distance smaller than $\omega$.
Mathematically, the two quantities, as functions of the bounded-confidence threshold $\omega$, are captured by the following two expressions:
\begin{equation}	
	\mathrm{LTCD}(\omega) = \frac{\langle{N_D}\rangle_{\omega}}{N}, \quad \mathrm{STCB}(\omega) = \sqrt{\sum_A \left(\frac{S_A}{N}\right)^2_{\omega}},
\end{equation}
where $N_D$ is the number of cultural domains in the final state of the Axelrod-type model, $N$ is the number of agents (and cultural vectors) and $S_A$ is the size of the A'th of connected components in the $\omega$-determined cultural graph. 
The average in the LTCD formula is taken over multiple simulations of the Axelrod-type model. 
The STCB quantity is calculated analytically, once the cultural connected components are found, based on the assumption of independent opinion-agreement within each connected component. 
An essential difference between the two quantities, reflected in the long-term/short-term distinction, consists of an idealized separation between two time-scales, in terms of the role that the SCV specified as input plays:
cultural vectors, together with the distances between them, are assumed to be dynamical by the LTCD definition and static by the STCB definition, 
such that one deals with dynamics of vectors and with dynamics on vectors in the two cases respectively.
The interested reader is referred to Refs.~\cite{Babeanu, Valori} for more details and remarks about the LTCD-STCB analysis.

For both the PG and the MPG models, the $\alpha$ and $\beta$ parameters are tuned in the manner described in Sec.~\ref{ModFit} for every value of the number of prototypes $k$, while the latter is simply iterated over. 
In Fig.~\ref{FigArtifDvC}, the LTCD-STCB plot is shown for the values $k=3$, $k=4$ and $k=5$, for the PG (left) and the MPG (right) models.
The value $k=2$ is omitted since the $\alpha$ and $\beta$ parameters could not be both tuned for MPG with two prototypes. 
All SCVs are generated using the cultural space of $\text{EBM}_{\text{n}}$, 
whose empirical SCVs also served for providing the AIVD and SIVD values in terms of which the tuning was conducted (Sec.~\ref{ModFit}).

When looking at Fig.~\ref{FigArtifDvC}, one should ask whether the universal, empirical patterns are reproduced by any of the six illustrated model scenarios.
Qualitatively, the patterns are defined first in terms of a higher compatibility between LTCD and STCB in the model-generated SCV than in the shuffled SCV and 
a higher compatibility in the shuffled SCV than in the random one, 
second in terms of the model-generated LTCD-STCB curve being close to the second diagonal.
These empirical features are visible in Fig.~\ref{FigEBM-Nom}.
It is clear that PG does not satisfy these criteria for any value of $k$.
Indeed, the model-generated curve is far below the second diagonal for most of the relevant interval and often below the shuffled curve. 
MPG, however, appears to satisfy all these criteria for all $k$ values, although for $k=3$ it is not obvious that the shuffled curve is indeed above the random one, 
due to the lack of points in the lower-left corner.
This has to do with the effective discreteness of the bounded-confidence threshold $\omega$ spectrum, due to the finite number of nominal features available --
in other words, it is meaningless to split the $\omega$ axis into intervals that are smaller than the nearest-neighbor spacing of the cultural space lattice. 
For a direct comparison with analogous empirical curves, one should use Fig.~\ref{FigEBM-Nom}, which shows the results of the LTCD-STCB analysis applied to $\text{EBM}_{\text{n}}$ data.
However, it is only meaningful to compare the qualitative nature of the empirical and the model curves, rather than the exact values, 
since, as discussed in Sec.~\ref{Disc}, neither model has a maximum-likelihood nature, due to a certain simplicity in the way prototypes are formalized and chosen here.
Still, MPG apparently does generate SCVs that are structurally similar to the empirical ones.
Thus, the notion of cultural prototypes, even if implemented in a simplistic way, 
can be used to reproduce the important, universal properties of empirical cultural states, as long as mixing of prototypes is in place.

\section{Discussion}
\label{Disc}

The purpose of this study was to develop a way of generating cultural states that reproduce the apparently universal properties of the empirical ones, namely those described by Ref.~\cite{Babeanu}.
This naturally calls for input from social science, in particular from social science theories that are intended to describe universal aspects of culture and society. 
There is an entire ``class'' of social science theories that appear relevant for this purpose, originating from either psychology or cultural anthropology~\cite{Thompson, Fiske, Triandis, Shweder, Graham},
some of them being explicit attempts at unifying social science.
All of them make use of cultural prototypes, although in somewhat different ways, under different names and numbers.
Moreover, they had all been overlooked by previous studies of cultural dynamics, on which Ref.~\cite{Babeanu} largely builds:
Ref.~\cite{Stivala} was the first study that connected quantitative studies of cultural dynamics with these theories, via the generic, formal notion of cultural prototypes. 
For creating an instructive and compact context, this work focused on one of these theories, namely on Plural Rationality Theory (PRT).

There are several aspects justifiying the focus on Plural Rationality Theory.
First, its informal notion of cultural bias matches very well the more formal notion of cultural prototype, in the manner used in Ref.~\cite{Stivala} and here.
Second, it is more appealing from a natural science perspective than the others, in particular from a physics and complex systems perspective.
This is largely due to various concepts that are qualitatively (and sometimes just implicitly) invoked by PRT, such as: energy landscapes, symmetry breaking, graph/network theory, dynamical systems, crossovers (possibly phase transitions), self-organization and fractals.
Third, it explicitly claims to provide some insight into how preferences form: preferences are formed in the process of building social relations, while different patterns of relations (and types of institutional settings) go along with different conglomerates of preferences (the cultural biases). 
Finally, this dualism between patterns of relations on one hand and cultural biases on the other hand comes along with distinguishing between a ``social plane'' and a ``cultural plane'' of interacting human systems, 
while acknowledging the dynamical nature of both, as well as the strong coupling and interdependency between the two. 
Thus, PRT seems to resonate well,
on one hand to research on social network structure and dynamics,
on the other hand to research on cultural structure and dynamics.

Up to now, little work has been done to explore either of these two connections.
While Ref.~\cite{Stivala} and the present work are the first steps in exploring the latter connection,
some steps have also been taken in exploring the former connection~\cite{Gross, Favre}.
Note, however, that Ref.~\cite{Stivala} refers to several theories similar to PRT, without explicitly mentioning PRT, 
that Ref.~\cite{Favre} focuses on a social theory similar to PRT, while still discussing a connection with PRT
and that Ref.~\cite{Gross} works with an earlier, more rudimentary version of PRT, which gave less importance to the notions of  ``way of life'', ``rationality'' and ``cultural bias''.
Although the coupling between social dynamics and cultural dynamics is recognized and studied by quantitative complex systems research (for instance Refs.~\cite{Pfau, Ngampruetikorn}), 
this has been carried out in isolation from PRT.

In loose terms, each rationality of PRT has, as a ``projection'' on the cultural plane, one distinct cultural bias. 
These cultural biases correspond to the cultural prototypes used in this study.
In agreement with Ref.~\cite{Stivala}, a cultural prototype is a combination of cultural traits, thus one point in cultural space -- 
the limitations of this assumptions are extensively discussed below. 
Relying on these notions, two stochastic, structural models of culture are developed and studied here: 
Prototype Generation (PG) and Mixed Prototype Generation (MPG). 
It is important that, regardless of which model is used, once the prototypes and the remaining free parameters (parameter $\beta$, for either PG or MPG) are specified, 
one implicitly defines a cultural space distribution (CSD): a probability mass function taking the cultural space as a support, as defined in Ref.~\cite{Babeanu}.
Generating a set of $N$ cultural vectors is then equivalent to selecting $N$ points at random according to this distribution. 
Thus, the resulting cultural states are generated in a non-uniformly random way, with non-uniformities depending on the prototypes and on other model specifications.

For this study, the usage of both stochastic models is restricted to cultural spaces constructed only from sets of nominal features. 
This is due to the assumption that every prototype picks one and only one trait in any feature, 
which from a PRT perspective means that, upon answering a question under the influence of one cultural bias, a respondent can only provide one specific answer.
In reality, even a specific cultural bias would generally point towards several answers, although with different probabilities, 
so it would be more realistic to say that every prototype corresponds to one probability distribution defined over that feature. 
Not allowing for this freedom makes this modeling paradigm incompatible to ordinal features, whose associated traits are by construction sorted along an axis,
in which case it is not reasonable to assume that a prototype points to one trait of a feature with full probability and to its nearest-neighbors with zero probability. 
Nonetheless, the paradigm is reasonably compatible with nominal features, in which case the distance between any two traits of one feature is anyway assumed to be the same. 

The current study belongs to a preliminary, simplistic paradigm which makes use of what one may call ``sharp prototypes''. 
A more realistic paradigm, which would account for the probabilistic nature of the cultural biases, would make use of what one may call ``diffuse prototypes''. 
Using sharp prototypes comes at the cost of not having enough flexibility to reproduce the empirical, feature-level frequency distributions, with either of the two models, 
since every prototype corresponds to a probability distribution entirely peaked on one trait.
Instead, using diffuse prototypes would allow this by enforcing, for every feature, that the empirical distribution is a linear combination of the prototype distributions.
Nonetheless, as shown in Sec.~\ref{ModFit}, both models are still able to reproduce the empirical average uniformity of the feature-level frequency distributions, namely the AIVD quantity. 
This is partly due to both models making some use of uniformly-random trait generation, independently of the prototypes.
This translates to a flat noise component in the probability distribution of every feature, which in a sense compensates for the rigid peaks of the sharp prototypes.
When also considering the results of Sec.~\ref{ModOut}, the usage of sharp prototypes restricted to nominal variables appears to be enough as a proof of concept.
This justifies further research towards the more sophisticated paradigm relying on diffuse prototypes. 
Although this is left for future studies, it is worth contemplating upon, in order to better understand the purpose, greater context and limitations of the current paradigm.

Working with diffuse prototypes should go hand in hand with a method of inferring them from data.
One can imagine doing this by applying a sensible clustering method on the empirical set of cultural vectors, 
followed by a sensible method of constructing one diffuse cultural prototype from every cluster, as a probabilistic entity that is representative of that cluster.
The main advantage of this approach is that once the prototypes are constructed and provided as input to a sensible stochastic model, the artificial SCVs generated with this model 
would be close-to-representative of the same distribution in cultural space as the empirical SCV on which the method is applied in the first place.
This means that the model would have a maximum-likelihood flavor, and could be used for generating synthetic data, which would also reproduce the feature-level frequency distributions.

By contrast, the approximation of sharp prototypes used here is too strong to be employed together with a method of inferring them from data.
Instead, sharp prototypes are being assigned to randomly chosen positions in the given cultural space. 
On one hand, the fact that the prototypes are randomly chosen makes any model symmetric up to any permutation of the traits of any feature, as long as all features are nominal, which is the case here, 
a symmetry which is broken by an empirical SCV and also by an artificial SCV generated from a specific choice of the prototypes.
On the other hand, the fact the prototypes are sharp does not allow for the exact frequency distribution of a specific feature to be reproduced, not even up to a permutation of the traits. 
Still, after parameter tuning, one should expect from a good model to provide a cultural space distribution whose rough ``shape'' is compatible with the empirical data, though the ``orientation'' 
and the structural details implied, for instance, by the feature-level distributions would not be compatible.
This should reflect in roughly reproducing the universal LTCD-STCB patterns emphasized in Ref.~\cite{Babeanu}: 
one one hand, the formulation of the LTCD and STCB observables is also symmetric up to permuting the traits of any feature, and thus independent of the ``orientation'';
on the other hand, the empirical, feature-level frequency distributions should heavily depend on the specific data set, thus being of little relevance for the universal patterns. 

There are various aspects that make the random generation of prototypes sensible for the purpose of the present work.
First, results are evaluated for various values of the number of prototypes $k$, which is considered a free parameter for both the PG and MPG model.
Second, the expected prototype-prototype distance is controlled for via parameter $\alpha$.
Third, for every choice of parameters, the prototypes are independently drawn for each realized cultural state in the set used for computing the model AIVD and SIVD quantities for fitting purposes.
These compensate somewhat for not inferring the prototypes from empirical data.

In order to give an example of how the sharp prototypes approximation can be pushed beyond its limits, 
it is worth recalling that fitting the MPG model is not possible for $k=2$ prototypes, as pointed out at the end of Sec.~\ref{ModFit}: 
the $\alpha$ parameter can be successfully tuned in terms of the AIVD only for small $\beta$ values, which do not allow for the subsequent fitting of the $\beta$ parameter in terms of the SIVD.
This is related to there being at least $q=3$ traits associated to every nominal feature selected from the Eurobarometer data set, 
while there are only two, prototype-induced peaks in the model probability distribution of every feature, on top of the uniform component.
Since the integrated probability of the uniform component cannot exceed $1/k$ by construction, all the distributions are bound to be relatively non-uniform, 
such that the empirical average uniformity is only attained for small-$\alpha$ (few coincidences between the prototype-induced peaks) and small-$\beta$ (large uniform component) combinations.
This does not hold for the PG model, as in this case the integrated probability of the uniform component can attain any value between 0 and 1. 
Nonetheless, if $k>2$, the fitting of the MPG leads to generated cultural states that reproduce much better the universal empirical patterns than PG.
This justifies considering MPG the successful model, while emphasizing the importance of the mixing ingredient, which validates the multiple self assumption.  

When thinking in terms of the feature-level probability distributions, it might seem that the MPG and PG models are not that different from each other. 
As mentioned above, for both models, if there are $k$ prototypes, the probability distribution of a certain feature would consist of $k$ peaks of equal probability contents and of a uniform component associated to the explicitly random trait generation.
Although the probability content of the uniform component of MPG is bounded from above, that of PG is not bounded in any way, 
so one might think that MPG is just a particular realization of PG. 
However, this reasoning is misleading, as it focuses on partial information encoded in the feature-level probability distributions, disregarding the rest of the information encoded in the complete cultural space distribution.  
With PG, a cultural vector whose trait, with respect to a certain feature, is generated under the probability peak of a certain prototype
will have its trait generated, with respect to another feature, under the well-determined probability peak of the same prototype or under the uniform component.
By contrast, with MPG, a cultural vector whose trait, with respect to a certain feature, is generated under the probability peak of a certain prototype,
will have its trait generated, with respect to another feature, under the probability peak of any prototype -- though with a higher likelihood under the peak of the dominating prototype -- or under the uniform component.
Thus, for the same choice of the prototypes and the same extent of explicitly random generation of traits (and consequently the same AIVD), PG implies a different level of cross-feature correlation and a different shape of the cultural space distribution than MPG.
This conceptually explains the impact of the mixing ingredient.
  
Although this study does not attempt at providing a complete mathematical theory of trait dynamics and formation, 
one can argue that the MPG model qualifies as a good effective,
\footnote{In this manuscript, ``effective description of'' stands for ``description of the effects of'', for ``approximate description'' or for ``phenomenological description'', as used in the Physics literature,
rather than for ``successful or ``efficacious''.} 
static description of (generic snapshots of) trait dynamics.
This static description is inspired by Plural Rationality Theory which, although originating in cultural anthropology, 
does seem to integrate notions of both psychology and of a (complex) systems based understanding of society. 
Although it is formulated in an a qualitative, informal way, Plural Rationality Theory and related research should be of use for developing a complete formal theory of trait dynamics, 
at least as a source of guidance and inspiration. 

\section{Summary and conclusions}
\label{Conc}

This study was dedicated to developing and testing a stochastic model for generating cultural states that would be structurally similar to the empirical ones.
The aim was to reproduce the universal, empirical properties pointed out in Ref.~\cite{Babeanu}, while relying on some social science hypothesis.
Following up on previous work, the idea of cultural prototypes was used for this purpose.
The study first tested the hypothesis that each cultural vector is a partial realization of one prototype and random for the rest, which is what was previously assumed. 
This turned out to be insufficient for reproducing the empirical patterns.
Instead, one has to assume that each cultural vector is a combination, or mixture of all prototypes, although still dominated by either of them,
which is what the MPG model encodes.
This additional, mixing ingredient is actually suggested by the same social science theories that inspired the prototypes idea in the first place.
In this specific, social science context, this aspect is often referred to as ``the multiple-self''.
These results provide indirect evidence for social science theories like PRT, that postulate, in one way or another, some notion of cultural prototypes, along with some associated notion of mixing.

Still, there is a certain rigidity in the way prototypes are currently formalized (Sec.~\ref{Disc}), 
related to the assumption that every prototype corresponds to one and only one value of every cultural variable, instead of corresponding to a probability distribution over the variable.
This makes the cultural space distribution induced by the successful, MPG model generally incompatible with the cultural space frequency distribution with respect to which it is fitted. 
As it stands, MPG is is far from being a maximum-likelihood type of model and thus cannot be used to generate synthetic data.
Nonetheless, this is arguably achievable once diffuse prototypes are used instead of sharp ones, while being inferred from the data rather than randomly chosen.
In this sense, this work can be seen as an important step towards a realistic, maximum-likelihood model of empirical cultural states, and towards generating synthetic sets of cultural vectors.
Moreover, MPG can be considered an effective description of the outcome of trait dynamics, since the generated cultural states seem to reproduce the generic structure of the empirical ones.
The LTCD-STCB analysis, used for validating this effective theory, could also be used for validating a more fundamental, dynamical theory of culture.
It appears likely that Plural Rationality Theory has more to say for aiding the development of such a theory.

{\bf Acknowledgments:} 

The authors are grateful to Maroussia Favre for her thoughtful comments on previous versions of this manuscript.
AIB also acknowledges discussions with Ulf Dieckmann, Gerard 't Hooft, Petter S\"{a}terskog, Frank Takes, Leandros Talman, Michael Thompson, Marco Verweij and Jorinde v.d. Vis.
DG acknowledges financial support from the Dutch Econophysics Foundation (Stichting Econophysics, Leiden, the  Netherlands). 
This work was also supported by the Netherlands Organization for Scientific Research (NWO/OCW).
The authors declare that there is no conflict of interest regarding the publication of this paper.

\bibliographystyle{unsrt}
\bibliography{Paper}{}

%\begin{thebibliography}{2}

%\end{thebibliography}

\appendix
		
\section{Controlling the generation of prototypes}
\label{AppGenProt}

This section describes the calculation of probabilities attached to sets of cultural prototypes employed by the PG and MPG models defined in Sec.~\ref{ModDesc}.
These probabilities are collectively controlled via a parameter ($\alpha$), which effectively dictates the expectation value of the average prototype-prototype cultural distance for one set of prototypes. 
The assignment of traits to prototypes is conducted independently for every feature, so the discussion is reduced to assigning probabilities to prototype-to-trait mappings at the level of a single feature. 
Furthermore, since prototype generation neglects empirical occurrence frequencies of specific traits, the problem is symmetric with respect to permutations of the traits, 
so the discussion is further reduced to assigning probabilities to ``topologies'' of prototype-to-trait mappings at the level of a single feature. 
Mathematically, such a topology is an ``integer partition''. 
Integer partitions turn out to be the mathematical objects to which elementary probabilities are to be assigned.
Sec.~\ref{AppIntPartProb} explains the procedure for assigning the probabilities to integer partitions, 
while Sec.~\ref{AppIntPartGen} explains the procedure for generating the integer partitions. 

\subsection{Integer partition probabilities}
\label{AppIntPartProb}

{\bf Let} $I_k$ be the set of all integer partitions of $k$ elements, where an integer partition of $k$ elements is an ordered sequence of integers that add up to $k$, also called ``parts''. 
Let the ordered sequence $(k_1,...,k_s) \in I_k$ be one generic element of this set, where $s$ counts the number of non-zero parts.
This notation implies that the parts are sorted for descending values $k_i \ge k_{i+1} \forall i \in\{1,..,s-1\}$ and that they add up to $k = \sum_{i=1}^s k_i$. 
For instance, $(3,2,2,1)$ is an integer partition of 8 elements with 4 parts.
For the purpose of this work, an element of the integer partition corresponds to one prototype.
For a specific choice of the prototypes and a specific feature, 
an integer partition is a representation of how the prototypes are distributed over the traits of this feature, up to a permutation of these traits.
Thus, when the fraction of traits that are randomly generated vanishes, the probabilities of the traits are just the normalized part sizes -- 
in the example above, the ordered sequence of probabilities associated to the traits would be $(\frac{3}{8},\frac{2}{8},\frac{2}{8},\frac{1}{8})$.
Random trait generation then simply introduces a uniform, noise component to the feature probability distribution, whose contribution increases with the fraction of traits that are randomly generated. 
Thus, the integer partition is in any case a proxy for the feature probability distribution, regardless of which stochastic model is used.

{\bf Let} $c(k_1,...,k_s)$ be the ``compactness'' of integer partition $(k_1,...,k_s)$, defined by: 
\begin{equation}
  c(k_1,...,k_s) = \sum_{i=1}^{s} \frac{k_i(k_i-1)}{2},
\end{equation}
which counts the number of pairs of elements belonging to the same part. 
For instance, the compactness of integer partition $(3,2,2,1)$ is $c(3,2,2,1)=3^2+1^2+1^2+0^2 = 11$. 
The compactness thus counts the prototype-prototype coincidences for one feature. 
In light of the above paragraph, a small compactness implies a high uniformity for the feature probability distribution and thus a high value of the associated (feature-level) AIVD contribution. 

{\bf Let} $I_k^q$ be the set of integer partitions of $k$ elements of at most $q$ parts (which implies that $I_k^q \subseteq I_k$).
This definition is needed for working with features with range $q < k$.
Furthermore, {\bf let} $c_{k,q}^{\text{min}}$ and $c_{k,q}^{\text{max}}$ be the minimal and maximal compactness values attainable by the elements of $I_k^q$. 
These notions are needed for normalizing generic compactness values.
They formally read: 
\begin{align}
    &	c_{k,q}^{\text{min}} = c(\lambda').(\lambda' \in I_k^q \land \nexists \lambda \in I_k^q.(c(\lambda) < c(\lambda'))), \nonumber \\
    &	c_{k,q}^{\text{max}} = c(\lambda').(\lambda' \in I_k^q \land \nexists \lambda \in I_k^q.(c(\lambda) > c(\lambda'))),
\end{align}
where the ``.'' (dot) notation stands for ``with the property that''. 

At this point, it is possible to define an non-normalized probability mass function parametrized by $\alpha$ over the discrete set of integer partitions $I_k^q$, 
function whose shape would depend on $\alpha$.
High $\alpha$ values correspond to integer partitions of high compactness values being favored over those of low compactness values, 
while low $\alpha$ values correspond to integer partitions of low compactness values being favored over those of high compactness values. 
For simplicity, the function is chosen to be monotonous when re-expressed in terms of compactness. 
A simple choice for such a function, denoted here by $\rho_{k,q}^{\alpha}$, is given by:
\begin{widetext}
\begin{equation}
	\label{IntPartProbNN}
  \rho_{k,q}^{\alpha}(\lambda) = \exp{\left\{\tan{\left[(2\alpha-1)\frac{\pi}{2}\right]}\frac{2c(\lambda) - c_{k,q}^{\text{max}} - c_{k,q}^{\text{min}}}{c_{k,q}^{\text{max}} - c_{k,q}^{\text{min}}}\right\}},
\end{equation}
\end{widetext}
where the inner fraction linearly maps the compactness $c(\lambda)$ from interval $[c_{k,q}^{\text{min}}, c_{k,q}^{\text{max}}]$ to interval $[-1,1]$, 
while the argument of the $\tan$ function linearly maps $\alpha$ from interval $(0,1)$ to interval $(-1,1)$, from where it is further mapped to $(-\infty,\infty)$ by the $\tan$ function.
In this manner, 
the function is increasing with $c(\lambda)$ for $\alpha > 0.5$ (implying a relatively low expectation value of average prototype-prototype separation),
the function is decreasing with $c(\lambda)$ for $\alpha < 0.5$ (implying a relatively high expectation value of average prototype-prototype separation)
and the function is a constant of $c(\lambda)$ for $\alpha = 0.5$. 
The actual probability $P_{k,q}^{\alpha}(\lambda)$ associated to integer partition $\lambda$ can then be obtained via the normalization:
\begin{equation}
	\label{IntPartProb}
  P_{k,q}^{\alpha}(\lambda) = \frac{\rho_{k,q}^{\alpha}(\lambda)}{\displaystyle\sum_{\lambda \in I_k^q} \rho_{k,q}^{\alpha}(\lambda)},
\end{equation}
with the sum in the denominator being taken over all integer partitions in $I_k^q$.

\subsection{Integer partition generation}
\label{AppIntPartGen}

{\bf Let} $I \eqd \{0^I, 1^I\} \cup I_1 \cup I_2 \cup ...$ be the set of all integer partitions of any size, together with a ``null'' element $0^I$ and a ``unity'' element $1^I$, 
which are meaningful in relation to the $\oplus$ operation defined below and are needed for keeping some of the following definitions compact and self-consistent.

{\bf Let} the integer partition ``merging'' $\oplus : I \times I \rightarrow I$, acting on two integer partitions of $k_a$ and $k_b$ elements, with $s_a$ and $s_b$ parts respectively, be defined in the following way:
\begin{equation}
  (k_1^a, ... k_{s_a}^a) \oplus (k_1^b, ... k_{s_b}^b) = (k_1, ... k_{s}),
\end{equation}
producing another integer partition of $k=k_a + k_b$ elements and $s=s_a+s_b$ parts, such that the sequence of parts in the resulting partition is a sorted merging of the two original sequences of parts. 
For instance: $(3,2,2,1) \oplus (4,2) = (4,3,2,2,2,1)$. 
Moreover, any integer partition $\lambda \in I$ satisfies $\lambda \oplus 0^I = 0^I$ and $\lambda \oplus 1^I = \lambda$.
 
{\bf Let} the integer partition ``multi-merging'' $\otimes: I \times \mathcal{P}(I) \rightarrow \mathcal{P}(I)$, where $\mathcal{P}(I)$ is the set of all subsets of $I$, be defined by:
\begin{equation}
  \alpha \otimes \{\alpha_1, ..., \alpha_{\sigma}\} = \{ \alpha \oplus \alpha_1, ..., \alpha \oplus \alpha_{\sigma} \},
\end{equation}
where $\alpha, \alpha_1, ..., \alpha_{\sigma} \in I$ are all integer partitions. 
The $\otimes$ operation produces a set of integer partitions of $\sigma$ elements from an initial set of integer partitions of the same size and another integer partition $\alpha$, 
by merging $\alpha$ with each element $\alpha_i$ in the initial set via the $\oplus$ operation.

Relying on the notions above, the following recursive definition of function $\text{sip}(k, m_L, m_V) : \mathbb{N} \times \mathbb{N}^{*} \times \mathbb{N}^{*} \rightarrow \mathcal{P}(I)$ encodes the procedure for generating the set of integer partitions of $k$ elements, of maximally $m_L$ parts, with maximal part value $m_V$:
\begin{widetext}
  \begin{equation}
    \text{sip}(k, m_L, m_V) = 
    \begin{cases} 
      \{1^I\} 										& k = 0,  \\ 
      \{0^I\} 										& k > m_L \cdot m_V, \\
      \{(m_V, ..., m_V)_{m_L \text{ entries}}\} 					& k = m_L \cdot m_V, \\
      \bigcup_{x \in \overline{1,\min(k,m_V)}} \left[ (x) \otimes \text{sip}(k-x, m_L-1, x) \right] 	& \text{else.}
    \end{cases}
  \end{equation}
\end{widetext}
definition inspired by Ref.~\cite{forum}, where the order of the four cases matters, in the sense that one case is considered only if none of the conditions of the above cases is valid.
The last line returns the set resulted from the reunion ``$\cup$'' of all sets of integer partitions of type $(x) \otimes \text{sip}(k-x, m_L-1, x)$, where $x$ spans the indicated interval. 
This general formulation, which also takes the maximal part value $m_V$ as argument, is required for a compact recursive definition.
But of actual interest for this work is the set of integer partitions of $k$ elements and maximal part value $q$, $I_k^q$, given by:
\begin{equation}
  I_k^q = \text{sip}(k,q,k) - \{0^I, 1^I\},
\end{equation}
where the last part of the expression takes out the null and/or the unity element, which might be present in the set of integer partitions as leftovers from the computation.
Here we explicitly show how the sip function works when calculating the set of integer partitions of 4 elements of maximally 3 parts, given by $I_4^3 = \text{sip}(4,3,4) - \{0^I, 1^I\}$, where:
\begin{widetext}
\begin{align}
  \text{sip}(4,3,4) & = \bigcup_{x \in \overline{1,4}} (x) \otimes \text{sip}(4-x, 2, x) \nonumber \\
		    & = [(1) \otimes \text{sip}(3, 2, 1)] \cup [(2) \otimes \text{sip}(2, 2, 2)] \cup [(3) \otimes \text{sip}(1, 2, 3)] \cup [(4) \otimes \text{sip}(0, 2, 4)] \nonumber \\
		    & = [(1) \otimes \{0^I\}] \cup [(2) \otimes \bigcup_{x \in \overline{1,2}} (x) \otimes \text{sip}(2-x, 1, x)] \cup [(3) \otimes (1) \otimes \text{sip}(0,1,1)] \cup [(4) \otimes \{1^I\}] \nonumber \\
		    & = \{0^I\} \cup [(2) \otimes [[(1) \otimes \text{sip}(1,1,1)] \cup [(2) \otimes \text{sip}(0,1,2)]]] \cup [(3) \otimes (1) \otimes \{1^I\}] \cup \{(4)\} \nonumber \\
		    & = \{0^I\} \cup [(2) \otimes [[(1) \otimes \{(1)\}] \cup [(2) \otimes \{1^I\}]]] \cup [(3) \otimes \{(1)\}] \cup \{(4)\} \nonumber \\
		    & = \{0^I\} \cup [(2) \otimes [\{(1,1)\} \cup \{(2)\}]] \cup \{(3,1)\} \cup \{(4)\} \nonumber \\
		    & = \{0^I\} \cup [(2) \otimes \{(1,1), (2)\}] \cup \{(3,1), (4)\} \nonumber \\
		    & = \{0^I\} \cup \{(2,1,1), (2,2)\} \cup \{(3,1), (4)\} \nonumber \\
		    & = \{0^I, (2,1,1), (2,2), (3,1), (4)\}. \nonumber
\end{align}
\end{widetext}
yealding $I_4^3 = \{(2,1,1), (2,2), (3,1), (4)\}$, which is the expected result.

\section{Analytic calculations of model average inter-vector distance}
\label{AppAnAIVD}

This section explains the analytic calculation for the expectation value of the average inter-vector distance (AIVD) for sets of cultural vectors generated using either the PG or MPG model. 
The first part of this section just gives the essential formulas --
Eqs. \eqref{AvAIVD} and \eqref{AvAIVD_q}) are common for the two models; the difference between the model becomes apparent when comparing Eq.~\ref{pi_PG} with Eq.~\ref{pi_MPG}. 
The second part gives the proof Eq. \eqref{IdeaAIVD}, which is the basis for Eq. \eqref{AvAIVD_q}.

The expectation value of the AIVD, as a function of the three model parameters $k, \alpha, \beta$ is given by the average over the feature-level expectation values:
  \begin{equation}
    \label{AvAIVD}
    \left<\aivd\right>_{\alpha, \beta}^k = \frac{1}{F} \sum_q n_q \left<\aivd\right>_{\alpha, \beta}^{k,q},
  \end{equation}
where the sum goes over all possible values ranges $q$ and $n_q$ is the number of features with range $q$, with $\sum_q n_q = F$ being implicitly satisfied, where $F$ is the number of features. 
Note that the feature-level contribution also depends on $q$. In turn, this contribution is given by:  
\begin{widetext}  
  \begin{equation}
    \label{AvAIVD_q} \hspace{-0.3cm}
    \left<\aivd\right>_{\alpha, \beta}^{k,q} = 1 - \sum_{(k_1,...,k_s)}^{I_k^q} P_{k,q}^{\alpha} (k_1,...,k_s)
    \left\{ \sum_{i=1}^{s} \left[ \pi_{\beta,F}^k \frac{k_i}{k} + \left( 1 - \pi_{\beta,F}^k \right) \frac{1}{q} \right]^2 + (q-s) \left(\frac{1-\pi_{\beta,F}^k}{q}\right)^2 \right\},
  \end{equation}
\end{widetext}
which is essentially a weighted averaging of Eq. \eqref{IdeaAIVD} over the set of integer partitions $(k_1,...,k_s) \in I_k^q$,
where the weights are the integer partition probabilities $P_{k,q}^{\alpha} (k_1,...,k_s)$.
These are calculated in the manner described in Sec.~\ref{AppIntPartProb},
while the integer partitions themselves are generated in the manner described in Sec.~\ref{AppIntPartGen}.
The set of $p_i$'s of Eq. \eqref{IdeaAIVD} depends on the integer partition in the manner illustrated between the braces of Eq. \eqref{AvAIVD_q}, 
where the first term accounts for the $s$ traits that are covered by the (non-zero) elements of the integer partition, 
namely those under the peak(s) of one (or more) prototype and under the flat noise component,
while the second term accounts for the remaining $q-s$ traits,
namely those that are only under the flat noise component.
The dependence on whether the PG or the MPG model is used is captured by $\pi_{\beta,F}$, which is the average fraction of traits directly copied from prototypes, given by:
  \begin{equation}
    \label{pi_PG}
    \pi_{\beta,F}^k = \frac{\text{round}(\beta F)}{F},
  \end{equation}
for PG, where the ``round'' function accounts for the fact that only integer numbers of traits can be copied, and:
  \begin{equation}
    \label{pi_MPG}
    \pi_{\beta,F}^k  = \frac{1}{|W_{k+1}^{\beta}|} \sum_{w}^{W_{k+1}^{\beta}} w,
  \end{equation}
for MPG, where $w$ iterates over all values of $W_{k+1}^{\beta}$, which is a large sequence of lowest MPG discrete weights (see Sec.~\ref{ModDesc}), 
which are numerically generated during a previous step, for each used combination of $(k,\beta)$ values. 
$|W_{k+1}^{\beta}|$ is the number of elements in this sequence of discrete weights. 
For this study, $|W_{k+1}^{\beta}| = 10^5$ elements were generated for every $(k,\beta)$ combination, 
which allows for a very precise numerical calculation of $\pi_{\beta,F}^k$ in the case of MPG.

\begin{figure*}
\centering
	\subfigure{\includegraphics[width=8cm]{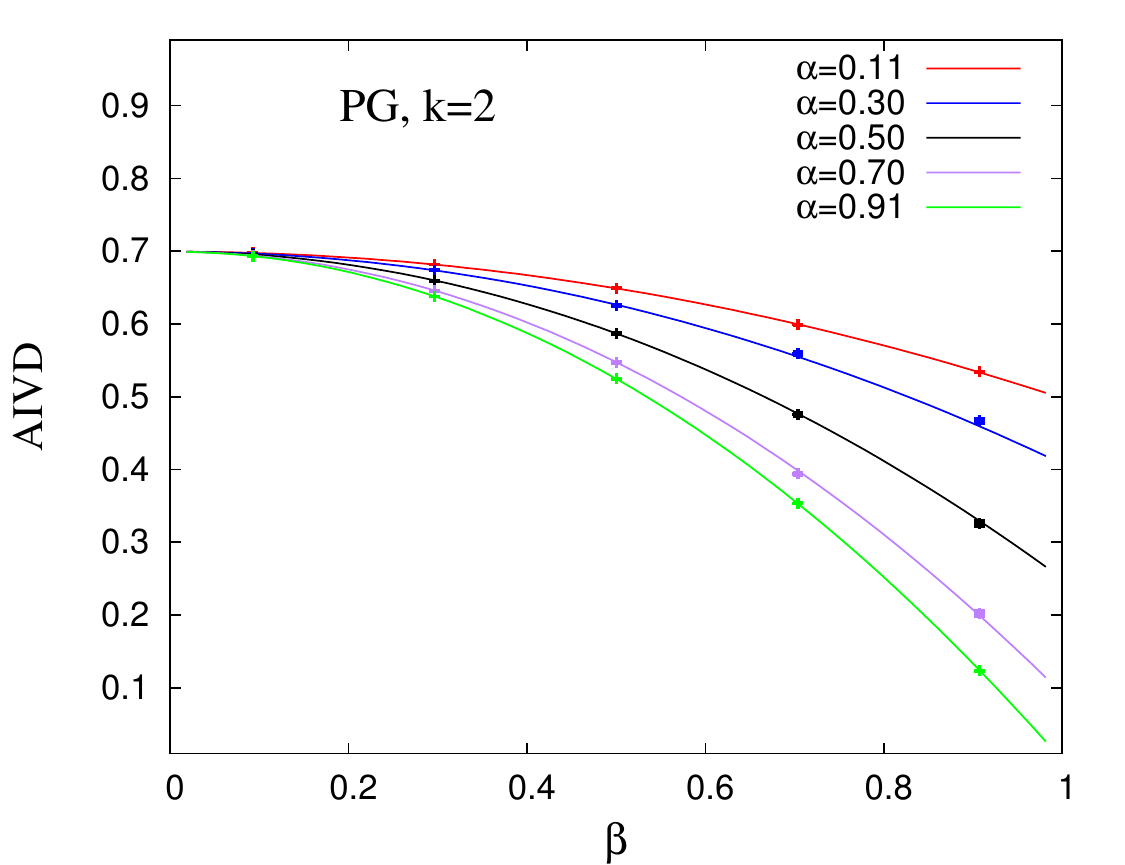}}
	\subfigure{\includegraphics[width=8cm]{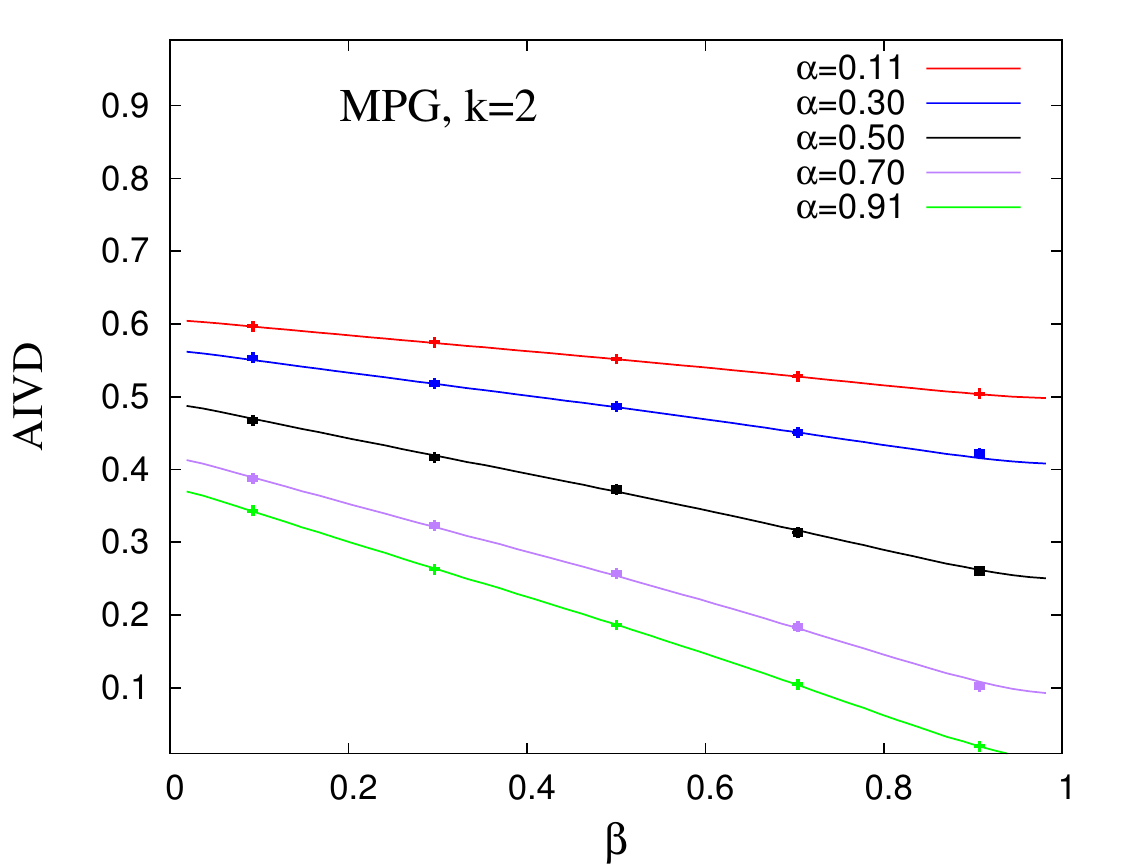}} \\
	\subfigure{\includegraphics[width=8cm]{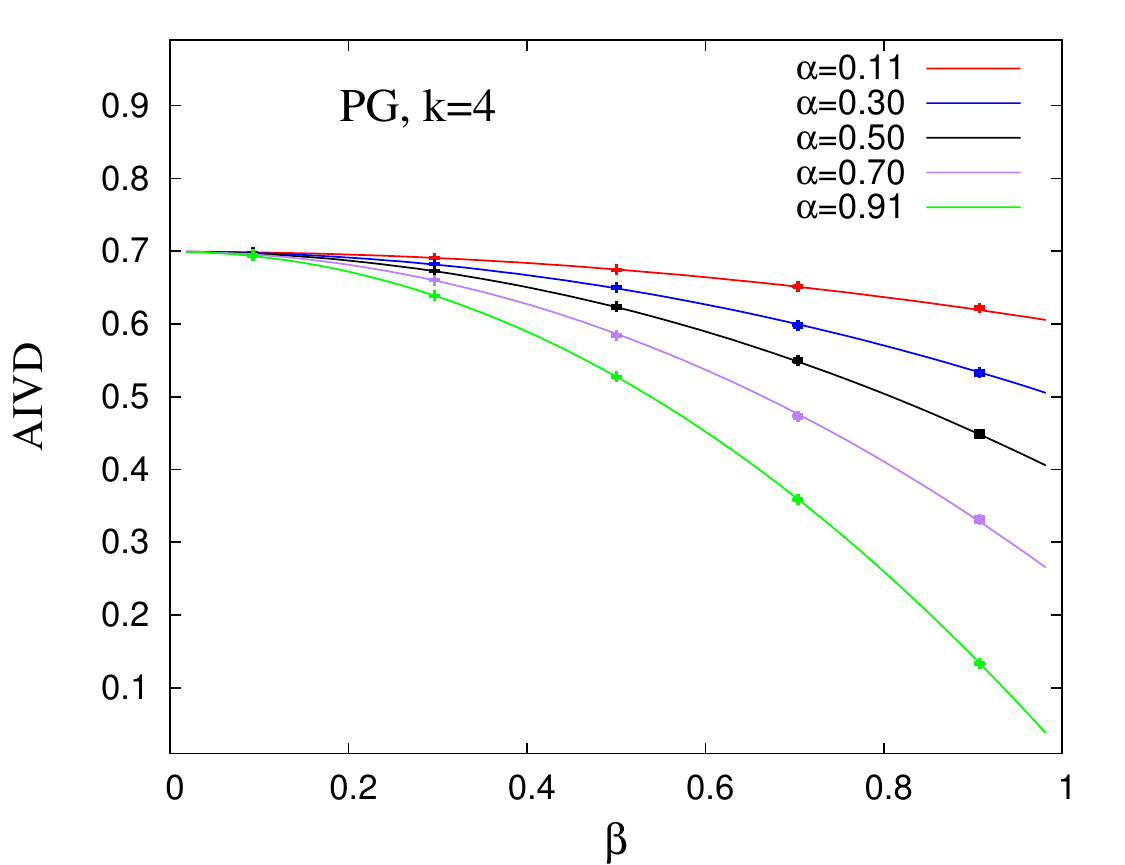}}
	\subfigure{\includegraphics[width=8cm]{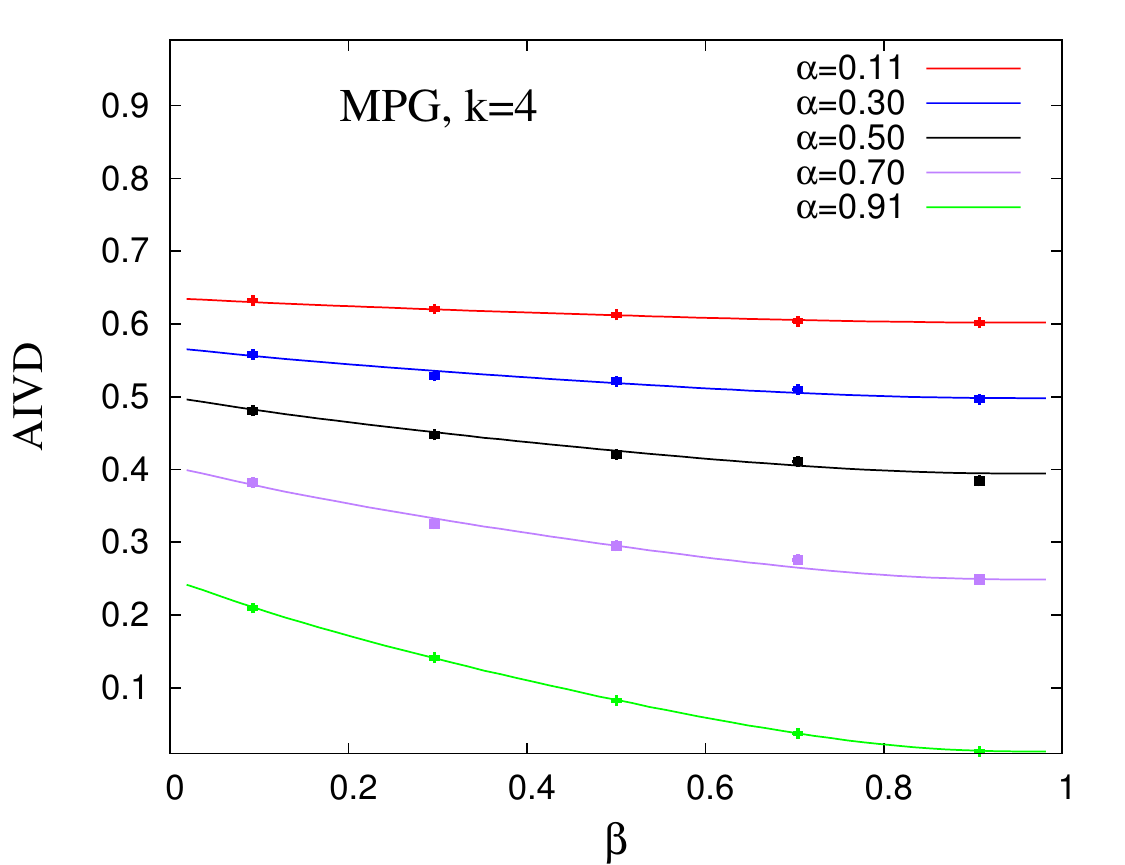}} \\
	\subfigure{\includegraphics[width=8cm]{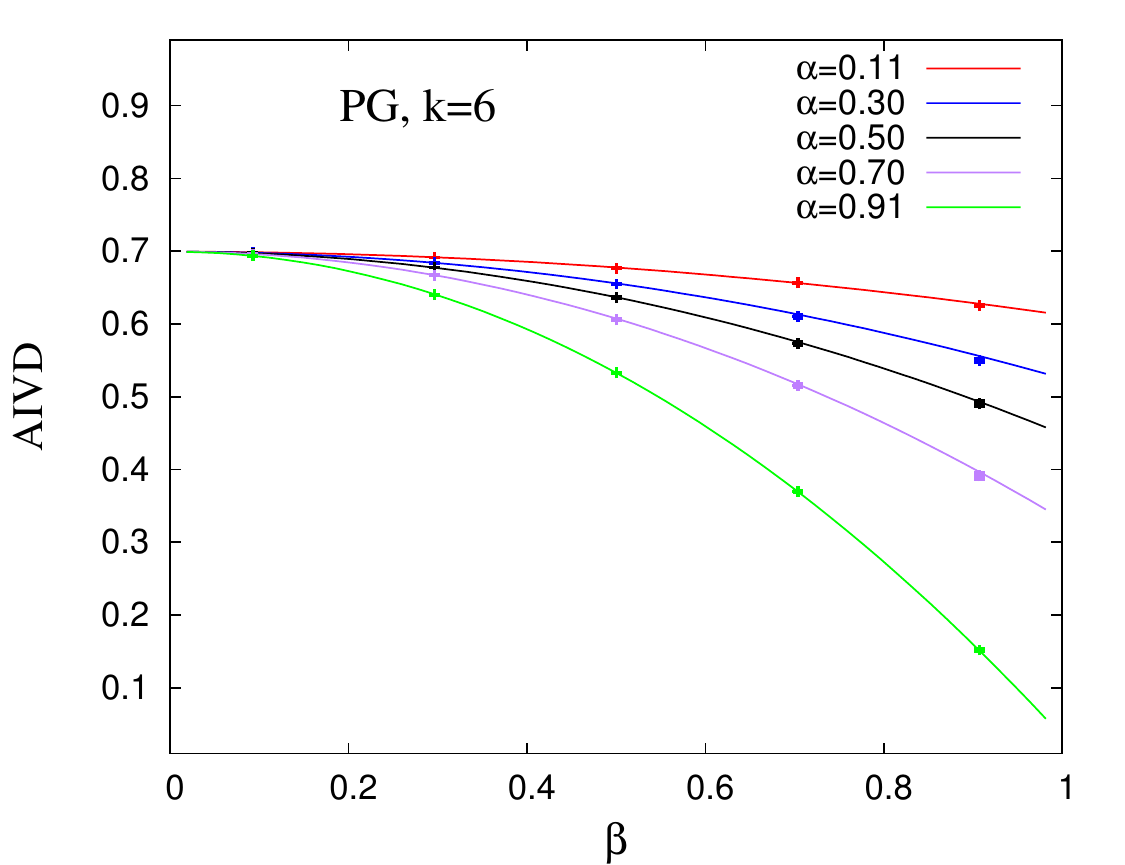}}
	\subfigure{\includegraphics[width=8cm]{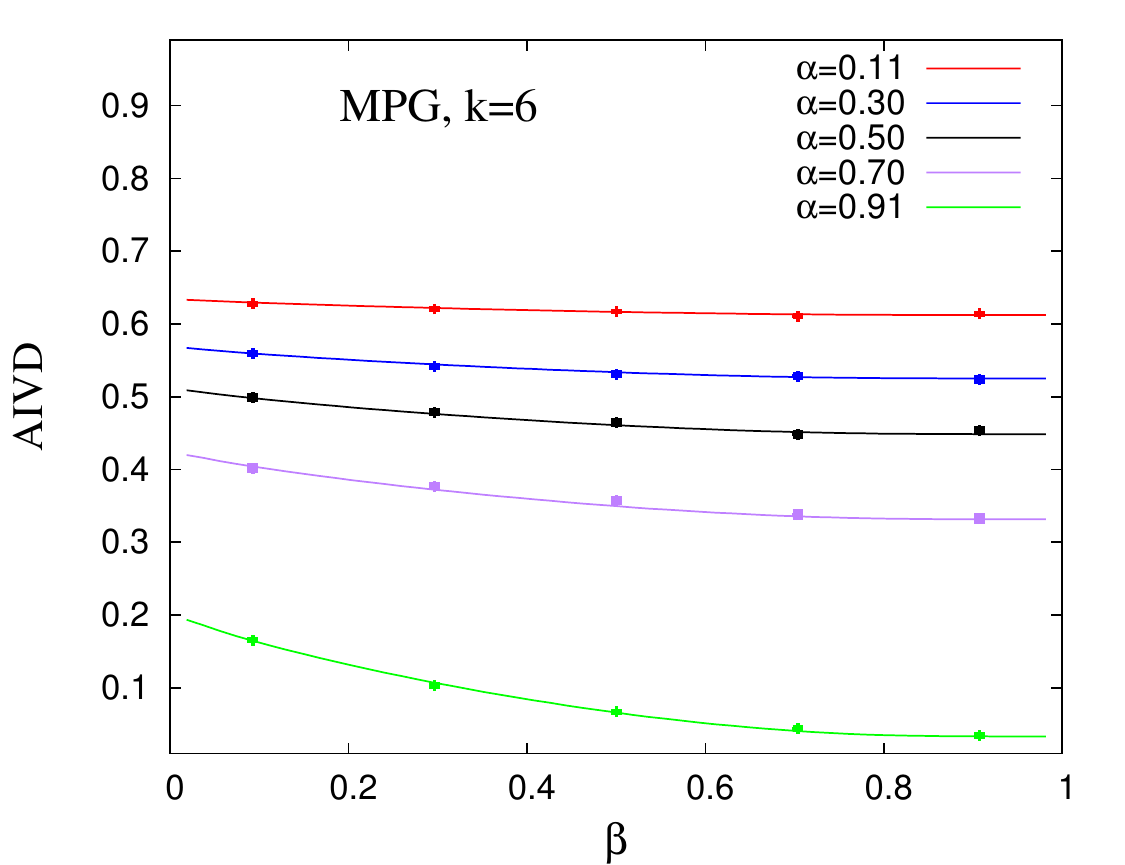}}
	\caption{Comparison between numerical (dots) and analytical (line) expected AIVD as a function of $\beta$, for the PG (left) and MPG (right) models,
						with $k=2$ (top), $k=4$ (center) and $k=6$ (bottom) prototypes, 
						for several values of $\alpha$ (legend).
}
\label{FigCAIVD}
\end{figure*} 
  
The consistency between the analytical AIVD calculation explained above and the numerical calculation is illustrated here via Fig.~\ref{FigCAIVD}.
The expected AIVD value is shown as a function of the $\beta$ parameter, for 5 values of the $\alpha$ parameter and 3 values of the $k$ parameter, for both the PG and MPG models.
The analytical values are shown by the lines, while the numerical ones are shown by the dots, which have small, almost indiscernible error bars attached. 
For the numerical case, 50 sets of $N=500$ cultural vectors are generated for each combination of parameters.
Note that the numerical profiles follow closely the analytical ones, with small deviations that are consistent with the expected fluctuations of the mean.
    
It is now worth presenting a proof of Eq. \eqref{IdeaAIVD}, on which Eq. \eqref{AvAIVD_q} is based. 
Consider a feature with $q$ traits and a set of a-priori probabilities $\{p_1,...,p_q\}$ attached to them.
Then, the entry of each cultural vector generated with respect to this feature is an independent, random choice from the $q$ traits, according to the probability mass function $(p_1,...,p_q)$.
Thus, the expected AIVD contribution from $N$ cultural vectors is given by: 
\begin{widetext}  
  \begin{align}
    \label{ProofAIVD_1}
    \langle \aivd\left(\{p_1,...,p_q\}\right) \rangle	& = 1 - \frac{2}{N(N-1)} \sum_{x_1,...,x_q}^{x_1+...+x_q=N} \sum_{i=1}^{q} \frac{x_i(x_i-1)}{2} f\left(N\text{, }_{p_1,...,p_q}^{x_1,...,x_q}\right), \nonumber \\
    \langle \aivd\left(\{p_1,...,p_q\}\right) \rangle	& = 1 - \frac{2}{N(N-1)} \sum_{i=1}^{q} \sum_{x_i}^{x_i \le N} \frac{x_i(x_i-1)}{2} \sum_{x_1,...,x_{i-1},x_{i+1},...,x_q}^{x_1+...+x_{i-1}+x_{i+1}+...+x_q=N-x_i} f\left(N\text{, }_{p_1,...,p_q}^{x_1,...,x_q}\right), \nonumber \\
    \langle \aivd\left(\{p_1,...,p_q\}\right) \rangle	& = 1 - \frac{2}{N(N-1)} \sum_{i=1}^{q} \sum_{x_i}^{x_i \le N} \frac{x_i(x_i-1)}{2} S_i.
%       
%					& = 1 - \sum_{i=1}^q p_i^2 \nonumber \\
  \end{align}
\end{widetext}
where $f\left(N\text{, }_{p_1,...,p_q}^{x_1,...,x_q}\right)$ denotes the probability that the N independent, random variables fill the $q$ traits with the frequency distribution $(x_1,...,x_q)$,
given the associated probability distribution $(p_1,...,p_q)$, where $\sum_{i-1}^q x_i = N$. 
This is conventionally called the multinominal distribution.
In the above derivation, $S_i$ stands for the summation over all elements of the multinominal except that which has a certain, $x_i$ number of entries for the $i$th trait, which can be further manipulated:
\begin{widetext}
  \begin{align}
    \label{ProofAIVD_2}
    S_i	& = \sum_{x_1,...,x_{i-1},x_{i+1},...,x_q}^{x_1+...+x_{i-1}+x_{i+1}+...+x_q=N-x_i} f\left(N\text{, }_{p_1,...,p_q}^{x_1,...,x_q}\right), \nonumber \\
    S_i	& = \sum_{x_1,...,x_{i-1},x_{i+1},...,x_q}^{x_1+...+x_{i-1}+x_{i+1}+...+x_q=N-x_i} \frac{N!}{x_1!...x_{i-1}!x_i!x_{i+1}!...x_q!} p_1^{x_1}...p_{i-1}^{x_{i-1}} p_{i}^{x_{i}} p_{i+1}^{x_{i+1}}...p_q^{x_q}, \nonumber \\
    S_i	& = p_i^{x_i}\frac{N!}{x_i!(N-x_i)!}\sum_{x_1,...,x_{i-1},x_{i+1},...,x_q}^{x_1+...+x_{i-1}+x_{i+1}+...+x_q=N-x_i} \frac{(N-x_i)!}{x_1!...x_{i-1}!x_{i+1}!...x_q!} p_1^{x_1}...p_{i-1}^{x_{i-1}} p_{i+1}^{x_{i+1}}...p_q^{x_q}, \nonumber \\
    S_i	& = \left(^N_{x_i}\right) p_i^{x_i} (1-p_i)^{N-x_i}.
  \end{align}
\end{widetext}
This shows that $S_i$ is just a term of the binomial distribution. 
By inserting the final expression of Eq. \eqref{ProofAIVD_2} in the final expression of \eqref{ProofAIVD_1}, one gets:
\begin{widetext}
\begin{align}
    \label{ProofAIVD_3}
        \langle \aivd\left(\{p_1,...,p_q\}\right) \rangle	& = 1 - \frac{1}{N(N-1)} \sum_{i=1}^{q} \sum_{x_i}^{x_i \le N} (x_i^2 - x_i) \left(^N_{x_i}\right) p_i^{x_i} (1-p_i)^{N-x_i}, \nonumber \\
        \langle \aivd\left(\{p_1,...,p_q\}\right) \rangle       & = 1 - \frac{1}{N(N-1)} \sum_{i=1}^{q} \left[Np_i(Np_i - p_i +1) - Np_i \right], \nonumber \\
        \langle \aivd\left(\{p_1,...,p_q\}\right) \rangle       & = 1 - \sum_{i=1}^{q} p_i^2. 
\end{align}
\end{widetext}
which concludes the proof of Eq. \eqref{IdeaAIVD}, after using the well known expressions for the first and second moments $\langle x_i \rangle$ and $\langle x_i^2 \rangle$ of the binomial distribution.
Note that the dependence on $N$ cancels out during the derivation.

Another, arguably shorter proof can be formulated with the aid of indicator functions of the type $\mathbb{I}_i(x)$, 
which gives $1$ if cultural vector $x$ is an entry of trait $i$ and gives $0$ otherwise. 
One can express the feature-level AIVD of one, generic set of cultural vectors in terms of indicator functions and write the expected, feature-level AIVD as an average of this expression. 
The $p_i^2$ part of Eq. \eqref{IdeaAIVD} then appears from an averaging of the $\mathbb{I}_i(x) \mathbb{I}_i(y)$ product, where $x$ and $y$ are two arbitrary cultural vectors.

\input{AppFitAlg.tex}

\end{document}

%% file: Sketch.pdf_tex
%% Creator: Inkscape inkscape 0.91, www.inkscape.org
%% PDF/EPS/PS + LaTeX output extension by Johan Engelen, 2010
%% Accompanies image file 'Sketch.pdf' (pdf, eps, ps)
%%
%% To include the image in your LaTeX document, write
%%   \input{<filename>.pdf_tex}
%%  instead of
%%   \includegraphics{<filename>.pdf}
%% To scale the image, write
%%   \def\svgwidth{<desired width>}
%%   \input{<filename>.pdf_tex}
%%  instead of
%%   \includegraphics[width=<desired width>]{<filename>.pdf}
%%
%% Images with a different path to the parent latex file can
%% be accessed with the `import' package (which may need to be
%% installed) using
%%   \usepackage{import}
%% in the preamble, and then including the image with
%%   \import{<path to file>}{<filename>.pdf_tex}
%% Alternatively, one can specify
%%   \graphicspath{{<path to file>/}}
%% 
%% For more information, please see info/svg-inkscape on CTAN:
%%   http://tug.ctan.org/tex-archive/info/svg-inkscape
%%
\begingroup%
  \makeatletter%
  \providecommand\color[2][]{%
    \errmessage{(Inkscape) Color is used for the text in Inkscape, but the package 'color.sty' is not loaded}%
    \renewcommand\color[2][]{}%
  }%
  \providecommand\transparent[1]{%
    \errmessage{(Inkscape) Transparency is used (non-zero) for the text in Inkscape, but the package 'transparent.sty' is not loaded}%
    \renewcommand\transparent[1]{}%
  }%
  \providecommand\rotatebox[2]{#2}%
  \ifx\svgwidth\undefined%
    \setlength{\unitlength}{567.45561342bp}%
    \ifx\svgscale\undefined%
      \relax%
    \else%
      \setlength{\unitlength}{\unitlength * \real{\svgscale}}%
    \fi%
  \else%
    \setlength{\unitlength}{\svgwidth}%
  \fi%
  \global\let\svgwidth\undefined%
  \global\let\svgscale\undefined%
  \makeatother%
  \begin{picture}(1,0.22130914)%
    \put(0,0){\includegraphics[width=\unitlength,page=1]{Sketch.pdf}}%
    \put(0.46090975,0.22552773){\color[rgb]{0,0,0}\makebox(0,0)[lt]{\begin{minipage}{0.06107656\unitlength}\raggedright $x_1$\end{minipage}}}%
    \put(0.48513964,0.22552773){\color[rgb]{0,0,0}\makebox(0,0)[lt]{\begin{minipage}{0.06107656\unitlength}\raggedright $x_2$\end{minipage}}}%
    \put(0.51456166,0.22552773){\color[rgb]{0,0,0}\makebox(0,0)[lt]{\begin{minipage}{0.06107656\unitlength}\raggedright $x_3$\end{minipage}}}%
    \put(0.63410365,0.22552773){\color[rgb]{0,0,0}\makebox(0,0)[lt]{\begin{minipage}{0.06107656\unitlength}\raggedright $x_N$\end{minipage}}}%
    \put(0.53701024,0.22552773){\color[rgb]{0,0,0}\makebox(0,0)[lt]{\begin{minipage}{0.06107656\unitlength}\raggedright $x_4$\end{minipage}}}%
    \put(0.56643226,0.22552773){\color[rgb]{0,0,0}\makebox(0,0)[lt]{\begin{minipage}{0.06107656\unitlength}\raggedright $x_5$\end{minipage}}}%
    \put(0,0){\includegraphics[width=\unitlength,page=2]{Sketch.pdf}}%
    \put(0.46195268,0.19474088){\color[rgb]{0,0.62745098,0}\makebox(0,0)[lt]{\begin{minipage}{0.01415836\unitlength}\raggedright C\end{minipage}}}%
    \put(0.46195268,0.14417753){\color[rgb]{0,0.62745098,0}\makebox(0,0)[lt]{\begin{minipage}{0.01415836\unitlength}\raggedright B\end{minipage}}}%
    \put(0.46195268,0.05626688){\color[rgb]{0,0.62745098,0}\makebox(0,0)[lt]{\begin{minipage}{0.01415836\unitlength}\raggedright C\end{minipage}}}%
    \put(0.46195268,0.11922135){\color[rgb]{0,0.62745098,0}\makebox(0,0)[lt]{\begin{minipage}{0.01415836\unitlength}\raggedright B\end{minipage}}}%
    \put(0.46195268,0.17100715){\color[rgb]{0,0,0}\makebox(0,0)[lt]{\begin{minipage}{0.01415836\unitlength}\raggedright B\end{minipage}}}%
    \put(0,0){\includegraphics[width=\unitlength,page=3]{Sketch.pdf}}%
    \put(0.4887823,0.16952757){\color[rgb]{0,0,1}\makebox(0,0)[lt]{\begin{minipage}{0.01601269\unitlength}\raggedright E\end{minipage}}}%
    \put(0.4887823,0.14417753){\color[rgb]{0,0,1}\makebox(0,0)[lt]{\begin{minipage}{0.01415836\unitlength}\raggedright C\end{minipage}}}%
    \put(0.48699365,0.05626688){\color[rgb]{0,0,1}\makebox(0,0)[lt]{\begin{minipage}{0.01415836\unitlength}\raggedright C\end{minipage}}}%
    \put(0.4887823,0.19635719){\color[rgb]{0,0,0}\makebox(0,0)[lt]{\begin{minipage}{0.01601269\unitlength}\raggedright A\end{minipage}}}%
    \put(0.4887823,0.1209252){\color[rgb]{0,0,0}\makebox(0,0)[lt]{\begin{minipage}{0.01990757\unitlength}\raggedright D\end{minipage}}}%
    \put(0,0){\includegraphics[width=\unitlength,page=4]{Sketch.pdf}}%
    \put(0.51382328,0.19474088){\color[rgb]{0,0,1}\makebox(0,0)[lt]{\begin{minipage}{0.01415836\unitlength}\raggedright B\end{minipage}}}%
    \put(0.51382328,0.16952757){\color[rgb]{0,0,1}\makebox(0,0)[lt]{\begin{minipage}{0.01601269\unitlength}\raggedright E\end{minipage}}}%
    \put(0.51382328,0.14417753){\color[rgb]{0,0,1}\makebox(0,0)[lt]{\begin{minipage}{0.01415836\unitlength}\raggedright C\end{minipage}}}%
    \put(0.51382328,0.11922135){\color[rgb]{0,0,1}\makebox(0,0)[lt]{\begin{minipage}{0.01415836\unitlength}\raggedright A\end{minipage}}}%
    \put(0.51382328,0.05626688){\color[rgb]{0,0,1}\makebox(0,0)[lt]{\begin{minipage}{0.01415836\unitlength}\raggedright C\end{minipage}}}%
    \put(0,0){\includegraphics[width=\unitlength,page=5]{Sketch.pdf}}%
    \put(0.5406529,0.19474088){\color[rgb]{0.78431373,0,0}\makebox(0,0)[lt]{\begin{minipage}{0.01415836\unitlength}\raggedright A\end{minipage}}}%
    \put(0.5406529,0.16952757){\color[rgb]{0.78431373,0,0}\makebox(0,0)[lt]{\begin{minipage}{0.01601269\unitlength}\raggedright D\end{minipage}}}%
    \put(0.5406529,0.14417753){\color[rgb]{0.78431373,0,0}\makebox(0,0)[lt]{\begin{minipage}{0.01415836\unitlength}\raggedright B\end{minipage}}}%
    \put(0.5406529,0.11922135){\color[rgb]{0.78431373,0,0}\makebox(0,0)[lt]{\begin{minipage}{0.01415836\unitlength}\raggedright C\end{minipage}}}%
    \put(0.5406529,0.05626688){\color[rgb]{0,0,0}\makebox(0,0)[lt]{\begin{minipage}{0.01415836\unitlength}\raggedright B\end{minipage}}}%
    \put(0,0){\includegraphics[width=\unitlength,page=6]{Sketch.pdf}}%
    \put(0.63902817,0.19474088){\color[rgb]{0.78431373,0,0}\makebox(0,0)[lt]{\begin{minipage}{0.01415836\unitlength}\raggedright A\end{minipage}}}%
    \put(0.63902817,0.16952757){\color[rgb]{0,0,0}\makebox(0,0)[lt]{\begin{minipage}{0.01601269\unitlength}\raggedright A\end{minipage}}}%
    \put(0.63902817,0.14417753){\color[rgb]{0.78431373,0,0}\makebox(0,0)[lt]{\begin{minipage}{0.01415836\unitlength}\raggedright B\end{minipage}}}%
    \put(0.63902817,0.11922135){\color[rgb]{0,0,0}\makebox(0,0)[lt]{\begin{minipage}{0.01415836\unitlength}\raggedright C\end{minipage}}}%
    \put(0.63902817,0.05626688){\color[rgb]{0.78431373,0,0}\makebox(0,0)[lt]{\begin{minipage}{0.01415836\unitlength}\raggedright A\end{minipage}}}%
    \put(0,0){\includegraphics[width=\unitlength,page=7]{Sketch.pdf}}%
    \put(0.56569388,0.19474088){\color[rgb]{0,0.62745098,0}\makebox(0,0)[lt]{\begin{minipage}{0.01415836\unitlength}\raggedright C\end{minipage}}}%
    \put(0.56569388,0.16952757){\color[rgb]{0,0.62745098,0}\makebox(0,0)[lt]{\begin{minipage}{0.01601269\unitlength}\raggedright A\end{minipage}}}%
    \put(0.56569388,0.14417753){\color[rgb]{0,0,0}\makebox(0,0)[lt]{\begin{minipage}{0.01415836\unitlength}\raggedright A\end{minipage}}}%
    \put(0.56569388,0.11922135){\color[rgb]{0,0.62745098,0}\makebox(0,0)[lt]{\begin{minipage}{0.01415836\unitlength}\raggedright B\end{minipage}}}%
    \put(0.56569388,0.05626688){\color[rgb]{0,0.62745098,0}\makebox(0,0)[lt]{\begin{minipage}{0.01415836\unitlength}\raggedright C\end{minipage}}}%
    \put(0.45419739,0.01783298){\color[rgb]{0,0,0}\makebox(0,0)[lt]{\begin{minipage}{0.14887816\unitlength}\raggedright {\bf PG} vectors\end{minipage}}}%
    \put(0,0){\includegraphics[width=\unitlength,page=8]{Sketch.pdf}}%
    \put(0.76873853,0.22541438){\color[rgb]{0,0,0}\makebox(0,0)[lt]{\begin{minipage}{0.06104567\unitlength}\raggedright $x_1$\end{minipage}}}%
    \put(0.79295617,0.22541438){\color[rgb]{0,0,0}\makebox(0,0)[lt]{\begin{minipage}{0.06104567\unitlength}\raggedright $x_2$\end{minipage}}}%
    \put(0.82236331,0.22541438){\color[rgb]{0,0,0}\makebox(0,0)[lt]{\begin{minipage}{0.06104567\unitlength}\raggedright $x_3$\end{minipage}}}%
    \put(0.94184485,0.22541438){\color[rgb]{0,0,0}\makebox(0,0)[lt]{\begin{minipage}{0.06104567\unitlength}\raggedright $x_N$\end{minipage}}}%
    \put(0.84480054,0.22541438){\color[rgb]{0,0,0}\makebox(0,0)[lt]{\begin{minipage}{0.06104567\unitlength}\raggedright $x_4$\end{minipage}}}%
    \put(0.87420768,0.22541438){\color[rgb]{0,0,0}\makebox(0,0)[lt]{\begin{minipage}{0.06104567\unitlength}\raggedright $x_5$\end{minipage}}}%
    \put(0,0){\includegraphics[width=\unitlength,page=9]{Sketch.pdf}}%
    \put(0.76978093,0.1946431){\color[rgb]{0,0.62745098,0}\makebox(0,0)[lt]{\begin{minipage}{0.0141512\unitlength}\raggedright C\end{minipage}}}%
    \put(0.76978093,0.14410532){\color[rgb]{0,0.62745098,0}\makebox(0,0)[lt]{\begin{minipage}{0.0141512\unitlength}\raggedright B\end{minipage}}}%
    \put(0.76978093,0.11916176){\color[rgb]{0,0.62745098,0}\makebox(0,0)[lt]{\begin{minipage}{0.0141512\unitlength}\raggedright B\end{minipage}}}%
    \put(0.76978093,0.16944254){\color[rgb]{0.78431373,0,0}\makebox(0,0)[lt]{\begin{minipage}{0.01600459\unitlength}\raggedright D\end{minipage}}}%
    \put(0.77156867,0.05623912){\color[rgb]{0.78431373,0,0}\makebox(0,0)[lt]{\begin{minipage}{0.0141512\unitlength}\raggedright A\end{minipage}}}%
    \put(0,0){\includegraphics[width=\unitlength,page=10]{Sketch.pdf}}%
    \put(0.79659698,0.16940382){\color[rgb]{0,0,1}\makebox(0,0)[lt]{\begin{minipage}{0.01600459\unitlength}\raggedright E\end{minipage}}}%
    \put(0.79659698,0.14410532){\color[rgb]{0,0,1}\makebox(0,0)[lt]{\begin{minipage}{0.0141512\unitlength}\raggedright C\end{minipage}}}%
    \put(0.79480925,0.05623912){\color[rgb]{0,0,1}\makebox(0,0)[lt]{\begin{minipage}{0.0141512\unitlength}\raggedright C\end{minipage}}}%
    \put(0.79659698,0.1946431){\color[rgb]{0,0.62745098,0}\makebox(0,0)[lt]{\begin{minipage}{0.0141512\unitlength}\raggedright C\end{minipage}}}%
    \put(0.79659698,0.11916176){\color[rgb]{0,0.62745098,0}\makebox(0,0)[lt]{\begin{minipage}{0.0141512\unitlength}\raggedright B\end{minipage}}}%
    \put(0,0){\includegraphics[width=\unitlength,page=11]{Sketch.pdf}}%
    \put(0.8216253,0.1946431){\color[rgb]{0,0,1}\makebox(0,0)[lt]{\begin{minipage}{0.0141512\unitlength}\raggedright B\end{minipage}}}%
    \put(0.8216253,0.16944254){\color[rgb]{0,0,1}\makebox(0,0)[lt]{\begin{minipage}{0.01600459\unitlength}\raggedright E\end{minipage}}}%
    \put(0.8216253,0.11916176){\color[rgb]{0,0,1}\makebox(0,0)[lt]{\begin{minipage}{0.0141512\unitlength}\raggedright A\end{minipage}}}%
    \put(0.8216253,0.05623912){\color[rgb]{0,0,1}\makebox(0,0)[lt]{\begin{minipage}{0.0141512\unitlength}\raggedright C\end{minipage}}}%
    \put(0.8216253,0.14410532){\color[rgb]{0.78431373,0,0}\makebox(0,0)[lt]{\begin{minipage}{0.0141512\unitlength}\raggedright B\end{minipage}}}%
    \put(0,0){\includegraphics[width=\unitlength,page=12]{Sketch.pdf}}%
    \put(0.87525741,0.16944254){\color[rgb]{0,0.62745098,0}\makebox(0,0)[lt]{\begin{minipage}{0.01600459\unitlength}\raggedright A\end{minipage}}}%
    \put(0.87346967,0.11916176){\color[rgb]{0,0.62745098,0}\makebox(0,0)[lt]{\begin{minipage}{0.0141512\unitlength}\raggedright B\end{minipage}}}%
    \put(0.87346967,0.05623912){\color[rgb]{0,0.62745098,0}\makebox(0,0)[lt]{\begin{minipage}{0.0141512\unitlength}\raggedright C\end{minipage}}}%
    \put(0.87346967,0.14410532){\color[rgb]{0,0,1}\makebox(0,0)[lt]{\begin{minipage}{0.0141512\unitlength}\raggedright C\end{minipage}}}%
    \put(0.87525741,0.1946431){\color[rgb]{0.78431373,0,0}\makebox(0,0)[lt]{\begin{minipage}{0.0141512\unitlength}\raggedright A\end{minipage}}}%
    \put(0,0){\includegraphics[width=\unitlength,page=13]{Sketch.pdf}}%
    \put(0.84844135,0.1946431){\color[rgb]{0.78431373,0,0}\makebox(0,0)[lt]{\begin{minipage}{0.0141512\unitlength}\raggedright A\end{minipage}}}%
    \put(0.84844135,0.16944254){\color[rgb]{0.78431373,0,0}\makebox(0,0)[lt]{\begin{minipage}{0.01600459\unitlength}\raggedright D\end{minipage}}}%
    \put(0.84844135,0.14410532){\color[rgb]{0.78431373,0,0}\makebox(0,0)[lt]{\begin{minipage}{0.0141512\unitlength}\raggedright B\end{minipage}}}%
    \put(0.84844135,0.11916176){\color[rgb]{0.78431373,0,0}\makebox(0,0)[lt]{\begin{minipage}{0.0141512\unitlength}\raggedright C\end{minipage}}}%
    \put(0.84844135,0.05623912){\color[rgb]{0,0.62745098,0}\makebox(0,0)[lt]{\begin{minipage}{0.0141512\unitlength}\raggedright C\end{minipage}}}%
    \put(0.84844135,0.11916176){\color[rgb]{0.78431373,0,0}\makebox(0,0)[lt]{\begin{minipage}{0.0141512\unitlength}\raggedright C\end{minipage}}}%
    \put(0,0){\includegraphics[width=\unitlength,page=14]{Sketch.pdf}}%
    \put(0.94676689,0.1946431){\color[rgb]{0.78431373,0,0}\makebox(0,0)[lt]{\begin{minipage}{0.0141512\unitlength}\raggedright A\end{minipage}}}%
    \put(0.94676689,0.16944254){\color[rgb]{0,0,0}\makebox(0,0)[lt]{\begin{minipage}{0.01600459\unitlength}\raggedright A\end{minipage}}}%
    \put(0.94676689,0.14410532){\color[rgb]{0.78431373,0,0}\makebox(0,0)[lt]{\begin{minipage}{0.0141512\unitlength}\raggedright B\end{minipage}}}%
    \put(0.94676689,0.05623912){\color[rgb]{0.78431373,0,0}\makebox(0,0)[lt]{\begin{minipage}{0.0141512\unitlength}\raggedright A\end{minipage}}}%
    \put(0.94676689,0.11916176){\color[rgb]{0,0,1}\makebox(0,0)[lt]{\begin{minipage}{0.0141512\unitlength}\raggedright A\end{minipage}}}%
    \put(0.76435381,0.01783298){\color[rgb]{0,0,0}\makebox(0,0)[lt]{\begin{minipage}{0.14887815\unitlength}\raggedright {\bf MPG} vectors\end{minipage}}}%
    \put(-0.00159162,0.20069845){\color[rgb]{0,0,0}\makebox(0,0)[lt]{\begin{minipage}{0.06883984\unitlength}\raggedright $Q_1$:\end{minipage}}}%
    \put(-0.00159162,0.17632197){\color[rgb]{0,0,0}\makebox(0,0)[lt]{\begin{minipage}{0.06883984\unitlength}\raggedright $Q_2$:\end{minipage}}}%
    \put(-0.00159162,0.06245065){\color[rgb]{0,0,0}\makebox(0,0)[lt]{\begin{minipage}{0.06883984\unitlength}\raggedright $Q_F$:\end{minipage}}}%
    \put(-0.00159162,0.15127332){\color[rgb]{0,0,0}\makebox(0,0)[lt]{\begin{minipage}{0.06883984\unitlength}\raggedright $Q_3$:\end{minipage}}}%
    \put(-0.00159162,0.12574331){\color[rgb]{0,0,0}\makebox(0,0)[lt]{\begin{minipage}{0.06883984\unitlength}\raggedright $Q_4$:\end{minipage}}}%
    \put(0,0){\includegraphics[width=\unitlength,page=15]{Sketch.pdf}}%
    \put(0.07671936,0.19511192){\color[rgb]{0,0,0}\makebox(0,0)[lt]{\begin{minipage}{0.01413701\unitlength}\raggedright A\end{minipage}}}%
    \put(0,0){\includegraphics[width=\unitlength,page=16]{Sketch.pdf}}%
    \put(0.07671936,0.16993664){\color[rgb]{0,0,0}\makebox(0,0)[lt]{\begin{minipage}{0.01598854\unitlength}\raggedright D\end{minipage}}}%
    \put(0,0){\includegraphics[width=\unitlength,page=17]{Sketch.pdf}}%
    \put(0.07671936,0.14462483){\color[rgb]{0,0,0}\makebox(0,0)[lt]{\begin{minipage}{0.01413701\unitlength}\raggedright B\end{minipage}}}%
    \put(0,0){\includegraphics[width=\unitlength,page=18]{Sketch.pdf}}%
    \put(0.07671936,0.11970628){\color[rgb]{0,0,0}\makebox(0,0)[lt]{\begin{minipage}{0.01413701\unitlength}\raggedright A\end{minipage}}}%
    \put(0,0){\includegraphics[width=\unitlength,page=19]{Sketch.pdf}}%
    \put(0.07671936,0.05684674){\color[rgb]{0,0,0}\makebox(0,0)[lt]{\begin{minipage}{0.01413701\unitlength}\raggedright C\end{minipage}}}%
    \put(0,0){\includegraphics[width=\unitlength,page=20]{Sketch.pdf}}%
    \put(0.10264082,0.19511192){\color[rgb]{0,0,0}\makebox(0,0)[lt]{\begin{minipage}{0.01413701\unitlength}\raggedright B\end{minipage}}}%
    \put(0,0){\includegraphics[width=\unitlength,page=21]{Sketch.pdf}}%
    \put(0.10264082,0.16993664){\color[rgb]{0,0,0}\makebox(0,0)[lt]{\begin{minipage}{0.01413701\unitlength}\raggedright B\end{minipage}}}%
    \put(0,0){\includegraphics[width=\unitlength,page=22]{Sketch.pdf}}%
    \put(0.10264082,0.14462483){\color[rgb]{0,0,0}\makebox(0,0)[lt]{\begin{minipage}{0.01413701\unitlength}\raggedright A\end{minipage}}}%
    \put(0,0){\includegraphics[width=\unitlength,page=23]{Sketch.pdf}}%
    \put(0.10264082,0.11970628){\color[rgb]{0,0,0}\makebox(0,0)[lt]{\begin{minipage}{0.01413701\unitlength}\raggedright A\end{minipage}}}%
    \put(0,0){\includegraphics[width=\unitlength,page=24]{Sketch.pdf}}%
    \put(0.10264082,0.05684674){\color[rgb]{0,0,0}\makebox(0,0)[lt]{\begin{minipage}{0.01413701\unitlength}\raggedright A\end{minipage}}}%
    \put(0,0){\includegraphics[width=\unitlength,page=25]{Sketch.pdf}}%
    \put(0.12856228,0.19511192){\color[rgb]{0,0,0}\makebox(0,0)[lt]{\begin{minipage}{0.01413701\unitlength}\raggedright A\end{minipage}}}%
    \put(0,0){\includegraphics[width=\unitlength,page=26]{Sketch.pdf}}%
    \put(0.12856228,0.16993664){\color[rgb]{0,0,0}\makebox(0,0)[lt]{\begin{minipage}{0.01413701\unitlength}\raggedright B\end{minipage}}}%
    \put(0,0){\includegraphics[width=\unitlength,page=27]{Sketch.pdf}}%
    \put(0.12856228,0.14462483){\color[rgb]{0,0,0}\makebox(0,0)[lt]{\begin{minipage}{0.01413701\unitlength}\raggedright C\end{minipage}}}%
    \put(0,0){\includegraphics[width=\unitlength,page=28]{Sketch.pdf}}%
    \put(0.12856228,0.11970628){\color[rgb]{0,0,0}\makebox(0,0)[lt]{\begin{minipage}{0.01413701\unitlength}\raggedright B\end{minipage}}}%
    \put(0,0){\includegraphics[width=\unitlength,page=29]{Sketch.pdf}}%
    \put(0.12856228,0.05684674){\color[rgb]{0,0,0}\makebox(0,0)[lt]{\begin{minipage}{0.01413701\unitlength}\raggedright C\end{minipage}}}%
    \put(0,0){\includegraphics[width=\unitlength,page=30]{Sketch.pdf}}%
    \put(0.19422998,0.19511192){\color[rgb]{0,0,0}\makebox(0,0)[lt]{\begin{minipage}{0.01413701\unitlength}\raggedright C\end{minipage}}}%
    \put(0,0){\includegraphics[width=\unitlength,page=31]{Sketch.pdf}}%
    \put(0.19422998,0.16993663){\color[rgb]{0,0,0}\makebox(0,0)[lt]{\begin{minipage}{0.01413701\unitlength}\raggedright C\end{minipage}}}%
    \put(0,0){\includegraphics[width=\unitlength,page=32]{Sketch.pdf}}%
    \put(0.19422998,0.14462483){\color[rgb]{0,0,0}\makebox(0,0)[lt]{\begin{minipage}{0.01413701\unitlength}\raggedright A\end{minipage}}}%
    \put(0,0){\includegraphics[width=\unitlength,page=33]{Sketch.pdf}}%
    \put(0.19422998,0.11970627){\color[rgb]{0,0,0}\makebox(0,0)[lt]{\begin{minipage}{0.01413701\unitlength}\raggedright E\end{minipage}}}%
    \put(0,0){\includegraphics[width=\unitlength,page=34]{Sketch.pdf}}%
    \put(0.19422998,0.05684674){\color[rgb]{0,0,0}\makebox(0,0)[lt]{\begin{minipage}{0.01413701\unitlength}\raggedright C\end{minipage}}}%
    \put(0,0){\includegraphics[width=\unitlength,page=35]{Sketch.pdf}}%
    \put(0.19422998,0.19511192){\color[rgb]{0,0,0}\makebox(0,0)[lt]{\begin{minipage}{0.01413701\unitlength}\raggedright C\end{minipage}}}%
    \put(0,0){\includegraphics[width=\unitlength,page=36]{Sketch.pdf}}%
    \put(0.19422998,0.16993663){\color[rgb]{0,0,0}\makebox(0,0)[lt]{\begin{minipage}{0.01413701\unitlength}\raggedright C\end{minipage}}}%
    \put(0,0){\includegraphics[width=\unitlength,page=37]{Sketch.pdf}}%
    \put(0.19422998,0.14462483){\color[rgb]{0,0,0}\makebox(0,0)[lt]{\begin{minipage}{0.01413701\unitlength}\raggedright A\end{minipage}}}%
    \put(0,0){\includegraphics[width=\unitlength,page=38]{Sketch.pdf}}%
    \put(0.19422998,0.11970627){\color[rgb]{0,0,0}\makebox(0,0)[lt]{\begin{minipage}{0.01413701\unitlength}\raggedright E\end{minipage}}}%
    \put(0,0){\includegraphics[width=\unitlength,page=39]{Sketch.pdf}}%
    \put(0.19422998,0.05684674){\color[rgb]{0,0,0}\makebox(0,0)[lt]{\begin{minipage}{0.01413701\unitlength}\raggedright C\end{minipage}}}%
    \put(0.07389205,0.22585235){\color[rgb]{0,0,0}\makebox(0,0)[lt]{\begin{minipage}{0.06098446\unitlength}\raggedright $x_1$\end{minipage}}}%
    \put(0.09808542,0.22585233){\color[rgb]{0,0,0}\makebox(0,0)[lt]{\begin{minipage}{0.06098445\unitlength}\raggedright $x_2$\end{minipage}}}%
    \put(0.12746307,0.22585233){\color[rgb]{0,0,0}\makebox(0,0)[lt]{\begin{minipage}{0.06098446\unitlength}\raggedright $x_3$\end{minipage}}}%
    \put(0.18967458,0.22585235){\color[rgb]{0,0,0}\makebox(0,0)[lt]{\begin{minipage}{0.06098446\unitlength}\raggedright $x_N$\end{minipage}}}%
    \put(0.07073128,0.01783298){\color[rgb]{0,0,0}\makebox(0,0)[lt]{\begin{minipage}{0.15984382\unitlength}\raggedright {\bf Empir.} vectors\end{minipage}}}%
    \put(0,0){\includegraphics[width=\unitlength,page=40]{Sketch.pdf}}%
    \put(0.29638846,0.19464308){\color[rgb]{0.78431373,0,0}\makebox(0,0)[lt]{\begin{minipage}{0.0141512\unitlength}\raggedright A\end{minipage}}}%
    \put(0,0){\includegraphics[width=\unitlength,page=41]{Sketch.pdf}}%
    \put(0.29638846,0.16944253){\color[rgb]{0.78431373,0,0}\makebox(0,0)[lt]{\begin{minipage}{0.01600459\unitlength}\raggedright D\end{minipage}}}%
    \put(0,0){\includegraphics[width=\unitlength,page=42]{Sketch.pdf}}%
    \put(0.29638846,0.14410533){\color[rgb]{0.78431373,0,0}\makebox(0,0)[lt]{\begin{minipage}{0.0141512\unitlength}\raggedright B\end{minipage}}}%
    \put(0,0){\includegraphics[width=\unitlength,page=43]{Sketch.pdf}}%
    \put(0.29638846,0.11916176){\color[rgb]{0.78431373,0,0}\makebox(0,0)[lt]{\begin{minipage}{0.0141512\unitlength}\raggedright C\end{minipage}}}%
    \put(0,0){\includegraphics[width=\unitlength,page=44]{Sketch.pdf}}%
    \put(0.29638846,0.05623913){\color[rgb]{0.78431373,0,0}\makebox(0,0)[lt]{\begin{minipage}{0.0141512\unitlength}\raggedright A\end{minipage}}}%
    \put(0,0){\includegraphics[width=\unitlength,page=45]{Sketch.pdf}}%
    \put(0.32320451,0.1946431){\color[rgb]{0,0.62745098,0}\makebox(0,0)[lt]{\begin{minipage}{0.0141512\unitlength}\raggedright C\end{minipage}}}%
    \put(0,0){\includegraphics[width=\unitlength,page=46]{Sketch.pdf}}%
    \put(0.32320451,0.16944254){\color[rgb]{0,0.62745098,0}\makebox(0,0)[lt]{\begin{minipage}{0.01600459\unitlength}\raggedright A\end{minipage}}}%
    \put(0,0){\includegraphics[width=\unitlength,page=47]{Sketch.pdf}}%
    \put(0.32320451,0.14410533){\color[rgb]{0,0.62745098,0}\makebox(0,0)[lt]{\begin{minipage}{0.0141512\unitlength}\raggedright B\end{minipage}}}%
    \put(0,0){\includegraphics[width=\unitlength,page=48]{Sketch.pdf}}%
    \put(0.32320451,0.11916176){\color[rgb]{0,0.62745098,0}\makebox(0,0)[lt]{\begin{minipage}{0.0141512\unitlength}\raggedright B\end{minipage}}}%
    \put(0,0){\includegraphics[width=\unitlength,page=49]{Sketch.pdf}}%
    \put(0.32320451,0.05623913){\color[rgb]{0,0.62745098,0}\makebox(0,0)[lt]{\begin{minipage}{0.0141512\unitlength}\raggedright C\end{minipage}}}%
    \put(0,0){\includegraphics[width=\unitlength,page=50]{Sketch.pdf}}%
    \put(0.35002056,0.1946431){\color[rgb]{0,0,1}\makebox(0,0)[lt]{\begin{minipage}{0.0141512\unitlength}\raggedright B\end{minipage}}}%
    \put(0,0){\includegraphics[width=\unitlength,page=51]{Sketch.pdf}}%
    \put(0.35002056,0.16944254){\color[rgb]{0,0,1}\makebox(0,0)[lt]{\begin{minipage}{0.01600459\unitlength}\raggedright E\end{minipage}}}%
    \put(0,0){\includegraphics[width=\unitlength,page=52]{Sketch.pdf}}%
    \put(0.35002056,0.14410533){\color[rgb]{0,0,1}\makebox(0,0)[lt]{\begin{minipage}{0.0141512\unitlength}\raggedright C\end{minipage}}}%
    \put(0,0){\includegraphics[width=\unitlength,page=53]{Sketch.pdf}}%
    \put(0.35002056,0.11916176){\color[rgb]{0,0,1}\makebox(0,0)[lt]{\begin{minipage}{0.0141512\unitlength}\raggedright A\end{minipage}}}%
    \put(0,0){\includegraphics[width=\unitlength,page=54]{Sketch.pdf}}%
    \put(0.35002056,0.05623913){\color[rgb]{0,0,1}\makebox(0,0)[lt]{\begin{minipage}{0.0141512\unitlength}\raggedright C\end{minipage}}}%
    \put(0.28905717,0.0178945){\color[rgb]{0,0,0}\makebox(0,0)[lt]{\begin{minipage}{0.09300579\unitlength}\raggedright prototypes\end{minipage}}}%
    \put(0.29118031,0.22585235){\color[rgb]{0.78431373,0,0}\makebox(0,0)[lt]{\begin{minipage}{0.06098446\unitlength}\raggedright $p_1$\end{minipage}}}%
    \put(0.31755362,0.22585233){\color[rgb]{0,0.62745098,0}\makebox(0,0)[lt]{\begin{minipage}{0.06098445\unitlength}\raggedright $p_2$\end{minipage}}}%
    \put(0.34475278,0.22585233){\color[rgb]{0,0,1}\makebox(0,0)[lt]{\begin{minipage}{0.06098445\unitlength}\raggedright $p_3$\end{minipage}}}%
  \end{picture}%
\endgroup%

%% file: AppFitAlg.tex
\section{Fitting algorithm}
\label{AppFitAlg}

This section explains the procedure used for simultaneously tuning the $\alpha$ and $\beta$ parameters of either of the two stochastic models of culture,
such that a match is obtained between the model and the empirical data, in terms of the averages of the $\aivd$ and $\sivd$ observables:
\begin{align}
	\label{FitProbForm}
	\langle\aivd(\alpha, \beta)\rangle & = \aivd_\emp, \\
	\langle\sivd(\alpha, \beta)\rangle & = \sivd_\emp, \nonumber
\end{align}
for a fixed number of prototypes $k$, assuming that either of the two equalities above is satisfied when there is an overlap between the uncertainty range associated to the quantity on the left side and that associated to the quantity on the right side.

There are multiple reasons why this problem is challenging:
\begin{itemize}
	\item an analytical formula for the $\langle\sivd(\alpha, \beta)\rangle$ quantity could not be found
	\item although an analytical formula for the $\langle\aivd(\alpha, \beta)\rangle$ quantity was found (Eqs. \eqref{AvAIVD} to \eqref{pi_MPG}) 
	\footnote{Which implies that the specific uncertainty range of $\langle\aivd(\alpha, \beta)\rangle$ has a null width.}, 
	it does not allow for inverting the function and for analytically solving the system
	\item the $\langle\sivd(\alpha, \beta)\rangle$, $\aivd_\emp$ and $\sivd_\emp$ quantities have non-vanishing uncertainty ranges attached to them
\end{itemize}

Assuming that there exists a unique solution to the above system, a numerical approach for solving it is in order.
The method used here relies on a nested, 2-level, adapted bisection method. 
The first (inner) level of the method takes care of fitting, via bisection, the first quantity for a fixed $\beta$ -- 
it finds the $\alpha$ value for which $\langle\aivd(\alpha, \beta)\rangle = \aivd_\emp$ is satisfied for a given $\beta$.
The second (outer) level of the method takes care of fitting, via bisection, the second quantity --
it finds the $\beta$ for which $\langle\sivd(\alpha(\beta), \beta)\rangle = \sivd_\emp$ is satisfied, 
where $\alpha(\beta)$ is provided by the first level.
This choice of assigning the $\aivd$ and $\sivd$ observables and the $\alpha$ and $\beta$ parameters to the two levels in this manner is numerically convenient for several reasons. 
First, the $\aivd$ can be much more easily computed via the analytical formula, such that assigning it to the first level, which is repeated multiple times (once for each value of $\beta$ that the second level samples) is more effective.
Second, the model $\aivd$ turns out to be relatively insensitive to $\beta$ for relatively many combinations of values for the $k$ and $\alpha$ parameters, 
such that fitting $\aivd$ in terms of $\alpha$ within the first level makes more sense.

In addition to adaptations required by the 2-level scheme, 
other adaptations with respect to the traditional bisection method are needed for allowing it to work with model and empirical uncertainties, 
as well as to enhance the numerical precision for the $\langle\sivd(\alpha, \beta)\rangle$ quantity when needed, to the extent needed.
Moreover, in addition to statistical errors originating directly in the empirical uncertainties of the $\aivd_\emp$ and $\sivd_\emp$ quantities and in the numerical uncertainty of the model $\sivd$ quantity, the second level of the method is also affected by ``systematic errors'' on $\langle\sivd(\alpha(\beta), \beta)\rangle$, 
originating in the fitting procedure at the first level, 
and indirectly in the empirical uncertainty of $\aivd_\emp$ -- which for all practical purposes can be assumed fixed, thus motivating using the term ``systematic'' for its propagation to the model $\sivd$ at the second level. 

In order to address all these challenges in a self consistent way, the method developed here turns out to be quite sophisticated, 
which is why it is explained in detail in the following four sections. 
Specifically, Sec.~\ref{AppFirLevFit} focuses on the first fitting level, Sec.~\ref{AppSecLevFit} focuses on the second fitting level, 
Sec.~\ref{AppUsFunc} describes how various sub-problems invoked by the previous two sections are addressed, 
while Sec.~\ref{AppFormUsg} describes how the tools presented in Sections~\ref{AppFirLevFit},~\ref{AppSecLevFit} and~\ref{AppUsFunc} are used 
for producing some of the results presented in Sections \ref{ModFit} and \ref{ModOut}..
The method is potentially of use for addressing other problems that are formally similar to the problem presented here, 
although certain adaptations might be needed. 

Since the method has mostly an algorithmic nature, much of it is explained via pseudocode, such that a few conventions that will be extensively used below and that are not necessarily standard are worth mentioning.
First, the ``='' symbol is used with double meaning: in a normal statement (such as ``$a = b$'') it is to be interpreted as an assignment (of the value of variable b to variable a); 
in the header of an {\bf if} or {\bf while} statement (such as ``{\bf if $a = b$}'') it is to be interpreted as a check (of whether the values of $a$ and $b$ are equal).
A variable is implicitly declared when it first appears, either on the left side of an assignment or in the header of a function definition (in which case it is also called an argument or function parameter); 
the scope of the variable is the part of the function below and to the right of the place where it first appears.
Functions are distinguished from each other through their names, their numbers of arguments and the types of those arguments 
\footnote{Sometimes this can be confusing, since the types of the arguments are only mentioned in the text before the definition of the function. 
In these cases however, the reader is guided by the names of the arguments, which in the function definition are kept as close as possible to those in the function call(s).} 
On the other hand, the arguments of a function are distinguished from each other via their order.
Some variables are actually ordered sequences of other variables, which in turn are denoted by $(x_1,..,x_n)$ notation. 
In the same spirit, an assignments of the type $X = (x_1,..,x_n)$ is referred to as a ``variable compression'', while one of the type $(x_1,..,x_n) = X$ is referred to as a ``variable decompression''. 
These allow for keeping the pseudocode compact, while still rigorous.
An uncertainty range refers to an interval $[x-\delta x, x+\delta x]$, where $x$ is a mean and $\delta x$ is an error relying (directly, or indirectly) on a standard mean error calculation, the uncertainty range being formally encoded by the sorted $(x,\delta x)$ sequence.
Note that the square brackets ``[,]'' are consistently used to denote an interval of real numbers, while the round brackets ``(,)'' are used to denote an ordered sequence of two or more elements. 
Finally, it is worth noting that the pseudocode relies heavily on function calls and on recursive definitions, and that there is a certain parallelism between the functions defined in Sec.~\ref{AppFirLevFit} 
and those defined in Sec.~\ref{AppSecLevFit}. 

\subsection{First level fitting}
\label{AppFirLevFit}

This section presents the algorithm part concerned with the first fitting level. 
The algorithm is split in three main functions: \textsc{Fit-1}, \textsc{Bisect-1}, \textsc{Displace-1}, 
all of them returning the same type of information. 
\textsc{Fit-1} always calls \textsc{Bisect-1}, while the latter may or may not call \textsc{Displace-1} at any stage, which in turn may or may not call \textsc{Bisect-1}. 
The pseudocode also invokes two constants, which are assumed to be known a-priori and available for use anywhere in these three functions. 
The first constant is $\delta\alpha$, which controls the desired resolution ($\delta\alpha$ is essentially a grid-spacing) in the $\alpha$ parameter, which is here set to the inverse of the number of features: $\delta\alpha = \frac{1}{F}$
\footnote{	There is no clear lower bound on $\delta\alpha$, regardless of which stochastic model is used, 
						but $\frac{1}{F}$ is a lower bound on $\delta\beta$ when PG is used, 
						so for simplicity the choice $\delta \alpha = \delta\beta = \frac{1}{F}$ is made.}. 
The second constant is $\aivd_\emp$, which stands for the $\aivd$ uncertainty range for the empirical data.

Function \textsc{Fit-1} acts as an interface for the first-level fitting,
which consists of tuning the $\alpha$ parameter, for given values of $\beta$ and $k$, such that the $\aivd$ quantity matches the empirical value.
Here, $\beta$ is a real number belonging to $[0,1]$ while $k$ is a strictly positive integer number.
The method returns the left ($\alpha_L$) and right ($\alpha_R$) margins of the tightest $\alpha$ interval found, together with the estimated $\alpha$ match within this interval assuming linearity ($\alpha_{\text{fit}}$) and an associated error ($\alpha_{\err}$).
It assumes that the empirical AIVD can actually be uniquely matched by varying $\alpha$, for the given values of $\beta$ and $k$.
The method essentially carries out some initializations (Lines 2,3), before passing the task to \textsc{Bisect-1}.

\begin{widetext}
  \begin{algorithmic}[1]
    \Statex
    \Function{Fit-1}{$\beta, k$}
      \State $(\alpha_L, \alpha_R) = \text{Init-1}(\delta\alpha)$	\Comment{initializing the $\alpha$-interval}
      \State $\aivd_L = \langle\aivd\rangle_{\alpha_L,\beta}^k$; $\aivd_R = \langle\aivd\rangle_{\alpha_R,\beta}^k$	\Comment{analytic calculations based on Eq. \eqref{AvAIVD}}
      \State \Return{\textsc{Bisect-1}($\alpha_L, \alpha_R, \aivd_L, \aivd_R, \beta, k$)}   
    \EndFunction
  \end{algorithmic}
\end{widetext}

Function \textsc{Bisect-1} is mostly a typical, recursive implementation of the bisection method.
This sequentially narrows down the $[\alpha_L, \alpha_R]$ interval, such that at each stage the empirical AIVD is contained, namely that $\min(\aivd_L,\aivd_R) < \aivd_\emp < \max(\aivd_L,\aivd_R)$ is satisfied, 
where the $\aivd_L, \aivd_R$ values correspond to the left and right margins of the $\alpha$ interval.
Here, $\alpha_L, \alpha_R, \aivd_L, \aivd_R$ are real numbers belonging to $[0,1]$ while $\beta$ and $k$ are of the same type as in \textsc{Fit-1}.
It returns the same type of information as \textsc{Fit-1}.
The method converges, the fitting being considered complete, when the interval has reached the $\delta\alpha$ resolution limit, 
in which case estimations for an ``ideal'' $\alpha$ inside this interval $\alpha_\fit$ and its error $\alpha_\err$ are made and returned together with the boundaries of the interval (lines 3-6).
Moreover, the method may also call \textsc{Displace-1} in case the $\aivd_M$ value corresponding to the computed midpoint $\alpha_M$ happens to fall within the $\aivd_\emp$ uncertainty range (lines 8-10) -- 
this is needed in order to keep the output format consistent and the final $\alpha$ interval relatively narrow.
Otherwise, the method decides to zoom in (by calling itself) on either the left or right halves of the interval, depending on the position of $\aivd_\emp$ with respect to $\aivd_L$, $\aivd_M$ and $\aivd_R$ (lines 11-16). 

\begin{widetext}
  \begin{algorithmic}[1]
    \Statex
    \Function{Bisect-1}{$\alpha_L, \alpha_R, \aivd_L, \aivd_R, \beta, k$} 	
      \State $\alpha_M = \textsc{Middle}(\alpha_L, \alpha_R, \delta\alpha)$	\Comment{computing midpoint on the $\alpha$ grid}
      \If{$\neg\textsc{Distinct}(\alpha_M, \alpha_L, \alpha_R)$}	
	\State $(\alpha_{\text{fit}}, \alpha_{\err}) = \textsc{InternFitLin-1}(\alpha_L, \alpha_R, \aivd_L, \aivd_R, \aivd_{\emp})$
	\State \Return{$(\alpha_L, \alpha_R, \alpha_{\text{fit}}, \alpha_{\err})$}  \Comment{fitting complete} 
      \EndIf
      \State $\aivd_M = \langle\aivd\rangle_{\alpha_M,\beta}^k$		\Comment{analytic calculations based on Eq. \eqref{AvAIVD}}
      \If{$\textsc{Match-1}(\aivd_M,\aivd_{\emp})$}
	\State \Return{\textsc{Displace-1}($\alpha_L, \alpha_R, \alpha_M, \aivd_L, \aivd_R, \beta, k$)}   
      \EndIf
      \If{$\textsc{Ord-1}(\aivd_L,\aivd_R) = \textsc{Ord-1}(\aivd_M,\aivd_{\emp})$}
	\State $\alpha_L=\alpha_M$; $\aivd_L=\aivd_M$			\Comment{selecting right interval}
      \Else
	\State $\alpha_R=\alpha_M$; $\aivd_R=\aivd_M$			\Comment{selecting left interval}
      \EndIf
      \State \Return{\textsc{Bisect-1}($\alpha_L, \alpha_R, \aivd_L, \aivd_R, \beta, k$)}   \Comment{zooming in on selected interval}
    \EndFunction
  \end{algorithmic}
\end{widetext}
  
Function \textsc{Displace-1} attempts to displace the midpoint $\alpha_M$ previously calculated at some stage in \textsc{Bisect-1}, 
in a way that its associated AIVD would fall outside the empirical uncertainty range. 
This function has all the arguments of \textsc{Bisect-1} and $\alpha_M$ as an additional one, which is a real number belonging to $[0,1]$.
It returns the same type of information as \textsc{Fit-1}.
The method first computes a ``secondary'' midpoint $\alpha_M'$ to the left of $\alpha_M$ and its corresponding $\aivd_M'$ value.
If the resolution limit $\delta\alpha$ is not reached and $\aivd_M'$ falls outside the $\aivd_\emp$ range, \textsc{Bisect-1} is applied further to the $[\alpha_M', \alpha_R]$ interval (lines 2-11). 
Otherwise, the analogous procedure is applied on the right side (12-21). 
If the procedure fails to provide a convenient, secondary midpoint on either side, the fitting is considered complete with the current $[\alpha_L, \alpha_R]$ interval and the $\alpha_\fit, \alpha_\err$ estimates made like in \textsc{Bisect-1} (lines 22-23).

\begin{widetext}
  \begin{algorithmic}[1]
    \Statex
    \Function{Displace-1}{$\alpha_L, \alpha_R, \alpha_M, \aivd_L, \aivd_R, \beta, k$}
      \State $\alpha_M'=\textsc{Middle}(\alpha_L, \alpha_M, \delta\alpha)$	\Comment{trying displacement to the left on the $\alpha$ grid}
      \If{$\textsc{Distinct}(\alpha_M', \alpha_L, \alpha_M)$}
	\State $\aivd_M'=\langle\aivd\rangle_{\alpha_M',\beta}^{k}$		\Comment{analytic calculations based on Eq. \eqref{AvAIVD}}
	\If{$\neg \textsc{Match-1}(\aivd_M', \aivd_{\emp})$}
	  \If{$\textsc{Ord-1}(\aivd_L,\aivd_R) = \textsc{Ord-1}(\aivd_M',\aivd_{\emp})$}
	    \State $\alpha_L = \alpha_M'$; $\aivd_L = \aivd_M'$
	    \State \Return{\textsc{Bisect-1}($\alpha_L, \alpha_R, \aivd_L, \aivd_R, \beta, k$)}   \Comment{zooming in on corrected interval}
	  \EndIf
	\EndIf
      \EndIf
      \State $\alpha_M'=\textsc{Middle}(\alpha_M, \alpha_R, \delta\alpha)$	\Comment{trying displacement to the right on the $\alpha$ grid}
      \If{$\textsc{Distinct}(\alpha_M', \alpha_M, \alpha_R)$}
	\State $\aivd_M'=\langle\aivd\rangle_{\alpha_M',\beta}^{k}$		\Comment{analytic calculations based on Eq. \eqref{AvAIVD}}
	\If{$\neg \textsc{Match-1}(\aivd_M', \aivd_{\emp})$}
	  \If{$\textsc{Ord-1}(\aivd_L,\aivd_R) \neq \textsc{Ord-1}(\aivd_M',\aivd_{\emp})$}
	    \State $\alpha_R = \alpha_M'$; $\aivd_R = \aivd_M'$
	    \State \Return{\textsc{Bisect-1}($\alpha_L, \alpha_R, \aivd_L, \aivd_R, \beta, k$)}   \Comment{zooming in on corrected interval}
	  \EndIf
	\EndIf
      \EndIf
      \State $(\alpha_{\text{fit}}, \alpha_{\err}) = \textsc{InternFitLin-1}(\alpha_L, \alpha_R, \aivd_L, \aivd_R, \aivd_{\emp})$
      \State \Return{$(\alpha_L, \alpha_R, \alpha_{\text{fit}}, \alpha_{\err})$}   \Comment{fitting complete}
    \EndFunction
  \end{algorithmic}
\end{widetext}
  
\subsection{Second level fitting}
\label{AppSecLevFit}
    
This section presents the algorithm part concerned with the second fitting level. 
Each of the three functions of the first fitting level (Sec.~\ref{AppFirLevFit}) has a correspondent here: \textsc{Fit-2}, \textsc{Bisect-2}, \textsc{Displace-2}, 
all of them returning the same type of information
\footnote{The type of information returned by the three functions at a second-level fitting is different than that of the three functions at the first-level fitting, and actually more complex.}, 
each of them having a similar, structure, purpose and role to the correspondent within the first fitting level.  
Additionally, this section presents the pseudocode for a fourth function, \textsc{NumSIVD}, which carries out the numerical SIVD calculations.
In addition to the two constants introduced at the first level, 
the second level pseudocode invokes two other constants, which are also assumed to be known a-priori and available for use anywhere in these four functions. 
First, $\delta\beta$ is the desired resolution in the $\beta$ parameter, which is here set to the inverse of the number of features: $\delta\beta = \frac{1}{F}$.
Second, $\sivd_\emp$ is the $\sivd$ uncertainty range for the empirical data.

In relation to the first three functions, the descriptions below attempt to mostly emphasize the elements that come in addition with respect to their first-level correspondents.
Some of these elements have a repetitive nature and are worth explaining before moving to the specific description of each function.  
First, the (generic) $\bar{\beta}_{\text{X}}$ notation (where ``X'' can stand for ``L'', ``R'' or ``M'') denotes the (generic) ``composite fitting information'' $\bar{\beta}_{\text{X}} = (\beta, \alpha_L, \alpha_R, \alpha_{\text{fit}}, \alpha_{\err})_{\text{X}}$,
which is a 5-tuple consisting of a $\beta$ value together with the associated four values returned by a (generic) call $\textsc{Fit-1}(\beta, k)$ for that specific $\beta$ and some arbitrary $k$. 
Second, whenever an ``$\sivd_{\text{X}}$'' variable appears in the first three functions (where ``X'' is again a generic label), except for $\sivd_\emp$,
it actually denotes the (generic) ``composite SIVD information'' $\sivd_{\text{X}} = ((\sivd_L^\fit, \sivd_L^\err), (\sivd_R^\fit, \sivd_R^\err))_{\text{X}}$, 
which is a pair of pairs of real numbers, each inner pair corresponding to a model SIVD uncertainty range associated to one margin of an $\alpha$ interval returned by a call to $\textsc{Fit-1}$, while both inner pairs have the same $\beta$. 
This schematically reads:
\begin{align}
	(\beta,\alpha_L) & \rightarrow (\sivd_L^\fit, \sivd_L^\err), \nonumber \\
	(\beta,\alpha_R) & \rightarrow (\sivd_R^\fit, \sivd_R^\err), \nonumber
\end{align}
Third, any (generic) call $\textsc{NumSIVD}(\bar{\beta},k)$ is necessarily preceded by an associated (generic) call $\textsc{Fit-1}(\beta, k)$ 
and by an associated (generic) variable compression $\bar{\beta} = (\beta, \alpha_L, \alpha_R, \alpha_{\text{fit}}, \alpha_{\err})$,
the last two being needed for producing the composite fitting information $\bar{\beta}$.
Fourth, whenever a piece of composite SIVD information appears in a call to \textsc{Ord-2} or \textsc{Match-2}, it is accompanied by an associated piece of composite fitting information, which allows for the mean, statistical error and systematic error of in the model SIVD to be all reconstructed within, for a given combination of $\beta$ and $k$.

Function \textsc{Fit-2} acts as an interface for the second-level fitting,
which consists of tuning the $\beta$ parameter, for a given value of $k$, such that the SIVD quantity matches the empirical value, relying on an underlying tuning of the $\alpha$ parameter in terms of the AIVD quantity (using \textsc{Fit-1}).
Here, $k$ is a strictly positive, integer number.
The method returns the composite fitting information associated to the left ($\bar{\beta}_L$) and right ($\bar{\beta}_R$) margins of the tightest $\beta$ interval found, together with the estimated $\beta$ match within this interval ($\beta_{\text{fit}}$) and its associated error ($\beta_{\err}$).
It assumes that the empirical SIVD can actually be uniquely matched by varying $\beta$ and $\alpha$, for the given value of $k$.
After checking that there exists a meaningful $[\beta_L, \beta_R]$ interval for which the first-level fitting is possible (lines 2,3),
the method conducts the numeric SIVD calculations on both sides of the interval (line 6), preceded, on each side, by the first level fitting and the decompression (lines 4,5, as explained above),
in order to finally pass the task to \textsc{Bisect-2}.

\begin{widetext}
  \begin{algorithmic}[1]
    \Statex
    \Function{Fit-2}{$k$}
      \State $(\beta_L, \beta_R) = \text{Init-2}(\delta\beta,k,\aivd_{\emp})$	\Comment{initializing the $\beta$-interval}
      \If{$\beta_L < \beta_R$}
				\State $(\alpha_L^L, \alpha_L^R, \alpha_L^{\text{fit}}, \alpha_L^{\err}) = \textsc{Fit-1}(\beta_L, k)$; $(\alpha_R^L, \alpha_R^R, \alpha_R^{\text{fit}}, \alpha_R^{\err}) = \textsc{Fit-1}(\beta_R, k)$
				\State $\bar{\beta}_L = (\beta_L, \alpha_L^L, \alpha_L^R, \alpha_L^{\text{fit}}, \alpha_L^{\err})$; $\bar{\beta}_R = (\beta_R, \alpha_R^L, \alpha_R^R, \alpha_R^{\text{fit}}, \alpha_R^{\err})$
				\State $\sivd_L = \textsc{NumSIVD}(\bar{\beta}_L, k)$; $\sivd_R = \textsc{NumSIVD}(\bar{\beta}_R, k)$		\Comment{numeric calculations}	
				\State \Return{\textsc{Bisect-2}($\bar{\beta}_L, \bar{\beta}_R, \sivd_L, \sivd_R, k$)}   
      \EndIf 
      \State \Return {\bf FittingImpossibleError}
    \EndFunction
  \end{algorithmic}
\end{widetext}

Function \textsc{Bisect-2} is another recursive implementation of the bisection method,
which sequentially narrows down the $[\beta_L, \beta_R]$ interval, such that at each stage the empirical SIVD is contained.
Here, $\bar{\beta}_L, \bar{\beta}_R$ are 5-tuples of real numbers encoding the left and right pieces of composite fitting information,
$\sivd_L, \sivd_R$ are the pairs of pairs of real numbers encoding the left-$\beta$ and right-$\beta$ pieces of composite SIVD information,
while $k$ is of the same type as in \textsc{Fit-2}.
It returns the same type of information as \textsc{Fit-2}.
Like \textsc{Bisect-1}, the function consists of a part concerned with convergence (lines 4-7), 
a part concerned with the jump to \textsc{Displace-2} (lines 11-13)
and a part concerned with choosing between the left and right $\beta$ subintervals and with zooming in on the chosen one (lines 14-19).
Note the additional statements concerned with decompressing the composite fitting information (line 2) and with preparing the numeric SIVD calculations at the midpoint (lines 8-9).

\begin{widetext}
  \begin{algorithmic}[1]
    \Statex
    \Function{Bisect-2}{$\bar{\beta}_L, \bar{\beta}_R, \sivd_L, \sivd_R, k$} 	
      \State $(\beta_L, \alpha_L^L, \alpha_L^R, \alpha_L^{\text{fit}}, \alpha_L^{\err}) = \bar{\beta}_L$; $(\beta_R, \alpha_R^L, \alpha_R^R, \alpha_R^{\text{fit}}, \alpha_R^{\err}) = \bar{\beta}_R$
      \State $\beta_M = \textsc{Middle}(\beta_L, \beta_R, \delta\beta)$	
      \If{$\neg\textsc{Distinct}(\beta_M, \beta_L, \beta_R)$}
	\State $(\beta_{\text{fit}}, \beta_{\err}) = \textsc{InternFitLin-2}(\bar{\beta}_L, \bar{\beta}_R, \sivd_L, \sivd_R, \sivd_{\emp})$	
	\State \Return{$(\bar{\beta}_L, \bar{\beta}_R, \beta_{\text{fit}}, \beta_{\err})$} 
      \EndIf						
      \State $(\alpha_M^L, \alpha_M^R, \alpha_M^{\text{fit}}, \alpha_M^{\err}) = \textsc{Fit-1}(\beta_M,k)$
      \State $\bar{\beta}_M = (\beta_M, \alpha_M^L, \alpha_M^R, \alpha_M^{\text{fit}}, \alpha_M^{\err})$
      \State $\sivd_M = \textsc{NumSIVD}(\bar{\beta}_M, k)$		\Comment{numeric calculations}
      \If{$\textsc{Match-2}(\bar{\beta}_M, \sivd_M, \sivd_{\emp})$}
	\State \Return{\textsc{Displace-2}($\bar{\beta}_L, \bar{\beta}_R, \bar{\beta}_M, \sivd_L, \sivd_R, k$)}   
      \EndIf
      \If{$\textsc{Ord-2}(\bar{\beta}_L,\bar{\beta}_R,\sivd_L,\sivd_R) = \textsc{Ord-2}(\bar{\beta}_M,\sivd_M,\sivd_{\emp})$}
	\State $\bar{\beta}_L=\bar{\beta}_M$; $\sivd_L=\sivd_M$			\Comment{selecting right interval}
      \Else
	\State $\bar{\beta}_R=\bar{\beta}_M$; $\sivd_R=\sivd_M$			\Comment{selecting left interval}
      \EndIf
      \State \Return{\textsc{Bisect-2}($\bar{\beta}_L, \bar{\beta}_R, \sivd_L, \sivd_R, k$)}   \Comment{zooming in on selected interval}
    \EndFunction
  \end{algorithmic}
\end{widetext}
  
Function \textsc{Displace-2} attempts to displace the midpoint $\beta_M$ previously calculated at some stage in \textsc{Bisect-2}, 
in a way that its associated SIVD uncertainty range does not overlap with the empirical one. 
This function has all the arguments of \textsc{Bisect-1} and $\bar{\beta}_M$ as an additional one, which is a 5-tuple of real numbers encoding the midpoint composite fitting information.
It returns the same type of information as \textsc{Fit-2}.
Like \textsc{Displace-1}, the function consists of a part that attempts a displacement to the left (lines 3-14), 
one that attempts a displacement to the right (lines 15-26) and one that takes care of the convergence (lines 27-28).
Note the additional statements concerned with decompressing the composite fitting information (line 2) and with preparing the numeric SIVD calculations for the left/right secondary midpoint (lines 5-6/17-18).

\begin{widetext}
  \begin{algorithmic}[1]
    \Statex
    \Function{Displace-2}{$\bar{\beta}_L, \bar{\beta}_R, \bar{\beta}_M, \sivd_L, \sivd_R, k$}	
      \State 	$(\beta_L, \alpha_L^L, \alpha_L^R, \alpha_L^{\text{fit}}, \alpha_L^{\err}) = \bar{\beta}_L$; 
		$(\beta_R, \alpha_R^L, \alpha_R^R, \alpha_R^{\text{fit}}, \alpha_R^{\err}) = \bar{\beta}_R$; 
		$(\beta_M, \alpha_M^L, \alpha_M^R, \alpha_M^{\text{fit}}, \alpha_M^{\err}) = \bar{\beta}_M$
      \State $\beta_M'=\textsc{Middle}(\beta_L, \beta_M, \delta\beta)$	\Comment{trying displacement to the left}
      \If{$\textsc{Distinct}(\beta_M', \beta_L, \beta_M)$}
	\State $(\dot{\alpha}_M^L, \dot{\alpha}_M^R, \dot{\alpha}_M^{\text{fit}}, \dot{\alpha}_M^{\err}) = \textsc{Fit-1}(\beta_M',k)$
	\State $\bar{\beta}_M' = (\beta_M', \dot{\alpha}_M^L, \dot{\alpha}_M^R, \dot{\alpha}_M^{\text{fit}}, \dot{\alpha}_M^{\err})$
	\State $\sivd_M'=\textsc{NumSIVD}(\bar{\beta}_M', k)$		\Comment{numeric calculations}
	\If{$\neg \textsc{Match-2}(\bar{\beta}_M',\sivd_M',\sivd_{\emp})$}
	  \If{$\textsc{Ord-2}(\bar{\beta}_L,\bar{\beta}_R,\sivd_L,\sivd_R) = \textsc{Ord-2}(\bar{\beta}_M', \sivd_M',\sivd_{\emp})$} 
	    \State $\bar{\beta}_L = \bar{\beta}_M'$; $\sivd_L = \sivd_M'$
	    \State \Return{\textsc{Bisect-2}($\bar{\beta}_L, \bar{\beta}_R, \sivd_L, \sivd_R, k$)}   	\Comment{zooming in on corrected interval}
	  \EndIf
	\EndIf
      \EndIf
      \State $\beta_M'=\textsc{Middle}(\beta_M, \beta_R, \delta\beta)$	\Comment{trying displacement to the right}
      \If{$\textsc{Distinct}(\beta_M', \beta_M, \beta_R)$}
	\State $(\dot{\alpha}_M^L, \dot{\alpha}_M^R, \dot{\alpha}_M^{\text{fit}}, \dot{\alpha}_M^{\err}) = \textsc{Fit-1}(\beta_M',k)$
	\State $\bar{\beta}_M' = (\beta_M', \dot{\alpha}_M^L, \dot{\alpha}_M^R, \dot{\alpha}_M^{\text{fit}}, \dot{\alpha}_M^{\err})$
	\State $\sivd_M'=\textsc{NumSIVD}(\bar{\beta}_M', k)$		\Comment{numeric calculations}
	\If{$\neg \textsc{Match-2}(\bar{\beta}_M',\sivd_M',\sivd_{\emp})$}
	  \If{$\textsc{Ord-2}(\bar{\beta}_L,\bar{\beta}_R,\sivd_L,\sivd_R) \neq \textsc{Ord-2}(\bar{\beta}_M',\sivd_M',\sivd_{\emp})$} 
	    \State $\bar{\beta}_R = \bar{\beta}_M'$; $\sivd_R = \sivd_M'$
	    \State \Return{\textsc{Bisect-2}($\bar{\beta}_L, \bar{\beta}_R, \sivd_L, \sivd_R, k$)}   	\Comment{zooming in on corrected interval}   
	  \EndIf
	\EndIf
      \EndIf
      \State $(\beta_{\text{fit}}, \beta_{\err}) = \textsc{InternFitLin-2}(\bar{\beta}_L, \bar{\beta}_R, \sivd_L, \sivd_R, \sivd_{\emp})$
      \State \Return{$(\bar{\beta}_L, \bar{\beta}_R, \beta_{\text{fit}}, \beta_{\err})$}   
    \EndFunction
  \end{algorithmic}
\end{widetext}

Function \textsc{NumSIVD} numerically generates a piece of composite SIVD information with a precision that is as high as possible.
Here, $\bar{\beta}$ is a 5-tuple of real numbers encoding a composite fitting information, while $k$ is a positive integer number.
One sequence of SIVD values is numerically generated (lines 4 and 11) for each of the two margins of the $\alpha$ interval (contained in $\bar{\beta}$), 
for the given $\beta$ (also contained in $\bar{\beta}$) and the given $k$.
An uncertainty range is obtained from each of the two sequences (lines 5 and 13).
These two uncertainty ranges are used together with the information in $\bar{\beta}$ to produce estimates for 
an average, a statistical error and a systematic error that are $\bar{\beta}$-specific rather than $(\alpha,\beta)$-specific (lines 6,7 and 14,15).
The number of SIVD values in the two sequences is increased and the calculations are repeated as long as the condition in line 9 remains true,  
namely as long as: 
the statistical error is higher than the systematic error, 
the desired separation between the model and empirical (statistical) uncertainty ranges is not reached and
the maximal SIVD sequence length is not reached. 
The desired separation and the SIVD sequence length are controlled via variables $s$ and $n$, initialized in line 2 
-- the initial values of these variables, as well as the upper bound on the latter are hard-coded, as visible in the pseudocode, 
and have been decided after some experimentation with \textsc{NumSIVD}, but they are not essential for the actual outcome. 
Also note the decompression of the composite fitting information (line 3) and the decompression of SIVD uncertainty ranges (lines 8 and 16).  

\begin{widetext}
  \begin{algorithmic}[1]
    \Statex
    \Function{NumSIVD}{$\bar{\beta}, k$}	
      \State $n=20$; $s=5$			\Comment{initial number of realizations and desired separation}
      \State $(\beta, \alpha_L, \alpha_R, \alpha_{\text{fit}}, \alpha_{\err}) = \bar{\beta}$
      \State $\sivd_{L}^{\text{seq}} = \textsc{GenSeqSIVD}(\alpha_L, \beta, k, n)$; $\sivd_{R}^{\text{seq}} = \textsc{GenSeqSIVD}(\alpha_R, \beta, k, n)$	
      \State $\sivd_{L} = \textsc{CompAvgErr}(\sivd_{L}^{\text{seq}})$; $\sivd_{R} = \textsc{CompAvgErr}(\sivd_{R}^{\text{seq}})$
      \State $\sivd = \textsc{Interpol}(\alpha_L, \alpha_R, \alpha_{\text{fit}}, \sivd_{L}, \sivd_{R})$
      \State $\sivd_{\syst} = \textsc{CompSystErr}(\alpha_L, \alpha_R, \alpha_{\err}, \sivd_{L}, \sivd_{R})$
      \State $(\sivd_{\avg}, \sivd_{\stat}) = \sivd$; $(\sivd_{\emp}^{\avg}, \sivd_{\emp}^{\stat}) = \sivd_{\emp}$	
      \While{$\sivd_{\stat} > \sivd_{\syst} \wedge (\sivd_{\stat} + \sivd_{\emp}^{\stat} > |\sivd_{\emp}^{\avg} - \sivd_{\avg}|/s) \wedge n < 350 $}
	\State $n = 2 \cdot n$
	\State $\sivd_{L}^{\text{tmpSeq}} = \textsc{GenSeqSIVD}(\alpha_L, \beta, k, n)$; $\sivd_{R}^{\text{tmpSeq}} = \textsc{GenSeqSIVD}(\alpha_R, \beta, k, n)$	
	\State $\sivd_{L}^{\text{seq}} = \textsc{Merge}(\sivd_{L}^{\text{seq}}, \sivd_{L}^{\text{tmpSeq}})$; $\sivd_{R}^{\text{seq}} = \textsc{Merge}(\sivd_{R}^{\text{seq}}, \sivd_{R}^{\text{tmpSeq}})$
	\State $\sivd_{L} = \textsc{CompAvgErr}(\sivd_{L}^{\text{seq}})$; $\sivd_{R} = \textsc{CompAvgErr}(\sivd_{R}^{\text{seq}})$
	\State $\sivd = \textsc{Interpol}(\alpha_L, \alpha_R, \alpha_{\text{fit}}, \sivd_{L}, \sivd_{R})$
	\State $\sivd_{\syst} = \textsc{CompSystErr}(\alpha_L, \alpha_R, \alpha_{\err}, \sivd_{L}, \sivd_{R})$
	\State $(\sivd_{\avg}, \sivd_{\stat}) = \sivd$	
      \EndWhile
      \State \Return{$(\sivd_{L}, \sivd_{R})$}
    \EndFunction
  \end{algorithmic}
\end{widetext}

\subsection{Used functions}
\label{AppUsFunc}

This section describes functions that are used by the pseudocode in sections~\ref{AppFirLevFit} or~\ref{AppSecLevFit} but are not described there. 
The following is a list of functions for which the pseudocode is also provided, following each text description.

Function \textsc{InterfitLin-1} fine-tunes the $\alpha$ parameter such that $\aivd_{\emp}$ is matched, relying on a linear approximation of the model $\aivd$ as a function of $\alpha$ within the $(\alpha_L, \alpha_R)$ interval, using the boundary values $\aivd_L$ and $\aivd_R$. 
Its arguments are of the same type as those of \textsc{InterFitLin} (described below), except that $\aivd_L$ and $\aivd_R$ are real numbers rather than uncertainty ranges. 
The output structure is entirely the same as that of \textsc{InterFitLin}.
It is essentially a first-level fitting interface for \textsc{InternFitLin}, which is called after specifying that the errors associated to $\aivd_L$ and $\aivd_R$ are zero. 

\begin{widetext}
  \begin{algorithmic}[1]
    \Statex
    \Function{InternFitLin-1}{$\alpha_L, \alpha_R, \aivd_L, \aivd_R, \aivd_{\emp}$}
			\State $\aivd_L' = (\aivd_L, 0)$; $\aivd_R' = (\aivd_R, 0)$
			\State \Return{\textsc{InternFitLin}($\alpha_L, \alpha_R, \aivd_L', \aivd_R', \aivd_{\emp}$)}   
    \EndFunction
  \end{algorithmic}
\end{widetext}
    
Function \textsc{InterfitLin-2} fine-tunes the $\beta$ parameter such that $\sivd_{\emp}$ is matched, relying on a linear approximation of the model $\sivd$ as a function of $\beta$ within the $[\beta_L, \beta_R]$ interval, using the boundary information stored in $\sivd_L$ and $\sivd_R$. 
Its arguments are of the same type as those of \textsc{InterFitLin} (described below), 
except that $\bar{\beta}_L$ and $\bar{\beta}_R$ are 5-tuples or real numbers rather than real numbers 
and $\sivd_L$ and $\sivd_R$ are pieces composite SIVD information rather than uncertainty ranges. 
The output structure is entirely the same as that of \textsc{InterFitLin}.
It is essentially a second-level fitting interface for \textsc{InternFitLin}, which is called after carrying out the following two operations:
computing the mean, statistical error and systematic error on each of the two margins of the $\beta$ interval, using the right combination of composite fitting information and composite SIVD information (lines 2,3);
compressing information into an SIVD uncertainty range for each of the two margins, after choosing the highest among the two errors for each margin.  

\begin{widetext}
	\begin{algorithmic}[1]
    \Statex
    \Function{InternFitLin-2}{$\bar{\beta}_L, \bar{\beta}_R, \sivd_L, \sivd_R, \sivd_{\emp}$}
			\State	$(\sivd_L^{\avg}, \sivd_L^{\stat}, \sivd_L^{\syst}) = \textsc{MeanStatSyst}(\bar{\beta}_L, \sivd_L)$
			\State	$(\sivd_R^{\avg}, \sivd_R^{\stat}, \sivd_R^{\syst}) = \textsc{MeanStatSyst}(\bar{\beta}_R, \sivd_R)$
			\State 	$\sivd_L' = (\sivd_L^{\avg}, \textsc{max}(\sivd_L^{\stat}, \sivd_L^{\syst}))$
			\State	$\sivd_R' = (\sivd_R^{\avg}, \textsc{max}(\sivd_R^{\stat}, \sivd_R^{\syst}))$
			\State \Return{\textsc{InternFitLin}($\beta_L, \beta_R, \sivd_L', \sivd_R', \sivd_{\emp}$)}   
    \EndFunction
  \end{algorithmic}
\end{widetext}

Function \textsc{Match-1} checks whether $\aivd$ (real value) falls within the uncertainty range specified by $\aivd_\emp$.
It acts as an interface for \textsc{Match} (described below) within the first-level fitting scheme.

\begin{widetext}
  \begin{algorithmic}[1]
    \Statex
    \Function{Match-1}{$\aivd, \aivd_{\emp}$}
			\State $\aivd' = (\aivd, 0)$
			\State \Return{\textsc{Match}($\aivd', \aivd_{\emp})$}
    \EndFunction
  \end{algorithmic}
\end{widetext}
  
Function \textsc{Match-2} checks whether there is an overlap between the model SIVD uncertainty range obtained from $\bar{\beta}$ (composite fitting information) and $\sivd$ (composite SIVD information) and the empirical one encoded by $\sivd_\emp$.
It acts as an interface for \textsc{Match} (described below) within the second-level fitting scheme.
  
\begin{widetext}  
  \begin{algorithmic}[1]
    \Statex
    \Function{Match-2}{$\bar{\beta},\sivd,\sivd_{\emp}$}
			\State $(\sivd_{\avg}, \sivd_{\stat}, \sivd_{\syst}) = \textsc{MeanStatSyst}(\bar{\beta},\sivd)$
			\State $\sivd' = (\sivd_{\avg}, \textsc{max}(\sivd_{\stat}, \sivd_{\syst}))$
			\State \Return{\textsc{Match}$(\sivd', \sivd_{\emp})$}
    \EndFunction
  \end{algorithmic}
\end{widetext}  
  
Function \textsc{Ord-1} (first version) checks whether $\aivd_L$ (real value) is smaller than $\aivd_R$ (real value), acting as an interface for \textsc{Ord} within the first-level fitting scheme.
  
\begin{widetext}  
  \begin{algorithmic}[1]
    \Statex
    \Function{Ord-1}{$\aivd_L, \aivd_R$}
			\State \Return{\textsc{Ord}$(\aivd_L, \aivd_R)$}
    \EndFunction
  \end{algorithmic}
\end{widetext}  
  
Function \textsc{Ord-1} (second version) checks whether $\aivd$ (real value) is smaller than the average stored in $\aivd_\emp$ (uncertainty range), 
acting as an interface for \textsc{Ord} within the first-level fitting scheme.

\begin{widetext}  
  \begin{algorithmic}[1]
    \Statex
    \Function{Ord-1}{$\aivd, \aivd_{\emp}$}
			\State $(\aivd_\emp^\avg, \aivd_\emp^\err) = \aivd_{\emp}$
			\State \Return{\textsc{Ord}$(\aivd, \aivd_\emp^\avg)$}
    \EndFunction
  \end{algorithmic}
\end{widetext}  

Function \textsc{Ord-2} (first version) checks whether 
the average stored in the SIVD uncertainty range obtained from $\bar{\beta}_L$ (composite fitting information) and $\sivd_L$ (composite SIVD information) is smaller than 
the average stored in that obtained from $\bar{\beta}_R$ (composite fitting information) and $\sivd_R$ (composite SIVD information),
acting as an interface for \textsc{Ord} within the second-level fitting scheme.
  
\begin{widetext}  
  \begin{algorithmic}[1]
    \Statex
    \Function{Ord-2}{$\bar{\beta}_L, \bar{\beta}_R, \sivd_L, \sivd_R$}
			\State $(\sivd_L^{\avg}, \sivd_L^{\stat}, \sivd_L^{\syst}) = \textsc{MeanStatSyst}(\bar{\beta}_L, \sivd_L)$
			\State $(\sivd_R^{\avg}, \sivd_R^{\stat}, \sivd_R^{\syst}) = \textsc{MeanStatSyst}(\bar{\beta}_R, \sivd_R)$
			\State \Return{\textsc{Ord}$(\sivd_L^{\avg}, \sivd_R^{\avg})$}
    \EndFunction
  \end{algorithmic}
\end{widetext}  
  
Function \textsc{Ord-2} (second version) checks whether 
the average stored in the SIVD uncertainty range obtained from $\bar{\beta}$ (composite fitting information) and $\sivd$ (composite SIVD information) is smaller than 
the average stored $\sivd_{\emp}$,
acting as an interface for \textsc{Ord} within the second-level fitting scheme.
  
\begin{widetext}  
  \begin{algorithmic}[1]
    \Statex
    \Function{Ord-2}{$\bar{\beta}, \sivd, \sivd_{\emp}$}
			\State $(\sivd_{\avg}, \sivd_{\stat}, \sivd_{\syst}) = \textsc{MeanStatSyst}(\bar{\beta}, \sivd)$
			\State $(\aivd_\emp^\avg, \aivd_\emp^\err) = \aivd_{\emp}$
			\State \Return{\textsc{Ord}$(\sivd_{\avg}, \aivd_\emp^\avg)$}
    \EndFunction
  \end{algorithmic}
\end{widetext}  
  
Function \textsc{MeanStatSyst} estimates a mean, a statistical error and a systematic error from a piece of composite fitting information and an associated piece of composite SIVD information, 
which are the two arguments of the function.
It returns the 3-tuple comprising of the three computed real numbers.
Note the decompression of composite fitting information (line 2) and the decompression of composite SIVD information (line 3). 

\begin{widetext}  
  \begin{algorithmic}[1]
		\Statex
		\Function{MeanStatSyst}{$\bar{\beta},\sivd$}
			\State $(\beta, \alpha_L, \alpha_R, \alpha_{\text{fit}}, \alpha_{\err}) = \bar{\beta}$
			\State $(\sivd_L, \sivd_R) = \sivd$
			\State $\sivd' = \textsc{Interpol}(\alpha_L, \alpha_R, \alpha_{\text{fit}}, \sivd_L, \sivd_R)$
			\State $\sivd_{\syst} = \textsc{CompSystErr}(\alpha_L, \alpha_R, \alpha_{\err}, \sivd_L, \sivd_R)$
			\State $(\sivd_{\avg}, \sivd_{\stat}) = \sivd'$
			\State \Return{$(\sivd_{\avg}, \sivd_{\stat}, \sivd_{\syst})$}
		\EndFunction
  \end{algorithmic}
\end{widetext}  
  
The following is a list of functions for which only text explanations are provided in schematic way, sometimes accompanied by figures.   
  
\begin{figure*}
\centering
  \subfigure[]{
		\label{Linear-a}
    \def\svgwidth{8.0cm}
    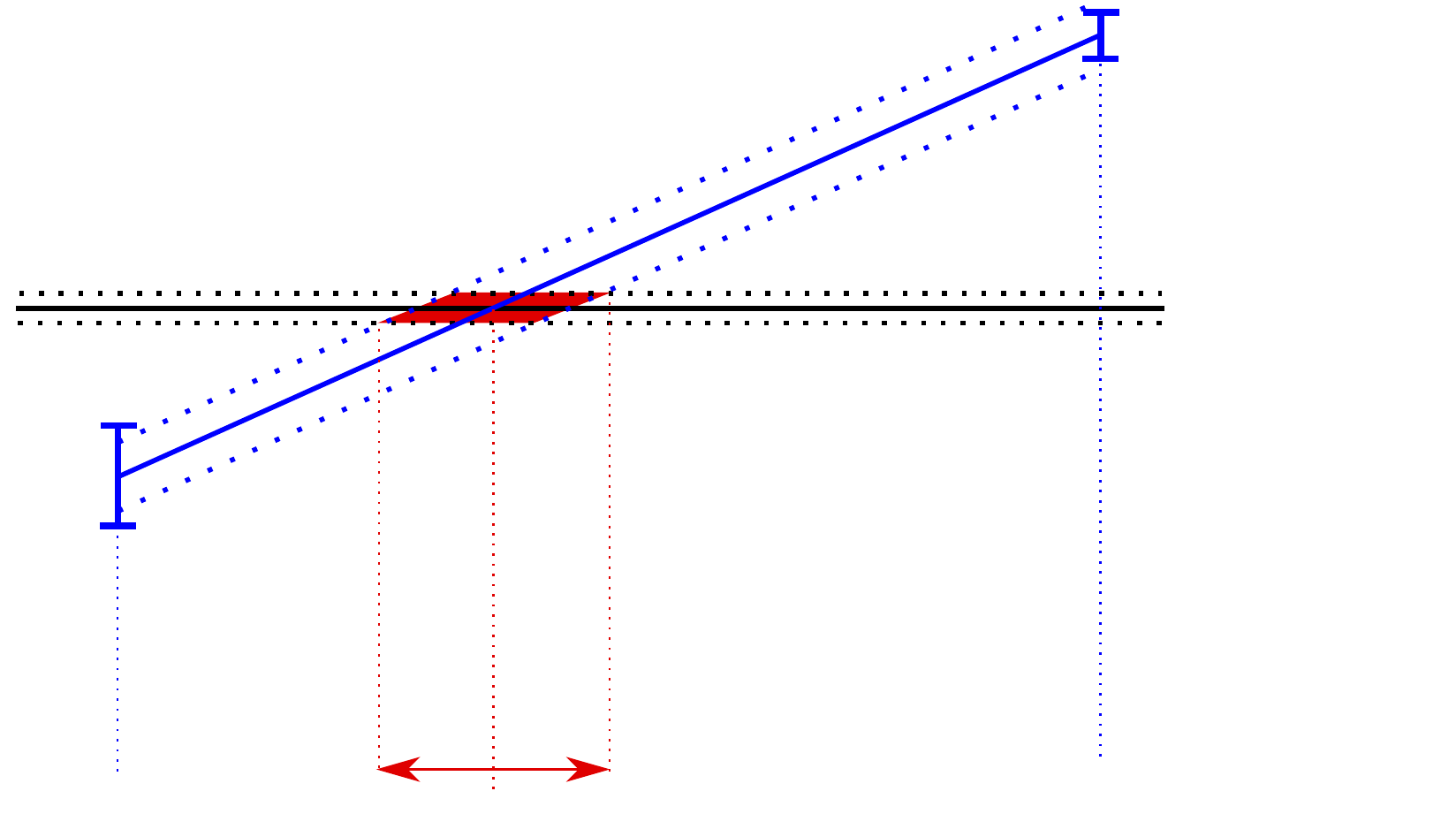
  }
  \subfigure[]{
		\label{Linear-b}
    \def\svgwidth{8.0cm}
    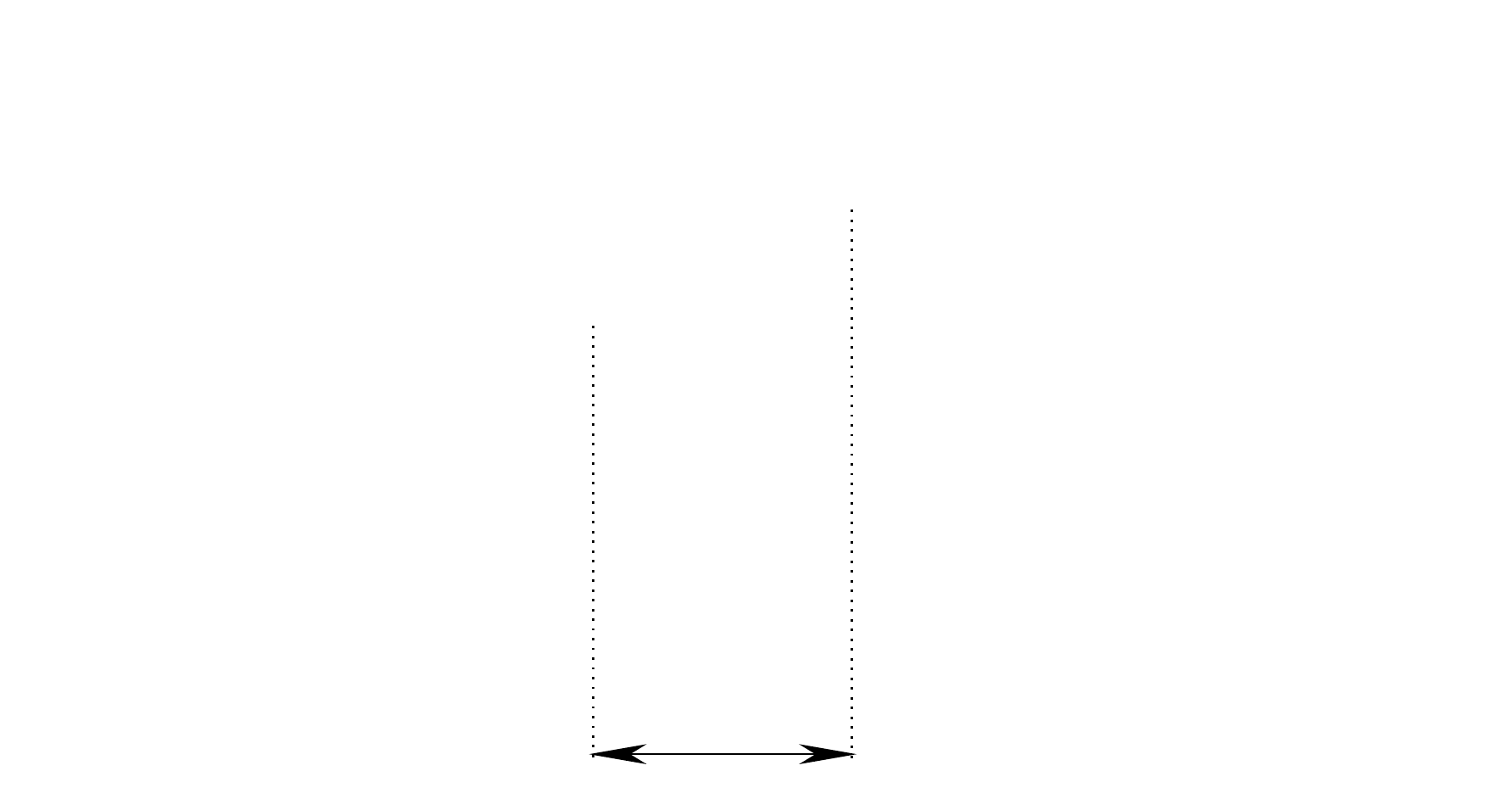 
  }
    \caption{Illustration of computation carried out by \textsc{InterFitLin} \protect\subref{Linear-a} and by \textsc{CompSystErr} \protect\subref{Linear-b}, with the output quantities highlighted in red.}
    \label{Linear} 
\end{figure*}

\begin{widetext}  
\begin{itemize}
  \item $\textsc{Init-1}(\delta\alpha)$: 
		\begin{itemize}
			\item gives the left and right boundaries of the largest possible interval for which the $\alpha$ parameter is compatible with the stochastic model in use, given the grid spacing $\delta\alpha$
			\item input: $\delta$ is a real number
			\item in practice it returns $(\delta\alpha, 1-\delta\alpha)$ regardless of whether PG or MPG is used  
		\end{itemize}
  \item $\textsc{Init-2}(\delta\beta, k, \aivd_{\emp})$: 
		\begin{itemize}
			\item gives the left and right boundaries of the largest possible interval, if any, for which the $\beta$ parameter allows for the (first level) fitting of $\aivd(\alpha)$ to successfully take place, given the grid spacing $\delta\beta$
			\item input: $\delta\beta$ is a real number, $k$ is a positive integer and $\aivd_\emp$ is an uncertainty range
			\item assumes that there exists at most one $\beta$ interval $[\beta_L,\beta_R]$ for which there exists an $\alpha$ such that $\langle\aivd\rangle_{\alpha,\beta}^k = \aivd_\emp$ is satisfied
			\item starts from the largest interval allowed by the model and independently adjusts each of the two boundaries via a branching algorithm, until the desired interval is reached
			\item returns two (incompatible) boundaries $\beta_L > \beta_R$ if such an interval does not exist
		\end{itemize}
  \item $\textsc{Middle}(l,r,\delta)$: 
    \begin{itemize}
        \item computes the value closest to the average between $l$ and $r$, on a grid of spacing $\delta$
        \item input: $l,r,\delta$ are all real numbers
        \item assumes that the interval length $l-r$ is equal to an integer times $\delta$
    \end{itemize}
    
  \item $\textsc{Distinct}(m,l,r)$:
    \begin{itemize}
        \item checks whether $m$ is different than both $l$ and $r$
        \item input: $m,l,r$ are all real numbers constrained constrained to a grid of constant spacing
    \end{itemize}
      
  \item $\textsc{InternFitLin}(p_L, p_R, O_L, O_R, O_{\emp})$:
		\begin{itemize}
			\item adjusts a parameter $p$ such that an observable $O$ attains a value compatible with the empirical in $O_{\emp}$ interval, assuming that $O$ is a linear function of $p$ within the $[p_L,p_R]$ interval
			\item input: 
				$p_L, p_R$ are real numbers, encoding the left and right boundaries of the interval; 
				$O_L$, $O_R$ are mean-error pairs of real numbers encoding the theoretical uncertainty ranges of the observable for the left and for the right boundaries;
				$O_{\emp}$ is a mean-error pair of real numbers encoding the empirical uncertainty range
			\item returns the value and associated error of the $p$ parameter resulting from this fitting process $(p_{\text{fit}}, p_{\err})$,
			computed based on geometrical considerations, in the manner illustrated in Fig.~\ref{Linear-a}
			\item $p_{\text{fit}}$ is calculated first by intersecting the theoretical line with the empirical one, disregarding all errors; 
			then, $p_{\text{err}}$ is calculated by assuming that the theoretical error is constant within the $[p_L,p_R]$ interval, with value given by interpolating the errors contained by $O_L$ and $O_R$ at $p_{\text{fit}}$
			\item $p_{\err}$ takes its origin both in the the empirical error as well as in the theoretical error, but also depends on the slope resulting from the linear approximation
		\end{itemize}
		
  \item $\textsc{Match}(r_1, r_2)$:
		\begin{itemize}
			\item checks whether there is an overlap between the uncertainty ranges encoded by $r_1$ and $r_2$
			\item input: $r_1, r_2$ are mean-error pairs of real numbers
		\end{itemize}
  
  \item $\textsc{Ord}(v_L, v_R)$:
		\begin{itemize}
			\item checks whether the condition $v_L < v_R$ is satisfied
			\item input: $v_L, v_R$ are real numbers
			\item assumes that $v_L \neq v_R$
		\end{itemize}
  \item $\textsc{GenSeqSIVD}(\alpha, \beta, k, n)$
		\begin{itemize}
			\item numerically generates a sequence of $n$ $\sivd$ values according to the respective stochastic model, subject to parameter values indicated by $k, \alpha, \beta$
			\item input: $\alpha, \beta$ are real numbers, while $k,n$ are positive integers 
		\end{itemize}
  \item $\textsc{Merge}(\sivd_1^{\text{seq}}, \sivd_2^{\text{seq}})$
		\begin{itemize}
			\item merges two sequences of (real) $\sivd$ values
			\item input: $\sivd_1^{\text{seq}}, \sivd_2^{\text{seq}}$ are both sequences of (real) $\sivd$ values
		\end{itemize}
  \item $\textsc{CompAvgErr}(\sivd^{\text{seq}})$
		\begin{itemize}
			\item computes the mean and standard error of the mean from $\sivd^{\text{seq}}$
			\item input: $\sivd^{\text{seq}}$ is a sequence of real $\sivd$ values
		\end{itemize}
  \item $\textsc{Interpol}(\alpha_L, \alpha_R, \alpha_{\text{fit}}, \sivd_{L}, \sivd_{R})$
		\begin{itemize}
			\item estimates the mean and error in $\sivd$ corresponding to $\alpha_{\text{fit}}$ based on the values attained for $\alpha_L$
			\item input: $\alpha_L, \alpha_R, \alpha_{\text{fit}}$ are real numbers, while $\sivd_{L}, \sivd_{R}$ are mean-error pairs of real numbers
			\item uses on a linear interpolation within the $[\alpha_L, \alpha_R]$ interval, separately for the mean and for the error
		\end{itemize}
  \item $\textsc{CompSystErr}(\alpha_L, \alpha_R, \alpha_{\err}, \sivd_{L}, \sivd_{R})$
		\begin{itemize}
			\item estimates the systematic error $\sivd^\syst$ of the SIVD quantity induced by the error $\alpha_{\text{err}}$ (associated to fitting the $\alpha$ parameter in terms of the AIVD quantity), assuming that SIVD is a linear function of $\alpha$ within the $[\alpha_L, \alpha_R]$ interval 
			\item input: $\alpha_L, \alpha_R, \alpha_{\text{err}}$ are real numbers while while $\sivd_{L}, \sivd_{R}$ are mean-error pairs of real numbers encoding the theoretical uncertainty ranges on the left and right boundaries
			\item $\sivd^\syst$ is computed based on geometrical considerations, in the manner illustrated in Fig.~\ref{Linear-b}
		\end{itemize}
\end{itemize}
\end{widetext}  

\subsection{Algorithm usage}
\label{AppFormUsg}

This section explain how the formalism presented throughout this document is effectively used for producing the results shown in Sections \ref{ModFit} and \ref{ModOut}.

First, the formalism is used for producing the plots showing the SIVD$(\beta)$ dependence (``Model fitting" section). 
For either PG or MPG, for a specific $k$ value and a specific $\beta$ on-grid value, the drawn model SIVD uncertainty range is obtained after the following computational steps:

\begin{widetext}  
\begin{algorithmic}[1]
	\Statex
	\State $(\alpha_L, \alpha_R, \alpha_{\text{fit}}, \alpha_{\text{err}}) = \textsc{Fit-1}(\beta,k)$	\Comment{executing 1st-level fitting}
	\State $\bar{\beta} = (\beta, \alpha_L, \alpha_R, \alpha_{\text{fit}}, \alpha_{\err})$			\Comment{creating composite fitting information}
	\State $\sivd = \textsc{NumSIVD}(\bar{\beta}, k)$							\Comment{numeric SIVD calculations}
	\State $(\sivd_{\text{avg}}, \sivd_{\text{stat}}, \sivd_{\text{syst}}) = \textsc{MeanStatSyst}(\bar{\beta}, \sivd)$
\end{algorithmic}
\end{widetext}  
which provides the values of the SIVD average $\sivd_{\text{avg}}$, the SIVD statistical error $\sivd_{\text{stat}}$ and the SIVD systematic error $\sivd_{\text{syst}}$.
One can then place a point at coordinates $(\beta,\sivd_{\text{avg}})$, within the respective $k$ curve, 
with an error bar given by the maximum between $\sivd_{\text{stat}}$ and $\sivd_{\text{syst}}$.

Second, the formalism is used for providing the best-fitting, on-grid values for the $\alpha$ and $\beta$ model parameters, 
which are used for generating sets of cultural vectors on which the LTCD-STCB analysis is applied (``Model outcomes" section). 
For either PG or MPG and for a specific $k$ value, the following procedure is followed:

\begin{widetext}  
\begin{algorithmic}[1]
	\Statex
	\State $(\bar{\beta}_L, \bar{\beta}_R, \beta_{\text{fit}}, \beta_{\text{err}}) = \textsc{Fit-2}(k)$	\Comment{Executing 2nd-level fitting}
	\State $(\beta_L, \alpha_L^L, \alpha_L^R, \alpha_L^{\text{fit}}, \alpha_L^{\text{err}}) = \bar{\beta}_L$ 	\Comment{Decompressing left-$\beta$ composite fitting information}
	\State $(\beta_R, \alpha_R^L, \alpha_R^R, \alpha_R^{\text{fit}}, \alpha_R^{\text{err}}) = \bar{\beta}_R$ 	\Comment{Decompressing right-$\beta$ composite fitting information}
	  \If{$\beta_{\text{fit}}-\beta_L < \beta_R - \beta_{\text{fit}}$}	
	    \State $\beta = \beta_L$			\Comment{choosing $\beta_L$, since it is closer to $\beta$}   
	    \If{$\alpha_L^{\text{fit}} - \alpha_L^L < \alpha_L^R - \alpha_L^{\text{fit}}$}	
	      \State $\alpha = \alpha_L^L$			\Comment{choosing $\alpha_L^L$, since it is closer to $\alpha_L^{\text{fit}}$}   
	    \Else
	      \State $\alpha = \alpha_L^R$			\Comment{choosing $\alpha_L^R$, since it is closer to $\alpha_L^{\text{fit}}$}
	    \EndIf
	  \Else
	    \State $\beta = \beta_R$			\Comment{choosing $\beta_R$, since it is closer to $\beta$}
	    \If{$\alpha_R^{\text{fit}} - \alpha_R^L < \alpha_R^R - \alpha_R^{\text{fit}}$}	
	      \State $\alpha = \alpha_R^L$			\Comment{choosing $\alpha_R^L$, since it is closer to $\alpha_R^{\text{fit}}$}   
	    \Else
	      \State $\alpha = \alpha_R^R$			\Comment{choosing $\alpha_R^R$, since it is closer to $\alpha_R^{\text{fit}}$}
	    \EndIf
	  \EndIf
\end{algorithmic}
\end{widetext}  
which provides the best on-grid values for the $(\alpha,\beta)$ pair.

%% file: InterFitLin.pdf_tex
%% Creator: Inkscape inkscape 0.48.4, www.inkscape.org
%% PDF/EPS/PS + LaTeX output extension by Johan Engelen, 2010
%% Accompanies image file 'InterFitLin.pdf' (pdf, eps, ps)
%%
%% To include the image in your LaTeX document, write
%%   \input{<filename>.pdf_tex}
%%  instead of
%%   \includegraphics{<filename>.pdf}
%% To scale the image, write
%%   \def\svgwidth{<desired width>}
%%   \input{<filename>.pdf_tex}
%%  instead of
%%   \includegraphics[width=<desired width>]{<filename>.pdf}
%%
%% Images with a different path to the parent latex file can
%% be accessed with the `import' package (which may need to be
%% installed) using
%%   \usepackage{import}
%% in the preamble, and then including the image with
%%   \import{<path to file>}{<filename>.pdf_tex}
%% Alternatively, one can specify
%%   \graphicspath{{<path to file>/}}
%% 
%% For more information, please see info/svg-inkscape on CTAN:
%%   http://tug.ctan.org/tex-archive/info/svg-inkscape
%%
\begingroup%
  \makeatletter%
  \providecommand\color[2][]{%
    \errmessage{(Inkscape) Color is used for the text in Inkscape, but the package 'color.sty' is not loaded}%
    \renewcommand\color[2][]{}%
  }%
  \providecommand\transparent[1]{%
    \errmessage{(Inkscape) Transparency is used (non-zero) for the text in Inkscape, but the package 'transparent.sty' is not loaded}%
    \renewcommand\transparent[1]{}%
  }%
  \providecommand\rotatebox[2]{#2}%
  \ifx\svgwidth\undefined%
    \setlength{\unitlength}{474.58534104bp}%
    \ifx\svgscale\undefined%
      \relax%
    \else%
      \setlength{\unitlength}{\unitlength * \real{\svgscale}}%
    \fi%
  \else%
    \setlength{\unitlength}{\svgwidth}%
  \fi%
  \global\let\svgwidth\undefined%
  \global\let\svgscale\undefined%
  \makeatother%
  \begin{picture}(1,0.55831323)%
    \put(0,0){\includegraphics[width=\unitlength]{InterFitLin.pdf}}%
    \put(-0.00558945,0.24261821){\color[rgb]{0,0,1}\makebox(0,0)[lt]{\begin{minipage}{0.21071023\unitlength}\raggedright $O_L$\end{minipage}}}%
    \put(0.7861037,0.55933827){\color[rgb]{0,0,1}\makebox(0,0)[lt]{\begin{minipage}{0.22858958\unitlength}\raggedright $O_R$\end{minipage}}}%
    \put(0.26910874,0.16714774){\color[rgb]{0.8745098,0,0}\makebox(0,0)[lt]{\begin{minipage}{0.08818147\unitlength}\raggedright $p_{\text{err}}$\end{minipage}}}%
    \put(0.35002147,0.16714774){\color[rgb]{0.8745098,0,0}\makebox(0,0)[lt]{\begin{minipage}{0.08818147\unitlength}\raggedright $p_{\text{err}}$\end{minipage}}}%
    \put(0.31630783,0.012065){\color[rgb]{0.8745098,0,0}\makebox(0,0)[lt]{\begin{minipage}{0.08818147\unitlength}\raggedright $p_{\text{fit}}$\end{minipage}}}%
    \put(0.06345553,0.01543636){\color[rgb]{0,0,1}\makebox(0,0)[lt]{\begin{minipage}{0.03098871\unitlength}\raggedright $p_L$\end{minipage}}}%
    \put(0.7410997,0.01543636){\color[rgb]{0,0,1}\makebox(0,0)[lt]{\begin{minipage}{0.03098871\unitlength}\raggedright $p_R$\end{minipage}}}%
    \put(0.5305249,0.32883298){\color[rgb]{0,0,0}\makebox(0,0)[lt]{\begin{minipage}{0.242784\unitlength}\raggedright $O_{\text{emp}}$\end{minipage}}}%
  \end{picture}%
\endgroup%

%% file: CompSystErr.pdf_tex
%% Creator: Inkscape inkscape 0.91, www.inkscape.org
%% PDF/EPS/PS + LaTeX output extension by Johan Engelen, 2010
%% Accompanies image file 'CompSystErr.pdf' (pdf, eps, ps)
%%
%% To include the image in your LaTeX document, write
%%   \input{<filename>.pdf_tex}
%%  instead of
%%   \includegraphics{<filename>.pdf}
%% To scale the image, write
%%   \def\svgwidth{<desired width>}
%%   \input{<filename>.pdf_tex}
%%  instead of
%%   \includegraphics[width=<desired width>]{<filename>.pdf}
%%
%% Images with a different path to the parent latex file can
%% be accessed with the `import' package (which may need to be
%% installed) using
%%   \usepackage{import}
%% in the preamble, and then including the image with
%%   \import{<path to file>}{<filename>.pdf_tex}
%% Alternatively, one can specify
%%   \graphicspath{{<path to file>/}}
%% 
%% For more information, please see info/svg-inkscape on CTAN:
%%   http://tug.ctan.org/tex-archive/info/svg-inkscape
%%
\begingroup%
  \makeatletter%
  \providecommand\color[2][]{%
    \errmessage{(Inkscape) Color is used for the text in Inkscape, but the package 'color.sty' is not loaded}%
    \renewcommand\color[2][]{}%
  }%
  \providecommand\transparent[1]{%
    \errmessage{(Inkscape) Transparency is used (non-zero) for the text in Inkscape, but the package 'transparent.sty' is not loaded}%
    \renewcommand\transparent[1]{}%
  }%
  \providecommand\rotatebox[2]{#2}%
  \ifx\svgwidth\undefined%
    \setlength{\unitlength}{513.42464741bp}%
    \ifx\svgscale\undefined%
      \relax%
    \else%
      \setlength{\unitlength}{\unitlength * \real{\svgscale}}%
    \fi%
  \else%
    \setlength{\unitlength}{\svgwidth}%
  \fi%
  \global\let\svgwidth\undefined%
  \global\let\svgscale\undefined%
  \makeatother%
  \begin{picture}(1,0.52808055)%
    \put(0,0){\includegraphics[width=\unitlength,page=1]{CompSystErr.pdf}}%
    \put(0.43353291,0.06874577){\color[rgb]{0,0,0}\makebox(0,0)[lt]{\begin{minipage}{0.19955955\unitlength}\raggedright $\alpha_{\text{err}}$\end{minipage}}}%
    \put(0,0){\includegraphics[width=\unitlength,page=2]{CompSystErr.pdf}}%
    \put(-0.00178843,0.22471298){\color[rgb]{0,0,1}\makebox(0,0)[lt]{\begin{minipage}{0.18426836\unitlength}\raggedright $\text{SIVD}_L$\end{minipage}}}%
    \put(0,0){\includegraphics[width=\unitlength,page=3]{CompSystErr.pdf}}%
    \put(0.14795645,0.01375309){\color[rgb]{0,0,1}\makebox(0,0)[lt]{\begin{minipage}{0.07238281\unitlength}\raggedright $\alpha_L$\end{minipage}}}%
    \put(0.82003203,0.51052485){\color[rgb]{0,0,1}\makebox(0,0)[lt]{\begin{minipage}{0.18426835\unitlength}\raggedright $\text{SIVD}_R$\end{minipage}}}%
    \put(0,0){\includegraphics[width=\unitlength,page=4]{CompSystErr.pdf}}%
    \put(0.77122222,0.01709199){\color[rgb]{0,0,1}\makebox(0,0)[lt]{\begin{minipage}{0.07238275\unitlength}\raggedright $\alpha_R$\end{minipage}}}%
    \put(0,0){\includegraphics[width=\unitlength,page=5]{CompSystErr.pdf}}%
    \put(0.14371434,0.37659453){\color[rgb]{0.91372549,0,0}\makebox(0,0)[lt]{\begin{minipage}{0.1394589\unitlength}\raggedright $\text{SIVD}^{\text{syst}}$\end{minipage}}}%
    \put(0,0){\includegraphics[width=\unitlength,page=6]{CompSystErr.pdf}}%
  \end{picture}%
\endgroup%